\documentclass[a4paper]{article}
\usepackage{amsmath, amsfonts, amsthm,amssymb}
\usepackage{mathrsfs}
\usepackage[margin=2cm]{geometry}
\usepackage{xcolor}
\usepackage{colortbl}
\usepackage{graphicx}
\usepackage{caption}
\usepackage{subfig}
\usepackage{multirow,makecell,multicol}
\usepackage{hhline}
\usepackage{stackengine}
\usepackage{lscape}
\usepackage[T1]{fontenc}

\newcommand\xrowht[2][0]{\addstackgap[.5\dimexpr#2\relax]{\vphantom{#1}}}
\allowdisplaybreaks
\makeatletter
\renewcommand*{\@fnsymbol}[1]{\ifcase#1 \or \@arabic{\numexpr#1 \relax}  \or \@arabic{\numexpr#1 \relax} \or \@arabic{\numexpr#1 \relax} \else \ * \fi}
\makeatother
\input{tcilatex}
\begin{document}

\title{Wave propagation in three-dimensional \\
fractional viscoelastic infinite solid body}
\author{Sla\dj an Jeli\'{c}\thanks{
Department of Physics, Faculty of Sciences, University of Novi Sad, Trg D.
Obradovi\'{c}a 4, 21000 Novi Sad, Serbia, sladjan.jelic@df.uns.ac.rs} \quad
Du\v{s}an Zorica\thanks{
Department of Physics, Faculty of Sciences, University of Novi Sad, Trg D.
Obradovi\'{c}a 4, 21000 Novi Sad, Serbia, dusan.zorica@df.uns.ac.rs} \ 
\thanks{%
Mathematical Institute, Serbian Academy of Arts and Sciences, Kneza Mihaila
36, 11000 Belgrade, Serbia, dusan\textunderscore zorica@mi.sanu.ac.rs} \ 
\thanks{%
Corresponding author}}
\maketitle

\begin{abstract}
\noindent In order to analyze the wave propagation in three-dimensional
isotropic and viscoelastic body, the Cauchy initial value problem on
unbounded domain is considered for the wave equation written as a system of
fractional partial differential equations consisting of equation of motion
of three-dimensional solid body, equation of strain, as well as of the
constitutive equation, obtained by generalizing the classical Hooke's law of
three-dimensional isotropic and elastic body by replacing Lam\'{e}
coefficients with the relaxation moduli to account for different memory
kernels corresponding to the propagation of compressive and shear waves. By
the use of the method of integral transforms, namely Laplace and Fourier
transforms, the displacement field, as a solution of the Cauchy initial
value problem, is expressed through Green's functions corresponding to both
compressive and shear wave propagation. Two approaches are adopted in
solving the Cauchy problem: in the first one the displacement field is
expressed through the scalar and vector fields, obtained as solutions of
wave equations which are consequences of decoupling the wave equation for
displacement field, while in the second approach the displacement field is
obtained by the action of the resolvent tensor on the initial conditions.
Fractional anti-Zener and Zener model I$^{{}^{+}}$ID.ID, as well as
fractional Burgers model VII, as representatives of models yielding infinite
and finite wave propagation speed, are used in numerical calculations and graphical representations of
Green's functions and its short and large time asymptotic behavior.

\end{abstract}

\section{Introduction}

The wave equation for three-dimensional isotropic and elastic body, written
in terms of displacement field $\boldsymbol{u}=\boldsymbol{u}\left( 
\boldsymbol{r},t\right) ,$ which is calculated at the point determined by
the radius vector $\boldsymbol{r}\in 
\mathbb{R}
^{3}$ at time instant $t>0,$ takes the form%
\begin{equation}
\frac{\mu }{\varrho }\Delta \boldsymbol{u}\left( \boldsymbol{r},t\right) +%
\frac{\lambda +\mu }{\varrho }\func{grad}\func{div}\boldsymbol{u}\left( 
\boldsymbol{r},t\right) +\boldsymbol{f}_{b}\left( \boldsymbol{r},t\right)
=\partial _{tt}^{2}\boldsymbol{u}\left( \boldsymbol{r},t\right) ,  \label{we}
\end{equation}%
representing an equation obtained by reduction of the system of equations
consisting of the equation of motion of deformable solid body%
\begin{equation}
\frac{1}{\varrho }\func{div}\boldsymbol{\hat{\sigma}}\left( \boldsymbol{r}%
,t\right) +\boldsymbol{f}_{b}\left( \boldsymbol{r},t\right) =\partial
_{tt}^{2}\boldsymbol{u}\left( \boldsymbol{r},t\right) ,  \label{EoM}
\end{equation}%
equation of infinitesimal strain%
\begin{equation}
\boldsymbol{\hat{\varepsilon}}\left( \boldsymbol{r},t\right) =\frac{1}{2}%
\left( \func{grad}\boldsymbol{u}\left( \boldsymbol{r},t\right) +\left( \func{%
grad}\boldsymbol{u}\left( \boldsymbol{r},t\right) \right) ^{\mathrm{T}%
}\right) ,  \label{strain}
\end{equation}%
and Hooke's law%
\begin{equation}
\boldsymbol{\hat{\sigma}}\left( \boldsymbol{r},t\right) =\lambda \mathrm{%
\limfunc{tr}}\boldsymbol{\hat{\varepsilon}}\left( \boldsymbol{r},t\right) 
\boldsymbol{\hat{I}}+2\mu \boldsymbol{\hat{\varepsilon}}\left( \boldsymbol{r}%
,t\right)  \label{hz}
\end{equation}%
as a constitutive equation, where $\lambda$ and $\mu $ are Lam\'{e}
coefficients corresponding to an isotropic and elastic body, $\varrho $ is
material density, while $\boldsymbol{f}_{b}$ are body forces per unit mass, $%
\boldsymbol{\hat{\sigma}}$ is the Cauchy stress tensor, $\boldsymbol{\hat{%
\varepsilon}}$ is the infinitesimal strain tensor, and $\boldsymbol{\hat{I}}$
is the identity tensor. The wave equation (\ref{we}), rewritten as 
\begin{equation}
c_{c}^{2}\func{grad}\func{div}\boldsymbol{u}\left( \boldsymbol{r},t\right)
-c_{s}^{2}\func{curl}\func{curl}\boldsymbol{u}\left( \boldsymbol{r},t\right)
+\boldsymbol{f}_{b}\left( \boldsymbol{r},t\right) =\partial _{tt}^{2}%
\boldsymbol{u}\left( \boldsymbol{r},t\right) ,  \label{we-rastavljen}
\end{equation}%
in terms of speeds of compressive and shear waves $c_{c}=\sqrt{\frac{\lambda
+2\mu }{\varrho }}$ and $c_{s}=\sqrt{\frac{\mu }{\varrho }}$ respectively,
decomposes by the application of divergence and curl operators into two
three-dimensional wave equations%
\begin{align}
c_{c}^{2}\,\Delta \varphi \left( \boldsymbol{r},t\right) -\partial
_{tt}^{2}\varphi \left( \boldsymbol{r},t\right) & =-\func{div}\boldsymbol{f}%
_{b}\left( \boldsymbol{r},t\right) ,\quad \text{with}\quad \varphi \left( 
\boldsymbol{r},t\right) =\func{div}\boldsymbol{u}\left( \boldsymbol{r}%
,t\right) ,  \label{we-fi} \\
c_{s}^{2}\,\Delta \boldsymbol{\Omega }\left( \boldsymbol{r},t\right)
-\partial _{tt}^{2}\boldsymbol{\Omega }\left( \boldsymbol{r},t\right) & =-%
\func{curl}\boldsymbol{f}_{b}\left( \boldsymbol{r},t\right) ,\quad \text{with%
}\quad \boldsymbol{\Omega }\left( \boldsymbol{r},t\right) =\func{curl}%
\boldsymbol{u}\left( \boldsymbol{r},t\right) ,  \label{we-omega}
\end{align}%
governing the propagation of compressive and shear waves respectively, see 
\cite{Arci}.

The reflection of three-dimensional elastic waves is considered in \cite%
{Wang-Chena-Wei-Li-2017-da} by analyzing the wave equations for both scalar
and vector fields in the case when there are no body forces, while
three-dimensional elastic wave propagation is analyzed in \cite%
{Meng-Deng-Xu-Zhang-Hou-2015-da,Ponnusamy-Rajagopal-2011-da,LimeiXua-2017-da}
using cylindrical coordinates. Further considerations of wave propagation in
periodic media are contained in \cite%
{ChenYueshengWangZhengdaoWang-2008-da,Dorn-Kochmann-2020-da,Drugan-2020-da}.

Coupling the mechanical and thermal effects by considering wave propagation
in three-dimensional thermo-elastic media is conducted in \cite%
{Singh-Tomar-2007-da} by examining the propagation of plane-waves and
considering the wave equations for scalar and vector fields, while in \cite%
{Yazdan-Morteza-2018-da} the wave propagation is considered in
thermo-elastic cylindrical media. Further, the reflection of
three-dimensional plane waves in a highly anisotropic thermo-elastic
material is analyzed in \cite{SINGH-SINGH-CHATTOPADHYAY-2021-da}. Numerical
methods are used in \cite{Ayzenberg-Stepanenkoa-2012-da} and \cite%
{Florent-Christophe-2021-da} for description of shock wave propagation, as
well as for three-dimensional elastic wave propagation on unbounded domain.

Aiming to generalize the classical three-dimensional wave equation for
isotropic and elastic body, given by (\ref{we}) or equivalently by (\ref%
{we-rastavljen}), to the three-dimensional wave equation for isotropic and
viscoelastic body, the approach arising from the considerations of
one-dimensional viscoelastic body is adopted, where the stress, instead by
Hooke's law $\sigma =E\varepsilon ,$ with $E$ being the Young modulus, can
be expressed in terms of strain through the relaxation modulus $\sigma _{sr}$
as%
\begin{equation}
\sigma \left( t\right) =\frac{\mathrm{d}}{\mathrm{d}t}\left( \sigma
_{sr}\left( t\right) \ast _{t}\varepsilon \left( t\right) \right) =\frac{%
\mathrm{d}}{\mathrm{d}t}\int_{0}^{t}\sigma _{sr}\left( t-t^{\prime }\right)
\varepsilon \left( t^{\prime }\right) \mathrm{d}t^{\prime },
\label{sigma=sr*epsilon}
\end{equation}%
where $\ast _{t}$ denotes the convolution in time, thus having different
material properties of viscoelastic body accounted by the choice of
different relaxation moduli. Note, the relaxation modulus is a function
accounting for the time evolution of stress having the strain in the
constitutive equation assumed as the Heaviside step function. Namely, the
classical Hooke's law (\ref{hz}), being of the form%
\begin{gather*}
\sigma _{11}=\left( \lambda +2\mu \right) \varepsilon _{11}+\lambda
\varepsilon _{22}+\lambda \varepsilon _{33},\quad \text{i.e.,}\quad \frac{1}{%
\varrho }\sigma _{11}=c_{c}^{2}\varepsilon _{11}+\left(
c_{c}^{2}-2c_{s}^{2}\right) \varepsilon _{22}+\left(
c_{c}^{2}-2c_{s}^{2}\right) \varepsilon _{33}, \\
\sigma _{22}=\lambda \varepsilon _{11}+\left( \lambda +2\mu \right)
\varepsilon _{22}+\lambda \varepsilon _{33},\quad \text{i.e.,}\quad \frac{1}{%
\varrho }\sigma _{22}=\left( c_{c}^{2}-2c_{s}^{2}\right) \varepsilon
_{11}+c_{c}^{2}\varepsilon _{22}+\left( c_{c}^{2}-2c_{s}^{2}\right)
\varepsilon _{33}, \\
\sigma _{33}=\lambda \varepsilon _{11}+\lambda \varepsilon _{22}+\left(
\lambda +2\mu \right) \varepsilon _{33},\quad \text{i.e.,}\quad \frac{1}{%
\varrho }\sigma _{33}=\left( c_{c}^{2}-2c_{s}^{2}\right) \varepsilon
_{11}+\left( c_{c}^{2}-2c_{s}^{2}\right) \varepsilon
_{22}+c_{c}^{2}\varepsilon _{33}, \\
\sigma _{12}=2\mu \varepsilon _{12},\quad \text{i.e.,}\quad \frac{1}{\varrho 
}\sigma _{12}=2c_{s}^{2}\varepsilon _{12}, \\
\sigma _{13}=2\mu \varepsilon _{13},\quad \text{i.e.,}\quad \frac{1}{\varrho 
}\sigma _{13}=2c_{s}^{2}\varepsilon _{13}, \\
\sigma _{23}=2\mu \varepsilon _{23},\quad \text{i.e.,}\quad \frac{1}{\varrho 
}\sigma _{23}=2c_{s}^{2}\varepsilon _{23},
\end{gather*}%
is generalized by employing two different relaxation moduli $\sigma
_{sr}^{\left( c\right) }$ and $\sigma _{sr}^{\left( s\right) }$ instead of
speeds of compressive and shear waves, $c_{c}$ and $c_{s}$, yielding%
\begin{align*}
\sigma _{11}& =\partial _{t}\left( \sigma _{sr}^{\left( c\right) }\ast
_{t}\varepsilon _{11}\right) +\partial _{t}\left( \sigma _{sr}^{\left(
c,s\right) }\ast _{t}\varepsilon _{22}\right) +\partial _{t}\left( \sigma
_{sr}^{\left( c,s\right) }\ast _{t}\varepsilon _{33}\right) ,\quad \sigma
_{12}=2\partial _{t}\left( \sigma _{sr}^{\left( s\right) }\ast
_{t}\varepsilon _{12}\right) , \\
\sigma _{22}& =\partial _{t}\left( \sigma _{sr}^{\left( c,s\right) }\ast
_{t}\varepsilon _{11}\right) +\partial _{t}\left( \sigma _{sr}^{\left(
c\right) }\ast _{t}\varepsilon _{22}\right) +\partial _{t}\left( \sigma
_{sr}^{\left( c,s\right) }\ast _{t}\varepsilon _{33}\right) ,\quad \sigma
_{13}=2\partial _{t}\left( \sigma _{sr}^{\left( s\right) }\ast
_{t}\varepsilon _{13}\right) , \\
\sigma _{33}& =\partial _{t}\left( \sigma _{sr}^{\left( c,s\right) }\ast
_{t}\varepsilon _{11}\right) +\partial _{t}\left( \sigma _{sr}^{\left(
c,s\right) }\ast _{t}\varepsilon _{22}\right) +\partial _{t}\left( \sigma
_{sr}^{\left( c\right) }\ast _{t}\varepsilon _{33}\right) ,\quad \sigma
_{23}=2\partial _{t}\left( \sigma _{sr}^{\left( s\right) }\ast
_{t}\varepsilon _{23}\right) ,
\end{align*}%
with $\sigma _{sr}^{\left( c,s\right) }=\sigma _{sr}^{\left( c\right)
}-2\sigma _{sr}^{\left( s\right) },$ and reducing to%
\begin{equation}
\boldsymbol{\hat{\sigma}}\left( \boldsymbol{r},t\right) =\partial _{t}\left(
\sigma _{sr}^{\left( c,s\right) }\left( t\right) \ast _{t}\mathrm{\limfunc{tr%
}}\boldsymbol{\hat{\varepsilon}}\left( \boldsymbol{r},t\right) \right) 
\boldsymbol{\hat{I}}+2\partial _{t}\left( \sigma _{sr}^{\left( s\right)
}\left( t\right) \ast _{t}\boldsymbol{\hat{\varepsilon}}\left( \boldsymbol{r}%
,t\right) \right) ,  \label{VEHook}
\end{equation}%
thus generalizing the characteristics of compressive and shear wave
propagation by taking into account viscous properties of isotropic body
besides the elastic ones, implying different type of compressive and shear
excitation propagation. The constitutive equation of three-dimensional
viscoelastic body (\ref{VEHook}) is further used along with the equation of
motion of deformable solid body (\ref{EoM}) and strain (\ref{strain}) in
order to formulate and solve the corresponding wave equation.

Linear fractional-order models of one-dimensional viscoelastic body, taking
the most general form 
\begin{equation}
\sum_{k=1}^{N_{1}}a_{k}\,_{0}\mathrm{I}_{t}^{\alpha _{k}}\sigma \left(
t\right) +\sum_{k=1}^{N_{2}}b_{k}\,_{0}\mathrm{D}_{t}^{\beta _{k}}\sigma
\left( t\right) =\sum_{k=1}^{M_{1}}c_{k}\,_{0}\mathrm{I}_{t}^{\mu
_{k}}\varepsilon \left( t\right) +\sum_{k=1}^{M_{2}}d_{k}\,_{0}\mathrm{D}%
_{t}^{\nu _{k}}\varepsilon \left( t\right) ,  \label{suma}
\end{equation}%
containing non-negative model parameters $a_{k},b_{k},c_{k},d_{k},$ as well
as fractional integration and differentiation orders $\alpha _{k},\beta
_{k},\mu _{k},\nu _{k},$ that can belong to both intervals $\left(
0,1\right) $ and $\left( 1,2\right) ,$ are among the constitutive models
that yield stress expressed in terms of strain through the relaxation
modulus according to expression (\ref{sigma=sr*epsilon}), where the
fractional integral of order $\mu >0$ and Riemann-Liouville fractional
derivative of order $n+\nu ,$ with $n\in \mathbb{N}_{0}$ and $\nu \in \left(
0,1\right) ,$ appearing in (\ref{suma}), are respectively defined by%
\begin{align*}
{}_{0}\mathrm{I}_{t}^{\mu }f\left( t\right) & =\frac{t^{\mu -1}}{\Gamma
\left( \mu \right) }\ast f\left( t\right) =\frac{1}{\Gamma (\mu )}%
\int_{0}^{t}\frac{f(t^{\prime })}{(t-t^{\prime })^{1-\mu }}\mathrm{d}%
t^{\prime }\quad \text{and} \\
{}_{0}\mathrm{D}_{t}^{n+\nu }f\left( t\right) & =\frac{\mathrm{d}^{n+1}}{%
\mathrm{d}t^{n+1}}\,{}_{0}\mathrm{I}_{t}^{1-\nu }f\left( t\right) =\frac{%
\mathrm{d}^{n+1}}{\mathrm{d}t^{n+1}}\left( \frac{t^{-\nu }}{\Gamma \left(
1-\nu \right) }\ast f\left( t\right) \right) ,
\end{align*}%
see \cite{TAFDE}. Namely, the application of Laplace transform with respect
to time, defined by%
\begin{equation}
\tilde{f}\left( s\right) =\mathcal{L}\left[ f\left( t\right) \right] \left(
s\right) =\int_{0}^{\infty }f\left( t\right) \mathrm{e}^{-st}\,\mathrm{d}%
t,\quad \func{Re}s>0,  \label{LT}
\end{equation}%
according to the Laplace transforms of fractional integral and
Riemann-Liouville fractional derivative 
\begin{align}
\mathcal{L}\left[ {}_{0}\mathrm{I}_{t}^{\mu }f\left( t\right) \right] \left(
s\right) & =\frac{1}{s^{\mu }}\tilde{f}(s),\quad \mu >0\quad \text{and}
\label{LT-frac-int} \\
\mathcal{L}\left[ {}_{0}\mathrm{D}_{t}^{n+\nu }f\left( t\right) \right]
\left( s\right) & =s^{n+\nu }\hat{f}(s)-\sum_{k=0}^{n}\left[ \frac{\mathrm{d}%
^{k}}{\mathrm{d}t^{k}}{}_{0}\mathrm{I}_{t}^{1-\nu }f\left( t\right) \right]
_{t=0}s^{n-k}=s^{n+\nu }\hat{f}(s),\quad \nu \in \left( 0,1\right) ,
\label{LT-frac-der}
\end{align}%
obtained assuming that function $f$ is bounded at zero, transforms the
constitutive model (\ref{suma}) into%
\begin{gather}
\Phi _{\sigma }\left( s\right) \tilde{\sigma}\left( s\right) =\Phi
_{\varepsilon }\left( s\right) \tilde{\varepsilon}\left( s\right) ,\quad 
\text{yielding}\quad \tilde{E}\left( s\right) =\frac{\tilde{\sigma}\left(
s\right) }{\tilde{\varepsilon}\left( s\right) }=\frac{\Phi _{\varepsilon }(s)%
}{\Phi _{\sigma }(s)},\quad \text{with}  \label{CM-Ld} \\
\Phi _{\sigma }\left( s\right) =\sum_{k=1}^{N_{1}}a_{k}\,\frac{1}{s^{\alpha
_{k}}}+\sum_{k=1}^{N_{2}}b_{k}\,s^{\beta _{k}}\quad \text{and}\quad \Phi
_{\varepsilon }\left( s\right) =\sum_{k=1}^{M_{1}}c_{k}\,\frac{1}{s^{\mu
_{k}}}+\sum_{k=1}^{M_{2}}d_{k}\,s^{\nu _{k}},  \label{Cm-Ld-2}
\end{gather}%
where $\tilde{E}$ denotes the complex modulus, implying the relaxation
modulus in Laplace domain to be of the form%
\begin{equation}
\tilde{\sigma}_{sr}\left( s\right) =\frac{1}{s}\frac{\Phi _{\varepsilon }(s)%
}{\Phi _{\sigma }(s)}=\frac{1}{s}\tilde{E}\left( s\right) ,  \label{sr-cr}
\end{equation}%
according to the constitutive model in Laplace domain (\ref{CM-Ld})$_{1}$,
since the Laplace transform of strain assumed as the Heaviside function $H\ $%
is $\tilde{\varepsilon}\left( x,s\right) =\mathcal{L}\left[ H\left( t\right) %
\right] \left( s\right) =\frac{1}{s},$ so that the constitutive model in
Laplace domain (\ref{CM-Ld})$_{1}$ is rewritten as%
\begin{equation*}
\tilde{\sigma}\left( s\right) =s\tilde{\sigma}_{sr}\left( s\right) \tilde{%
\varepsilon}\left( s\right) ,
\end{equation*}%
becoming%
\begin{align}
\sigma \left( t\right) & =\mathcal{L}^{-1}\left[ s\tilde{\sigma}_{sr}\left(
s\right) \right] \left( t\right) \ast \varepsilon \left( t\right) =\left(
\sigma _{sr/g}\,\delta \left( t\right) +\dot{\sigma}_{sr}\left( t\right)
\right) \ast \varepsilon \left( t\right) ,\quad \text{i.e.,}  \notag \\
\sigma \left( t\right) & =\sigma _{sr/g}\,\varepsilon \left( t\right) +\dot{%
\sigma}_{sr}\left( t\right) \ast \varepsilon \left( t\right) ,
\label{cosnt-eq}
\end{align}%
where the glass modulus, defined by%
\begin{equation*}
\sigma _{sr/g}=\sigma _{sr}\left( 0\right) =\lim_{s\rightarrow \infty }s%
\tilde{\sigma}_{sr}\left( s\right) ,  \label{glass-modulus}
\end{equation*}%
is obtained according to the Tauberian theorem, implying the final form 
\begin{equation*}
\sigma \left( t\right) =\sigma _{sr/g}\,\varepsilon \left( t\right) +\frac{%
\mathrm{d}}{\mathrm{d}t}\left( \sigma _{sr}\left( t\right) \ast \varepsilon
\left( t\right) \right) -\sigma _{sr/g}\,\varepsilon \left( t\right) =\frac{%
\mathrm{d}}{\mathrm{d}t}\left( \sigma _{sr}\left( t\right) \ast \varepsilon
\left( t\right) \right)
\end{equation*}%
of the constitutive equation (\ref{cosnt-eq}), according to the derivative
of a convolution%
\begin{equation*}
\frac{\mathrm{d}}{\mathrm{d}t}\left( f\left( t\right) \ast g\left( t\right)
\right) =\dot{f}\left( t\right) \ast g\left( t\right) +f\left( 0\right)
g\left( t\right) .
\end{equation*}

Actually, the glass modulus proves to be connected with the wave propagation
speed by%
\begin{equation}
c=\sqrt{\frac{\sigma _{sr/g}}{\varrho }},  \label{ce-iks}
\end{equation}%
which is accounted for one-dimensional wave equations in \cite{Mai-10},
obtained for linear fractional constitutive equations containing
Riemann-Liouville derivatives of orders in interval $\left( 0,1\right) $,
and further extended in \cite{KOZ19} for the case of distributed-order
models, where the integration is performed with respect to orders of
fractional derivatives in interval $\left( 0,1\right) $, as well as
extending to thermodynamically consistent fractional Burgers models
containing fractional derivatives of orders in interval $\left( 1,2\right) $
as well, having the wave propagation considered on infinite domain in \cite%
{OZO} and on finite domain in \cite{SD}. Micro-local analysis of the
fractional Zener and distributed-order wave equations, yielding support of
fundamental solution in the cone defined by wave propagation speed, is
conducted in \cite{OparnicaBroucke,OparnicaBroucke1}.

Linear fractional-order models containing only Riemann-Liouville fractional
derivatives of orders in interval $\left( 0,1\right) $, obtained from the
constitutive model (\ref{suma}) for $a_{k}=0,$ with $k=1,\ldots ,N_{1},$ and 
$c_{k}=0,$ with $k=1,\ldots ,M_{1},$ thus taking the form%
\begin{equation}
\sum_{i=1}^{n}a_{i}\,{}_{0}\mathrm{D}_{t}^{\alpha _{i}}\sigma
(x,t)=\sum_{j=1}^{m}b_{j}\,{}_{0}\mathrm{D}_{t}^{\beta _{j}}\varepsilon
(x,t),  \label{kejsovi}
\end{equation}%
prove to have four thermodynamically consistent cases 
\begin{gather}
\sum_{i=1}^{n}a_{i}\,{}_{0}\mathrm{D}_{t}^{\alpha _{i}}\sigma \left(
t\right) =\sum_{i=1}^{n}b_{i}\,{}_{0}\mathrm{D}_{t}^{\alpha _{i}}\varepsilon
\left( t\right) ,  \label{Case 1} \\
\sum_{i=1}^{n}a_{i}\,{}_{0}\mathrm{D}_{t}^{\alpha _{i}}\sigma \left(
t\right) =\sum_{i=1}^{n}b_{i}\,{}_{0}\mathrm{D}_{t}^{\alpha _{i}}\varepsilon
\left( t\right) +\sum_{i=n+1}^{m}b_{i}\,{}_{0}\mathrm{D}_{t}^{\beta
_{i}}\varepsilon \left( t\right) ,  \label{Case 2} \\
\sum_{i=1}^{n-m}a_{i}\,{}_{0}\mathrm{D}_{t}^{\alpha _{i}}\sigma \left(
t\right) +\sum_{i=n-m+1}^{n}a_{i}\,{}_{0}\mathrm{D}_{t}^{\alpha _{i}}\sigma
\left( t\right) =\sum_{j=1}^{m}b_{j}\,{}_{0}\mathrm{D}_{t}^{\alpha
_{n-m+j}}\varepsilon \left( t\right) ,  \label{Case 3} \\
\sum_{i=1}^{n}a_{i}\,{}_{0}\mathrm{D}_{t}^{\alpha _{i}}\sigma \left(
t\right) =\sum_{j=1}^{m}b_{j}\,{}_{0}\mathrm{D}_{t}^{\beta _{j}}\varepsilon
\left( t\right) ,  \label{Case 4}
\end{gather}
as derived in \cite{AKOZ} and listed in Supplementary material. By examining
the large-time asymptotics of creep compliance, one finds that models
belonging to Case I and II, given by (\ref{Case 1}) and (\ref{Case 2}),
display solid-like behavior, since the creep compliance takes the finite
value for infinite time, while models belonging to Case III and IV, given by
(\ref{Case 3}) and (\ref{Case 4}), display fluid-like behavior, since the
creep compliance tends to infinity as time tends to infinity. On the other
hand, models belonging to Case I and III yield finite wave propagation
speed, since the relaxation modulus tends to a finite value as time tends to
zero, while models belonging to Case II and IV yield infinite wave
propagation speed, since the relaxation modulus tends to a infinite value as
time tends to zero. These model properties are used in \cite{ZO} in proving
dissipativity of fractional wave equations using a priori energy estimates.

On the other hand, the representatives of models containing only
Riemann-Liouville fractional derivatives of orders in both intervals $\left(
0,1\right) $ and $\left( 1,2\right) $ are fractional Burgers models, derived
and checked for thermodynamical consistency in \cite{OZ-1}, which group into
two thermodynamically consistent classes 
\begin{equation}
\begin{split}  \label{burgersi}
\left( 1+a_{1}\,{}_{0}\mathrm{D}_{t}^{\alpha }+a_{2}\,{}_{0}\mathrm{D}%
_{t}^{\beta }+a_{3}\,{}_{0}\mathrm{D}_{t}^{\gamma }\right) \sigma \left(
x,t\right) =\left( b_{1}\,{}_{0}\mathrm{D}_{t}^{\mu }+b_{2}\,{}_{0}\mathrm{D}%
_{t}^{\mu +\eta }\right) \varepsilon \left( x,t\right) \quad \text{and} \\
\left( 1+a_{1}\,{}_{0}\mathrm{D}_{t}^{\alpha }+a_{2}\,{}_{0}\mathrm{D}%
_{t}^{\beta }+a_{3}\,{}_{0}\mathrm{D}_{t}^{\beta +\eta }\right) \sigma
\left( x,t\right) =\left( b_{1}\,{}_{0}\mathrm{D}_{t}^{\beta }+b_{2}\,{}_{0}%
\mathrm{D}_{t}^{\beta +\eta }\right) \varepsilon \left( x,t\right) ,
\end{split}%
\end{equation}
such that for models of the first class the highest fractional
differentiation order of strain is $\mu +\eta \in \left[ 1,2\right] ,$ with $%
\eta \in \left\{ \alpha ,\beta \right\} ,$ while the highest fractional
differentiation order of stress is either $\gamma \in \left[ 0,1\right] $ in
the case of Model I, with $0\leqslant \alpha \leqslant \beta \leqslant
\gamma \leqslant \mu \leqslant 1$ and $\eta \in \left\{ \alpha ,\beta
,\gamma \right\} ,$ or $\gamma \in \left[ 1,2\right] $ in the case of Models
II - V, with $0\leqslant \alpha \leqslant \beta \leqslant \mu \leqslant 1$
and $\left( \eta ,\gamma \right) \in \left\{ \left( \alpha ,2\alpha \right)
,\left( \alpha ,\alpha +\beta \right) ,\left( \beta ,\alpha +\beta \right)
,\left( \beta ,2\beta \right) \right\} $, while for models of the second
class it holds $0\leqslant \alpha \leqslant \beta \leqslant 1$ and $\beta
+\eta \in \left[ 1,2\right] ,$ with $\eta =\alpha ,$ in the case of Model
VI; $\eta =\beta $ in the case of Model VII; and $\alpha =\eta =\beta ,$ $%
\bar{a}_{1}=a_{1}+a_{2},$ and $\bar{a}_{2}=a_{3}$ in the case of Model VIII,
see Supplementary material for details.

In constitutive equations of Models I - V, the orders of fractional
derivatives are ordered from lowest to highest, separately for those in the
interval $[0,1]$ and separately for those in the interval $[1,2]$. Model I
contains only one derivative acting on strain having order in interval $%
\left[ 1,2\right] $, obtained as a sum of the highest differentiation order
in interval $[0,1]$ and any of the other three differentiation orders.
Models II - V contain two fractional derivatives having orders in interval $%
\left[ 0,1\right] $ and another one having order in interval $\left[ 1,2%
\right] $, all acting on stress, where the highest order is obtained either
as a sum of the orders in interval $\left[ 0,1\right] $ or as a double value
of any of them. Moreover, Models II - V contain one fractional derivative of
order in interval $\left[ 0,1\right] $ and one of order in interval $\left[
1,2\right] $, both acting on strain, where the second one is obtained as a
sum of the first one and any remaining order up to one.

Models VI and VII contain two fractional derivatives of orders in interval $%
\left[ 0,1\right] $ and another one of order in interval $\left[ 1,2\right] $%
, all acting on stress, as well as one fractional derivative of order in
interval $\left[ 0,1\right] $ and one of order in interval $\left[ 1,2\right]
$, both acting on strain, such that the highest orders of fractional
derivatives acting on strain coincide with the highest orders of fractional
derivatives acting on stress regardless of the interval they belong to.
Fractional derivative orders in interval $[1,2]$ are obtained either as a
sum of orders from interval $\left[ 0,1\right] $, or as a double value of
the higher one. Model VIII contains the same two fractional derivatives
acting on both stress and strain, the first one having order in the interval 
$[\frac{1}{2},1]$ and the second one having the order obtained as a double
value of the first one.

In \cite{OZ-2}, the responses of fractional Burgers models are examined in
creep and stress relaxation tests, so that the considerations of large-time
asymptotics of creep compliance categorize these models into fluid-like
models, since the equilibrium compliance tends to infinity, while the
considerations of short-time asymptotic behavior of relaxation modulus
separate these models into two classes, such that the mechanical disturbance
propagation speed has an infinite value for models of the first class, since
the glass modulus has an infinite value, while the propagation speed of
mechanical disturbance for models of the second class has a finite value,
since the glass modulus has a finite value, as well.  \begin{table}[h]
 \begin{center}
 \begin{tabular}{|c|c|} 
 \hline \xrowht{20pt} 

Model & Constitutive equation  \\ \hline \xrowht{20pt}

\textbf{ID.ID} &  $\left( a_{1}\,_{0}\mathrm{I}_{t}^{\alpha }+a_{2}\,_{0}\mathrm{D}_{t}^{\beta}\right) \sigma \left( t\right) =\left( b_{1}\,_{0}\mathrm{I}_{t}^{\mu
}+a_{2}\,_{0}\mathrm{D}_{t}^{\alpha +\beta -\mu }\right) \varepsilon \left(t\right) $ \\ \hline \xrowht{20pt}

\textbf{ID.DD}$^{{}^{+}}$ & $
\left( a_{1}\,_{0}\mathrm{I}_{t}^{\alpha }+a_{2}\,_{0}\mathrm{D}_{t}^{\beta
}\right) \sigma \left( t\right) =\left( b_{1}\,_{0}\mathrm{D}_{t}^{\mu
}+b_{2}\,_{0}\mathrm{D}_{t}^{\alpha +\beta +\mu }\right) \varepsilon \left(
t\right)$ \\ \hline \xrowht{20pt}

\textbf{IID.IID} & $\left( a_{1}\,_{0}\mathrm{I}_{t}^{\alpha }+a_{2}\,_{0}\mathrm{I}_{t}^{\beta
}+a_{3}\,_{0}\mathrm{D}_{t}^{\gamma }\right) \sigma \left( t\right) =\left(
b_{1}\,_{0}\mathrm{I}_{t}^{\alpha +\gamma -\eta }+b_{2}\,_{0}\mathrm{I}%
_{t}^{\beta +\gamma -\eta }+b_{3}\,_{0}\mathrm{D}_{t}^{\eta }\right)
\varepsilon \left( t\right) $ \\ \hline \xrowht{20pt}

\textbf{IDD.IDD} & $
\left( a_{1}\,_{0}\mathrm{I}_{t}^{\alpha }+a_{2}\,_{0}\mathrm{D}_{t}^{\beta
}+a_{3}\,_{0}\mathrm{D}_{t}^{\gamma }\right) \sigma \left( t\right) =\left(
b_{1}\,_{0}\mathrm{I}_{t}^{\mu }+b_{2}\,_{0}\mathrm{D}_{t}^{\alpha +\beta
-\mu }+b_{3}\,_{0}\mathrm{D}_{t}^{\alpha +\gamma -\mu }\right) \varepsilon
\left( t\right)
$  \\ \hline \xrowht{20pt}

\textbf{IID.IDD}  & $
\left( a_{1}\,_{0}\mathrm{I}_{t}^{\alpha }+a_{2}\,_{0}\mathrm{I}_{t}^{\beta
}+a_{3}\,_{0}\mathrm{D}_{t}^{\gamma }\right) \sigma \left( t\right) =\left(
b_{1}\,_{0}\mathrm{I}_{t}^{\mu }+b_{2}\,_{0}\mathrm{D}_{t}^{\nu }+b_{3}\,_{0}%
\mathrm{D}_{t}^{\alpha +\gamma -\mu }\right) \varepsilon \left( t\right)
$ \\ \hline \xrowht{20pt}

\textbf{I}$^{{}^{+}}$\textbf{ID.I}$^{{}^{+}}$\textbf{ID} & $
\left( a_{1}\,_{0}\mathrm{I}_{t}^{1+\alpha }+a_{2}\,_{0}\mathrm{I}_{t}^{%
\frac{1+\alpha -\gamma }{2}}+a_{3}\,_{0}\mathrm{D}_{t}^{\gamma }\right)
\sigma \left( t\right) =\left( b_{1}\,_{0}\mathrm{I}_{t}^{1+\mu }+b_{2}\,_{0}%
\mathrm{I}_{t}^{\frac{1+\mu -\left( \alpha +\gamma -\mu \right) }{2}%
}+b_{3}\,_{0}\mathrm{D}_{t}^{\alpha +\gamma -\mu }\right) \varepsilon \left(
t\right)
$ \\ \hline \xrowht{20pt}

\textbf{IDD}$^{{}^{+}}$\textbf{.IDD}$^{{}^{+}}$ & $
\left( a_{1}\,_{0}\mathrm{I}_{t}^{\alpha }+a_{2}\,_{0}\mathrm{D}_{t}^{\frac{%
1+\gamma -\alpha }{2}}+a_{3}\,_{0}\mathrm{D}_{t}^{1+\gamma }\right) \sigma
\left( t\right) =\left( b_{1}\,_{0}\mathrm{I}_{t}^{\alpha +\gamma -\eta
}+b_{2}\,_{0}\mathrm{D}_{t}^{\frac{1+\eta -\left( \alpha +\gamma -\eta
\right) }{2}}+b_{3}\,_{0}\mathrm{D}_{t}^{1+\eta }\right) \varepsilon \left(
t\right) $  \\ \hline \xrowht{20pt}

\textbf{I}$^{{}^{+}}$\textbf{ID.IDD}$^{{}^{+}}$ & $
\left( a_{1}\,_{0}\mathrm{I}_{t}^{1+\alpha }+a_{2}\,_{0}\mathrm{I}_{t}^{%
\frac{1+\alpha -\gamma }{2}}+a_{3}\,_{0}\mathrm{D}_{t}^{\gamma }\right)
\sigma \left( t\right) =\left( b_{1}\,_{0}\mathrm{I}_{t}^{\alpha +\gamma
-\eta }+b_{2}\,_{0}\mathrm{D}_{t}^{\frac{1+\eta -\left( \alpha +\gamma -\eta
\right) }{2}}+b_{3}\,_{0}\mathrm{D}_{t}^{1+\eta }\right) \varepsilon \left(
t\right) $  \\ \hline \xrowht{20pt}

\textbf{IID.ID} & $
\left( a_{1}\,_{0}\mathrm{I}_{t}^{\alpha +\beta -\gamma }+a_{2}\,_{0}\mathrm{%
I}_{t}^{\nu }+a_{3}\,_{0}\mathrm{D}_{t}^{\gamma }\right) \sigma \left(
t\right) =\left( b_{1}\,_{0}\mathrm{I}_{t}^{\alpha }+b_{2}\,_{0}\mathrm{D}%
_{t}^{\beta }\right) \varepsilon \left( t\right) $  \\ \hline \xrowht{20pt}

\textbf{IDD.DD}$^{{}^{+}}$ & $
\left( a_{1}\,_{0}\mathrm{I}_{t}^{\alpha }+a_{2}\,_{0}\mathrm{D}_{t}^{\beta
}+a_{3}\,_{0}\mathrm{D}_{t}^{\gamma }\right) \sigma \left( t\right) =\left(
b_{1}\,_{0}\mathrm{D}_{t}^{\mu }+b_{2}\,_{0}\mathrm{D}_{t}^{\alpha +\beta
+\mu }\right) \varepsilon \left( t\right)
$  \\ \hline \xrowht{20pt}

 \textbf{I}$^{{}^{+}}$\textbf{ID.ID} & $
\left( a_{1}\,_{0}\mathrm{I}_{t}^{\alpha +\beta +\nu }+a_{2}\,_{0}\mathrm{I}%
_{t}^{\nu }+a_{3}\,_{0}\mathrm{D}_{t}^{\alpha +\beta -\nu }\right) \sigma
\left( t\right) =\left( b_{1}\,_{0}\mathrm{I}_{t}^{\alpha }+b_{2}\,_{0}%
\mathrm{D}_{t}^{\beta }\right) \varepsilon \left( t\right) $  \\ \hline \xrowht{20pt}

\textbf{IDD}$^{{}^{+}}$\textbf{.DD}$^{{}^{+}}$  & $
\left( a_{1}\,_{0}\mathrm{I}_{t}^{\alpha }+a_{2}\,_{0}\mathrm{D}_{t}^{\beta
}+a_{3}\,_{0}\mathrm{D}_{t}^{\alpha +2\beta }\right) \sigma \left( t\right)
=\left( b_{1}\,_{0}\mathrm{D}_{t}^{\mu }+b_{2}\,_{0}\mathrm{D}_{t}^{\alpha
+\beta +\mu }\right) \varepsilon \left( t\right)
$  \\ \hline \xrowht{20pt}

\textbf{ID.IDD} & $
\left( a_{1}\,_{0}\mathrm{I}_{t}^{\mu }+a_{2}\,_{0}\mathrm{D}_{t}^{\nu
}\right) \sigma \left( t\right) =\left( b_{1}\,_{0}\mathrm{I}_{t}^{\alpha
}+b_{2}\,_{0}\mathrm{D}_{t}^{\beta }+b_{3}\,_{0}\mathrm{D}_{t}^{\mu +\nu
-\alpha }\right) \varepsilon \left( t\right) $  \\ \hline \xrowht{20pt}

 \textbf{ID.DDD}$^{{}^{+}}$ & $
\left( a_{1}\,_{0}\mathrm{I}_{t}^{\alpha }+a_{2}\,_{0}\mathrm{D}_{t}^{\beta
}\right) \sigma \left( t\right) =\left( b_{1}\,_{0}\mathrm{D}_{t}^{\mu
}+b_{2}\,_{0}\mathrm{D}_{t}^{\nu }+b_{3}\,_{0}\mathrm{D}_{t}^{\alpha +\beta
+\nu }\right) \varepsilon \left( t\right)
$  \\ \hline \xrowht{20pt}

 \textbf{ID.IDD}$^{{}^{+}}$ & $
\left( a_{1}\,_{0}\mathrm{I}_{t}^{\alpha }+a_{2}\,_{0}\mathrm{D}_{t}^{\beta
}\right) \sigma \left( t\right) =\left( b_{1}\,_{0}\mathrm{I}_{t}^{\alpha
+\beta -\nu }+b_{2}\,_{0}\mathrm{D}_{t}^{\nu }+b_{3}\,_{0}\mathrm{D}%
_{t}^{\alpha +\beta +\nu }\right) \varepsilon \left( t\right)
$  \\ \hline

 \end{tabular}
 \end{center}
 \caption{Symmetric and asymmetric fractional anti-Zener and Zener constitutive models.}
 \label{tbl-5}
 \end{table}

The linear fractional-order models containing both fractional integrals and
Riemann-Liouville fractional derivatives of orders in both intervals $\left(
0,1\right) $ and $\left( 1,2\right) $ are fractional anti-Zener and Zener
constitutive models, derived and checked for thermodynamical consistency in 
\cite{SD-1}, grouping into the classes of symmetric and asymmetric models,
where the symmetric models have the same number of operators, either
fractional integrals and/or Riemann-Liouville fractional derivatives, acting
on both stress and strain, see Table \ref{tbl-5} for the list of models and
Supplementary material for details. It is proved in \cite{SD-2} that
fractional anti-Zener and Zener models describe fluid-like solid body having
the wave propagation speed infinite. Moreover, in \cite{SD-2} energy balance
properties of viscoelastic body are examined a priori by splitting the power
per unit volume into a term being a time derivative of a quantity that can
be interpreted as the energy per unit volume and a term representing
dissipated power per unit volume. These results are applied in \cite{SD-4}
for investigating the energy balance properties of a fractional anti-Zener
and Zener type body subject to a harmonic strain.

Fractional generalization of the wave equation for damped waves is conducted
in \cite{Luchko-2013-da}, while considering the three-dimensional deformable
body, fractional Zener wave equation is considered in \cite{NH}, while in 
\cite{OparnicaSuli} existence and uniqueness of solution to the
three-dimensional fractional Zener wave equation is proved. Numerical
methods, including finite element analysis, are used in \cite%
{Pshenichnov-2010-da,Woo-Sung-2005-da} in order to analyze transient wave
processes in linearly viscoelastic solids, while in \cite%
{viscoacustic-2019-da,Sweilam-Hasan-2022-da,Tripathi-Espindola-Pintona-2019-da,Zhou-Wang-Zhang-2021-da}
different numerical methods are employed in order to obtain solutions to
proposed models. Also, different methods for the discretization of
stochastic fractional wave equation are used in \cite{Liu-2023-da}.

Various applications of viscoelastic wave equations include modeling of
ultrasonic waves used to estimate the distribution of a damage in a
material, see \cite{Ryuzono-Yashiro-2022-da}, three-dimensional solid
thermoviscoelastic material, see \cite{Sarkar-Banerjee-Shaw-2021-da}, wave
propagation after a tsunami, see \cite{tsunami-2022-da}, seismic waves, see 
\cite{Wang-Xing-Zhu-Zhou-Shi-da}, stratified porous media, as done in \cite%
{WENFEI-ZHANG-2005-da}, and many others.

In order to examine the torsional waves characteristic for human cervical
tissue, a three-dimensional wave equation is examined in \cite%
{Callejas-Melchor-Faris-Rus-2021-da}, while the research about ultrasound
waves traveling through biological tissues is carried out in \cite%
{Cardoso-Cowin-2012-da}. In \cite%
{WavePropagationinViscoelasticMaterials-2018-da}, considering shear modulus,
three-dimensional wave propagation is analyzed in viscoelastic materials,
having importance in, for example, modeling human tissues.

\section{Displacement field}

The method of integral transforms, Fourier transform with respect to spatial
coordinates, defined by%
\begin{equation*}
\bar{f}\left( \boldsymbol{k}\right) =\mathcal{F}\left[ f\left( \boldsymbol{r}%
\right) \right] \left( \boldsymbol{k}\right) =\int_{%
\mathbb{R}
^{3}}f\left( \boldsymbol{r}\right) \mathrm{e}^{-\mathrm{i}\boldsymbol{k\cdot
r}}\mathrm{d}_{\boldsymbol{r}}V,\quad \boldsymbol{k}\in 
\mathbb{R}
^{3},
\end{equation*}%
and Laplace transform with respect to time, defined by (\ref{LT}),$\ $is
used in order to solve the Cauchy initial-value problem for the
three-dimensional wave equation of viscoelastic body, considered as the
system of equations consisting of the equation of motion of deformable solid
body (\ref{EoM}), strain (\ref{strain}), and constitutive equation (\ref%
{VEHook}), where the relaxation moduli $\sigma _{sr}^{\left( c\right) }$ and 
$\sigma _{sr}^{\left( s\right) }$ define material response of viscoelastic
body, i.e., in order to solve the system of equations 
\begin{gather}
\frac{1}{\varrho }\func{div}\boldsymbol{\hat{\sigma}}\left( \boldsymbol{r}%
,t\right) +\boldsymbol{f}_{b}\left( \boldsymbol{r},t\right) =\partial
_{tt}^{2}\boldsymbol{u}\left( \boldsymbol{r},t\right) ,  \label{SoE-gen-1} \\
\boldsymbol{\hat{\varepsilon}}\left( \boldsymbol{r},t\right) =\frac{1}{2}%
\left( \func{grad}\boldsymbol{u}\left( \boldsymbol{r},t\right) +\left( \func{%
grad}\boldsymbol{u}\left( \boldsymbol{r},t\right) \right) ^{\mathrm{T}%
}\right) ,  \label{SoE-gen-2} \\
\boldsymbol{\hat{\sigma}}\left( \boldsymbol{r},t\right) =\partial _{t}\left(
\sigma _{sr}^{\left( c,s\right) }\left( t\right) \ast _{t}\mathrm{\limfunc{tr%
}}\boldsymbol{\hat{\varepsilon}}\left( \boldsymbol{r},t\right) \right) 
\boldsymbol{\hat{I}}+2\partial _{t}\left( \sigma _{sr}^{\left( s\right)
}\left( t\right) \ast _{t}\boldsymbol{\hat{\varepsilon}}\left( \boldsymbol{r}%
,t\right) \right) ,  \label{SoE-gen-3}
\end{gather}%
on $\boldsymbol{r}\in 
\mathbb{R}
^{3},$ for $t>0,$ subject to initial conditions%
\begin{equation}
\boldsymbol{u}\left( \boldsymbol{r},0\right) =\boldsymbol{u}_{0}\left( 
\boldsymbol{r}\right) \quad \text{and}\quad \partial _{t}\boldsymbol{u}%
\left( \boldsymbol{r},0\right) =\boldsymbol{v}_{0}\left( \boldsymbol{r}%
\right)  \label{poc-usl}
\end{equation}%
and assuming that all components of displacement field, as well as of the
Cauchy stress tensor, vanish as any of the coordinates tend to both positive
and negative infinity. Namely, the application of Fourier and Laplace
transforms, applied to the previously mentioned system of equations, yields%
\begin{gather}
\frac{1}{\varrho }\mathrm{i}\boldsymbol{k\,\bar{\tilde{\hat{\sigma}}}}\left( 
\boldsymbol{k},s\right) +\boldsymbol{\bar{\tilde{f}}}_{b}\left( \boldsymbol{k%
},s\right) =s^{2}\boldsymbol{\bar{\tilde{u}}}\left( \boldsymbol{k},s\right)
-s\boldsymbol{\bar{u}}_{0}\left( \boldsymbol{k}\right) -\boldsymbol{\bar{v}}%
_{0}\left( \boldsymbol{k}\right) ,  \label{SoE-FL-1} \\
\boldsymbol{\bar{\tilde{\hat{\varepsilon}}}}\left( \boldsymbol{k},s\right) =%
\frac{1}{2}\left( \mathrm{i}\boldsymbol{k}\otimes \boldsymbol{\bar{\tilde{u}}%
}\left( \boldsymbol{k},s\right) +\boldsymbol{\bar{\tilde{u}}}\left( 
\boldsymbol{k},s\right) \otimes \left( \mathrm{i}\boldsymbol{k\,}\right)
\right) ,  \label{SoE-FL-2} \\
\boldsymbol{\bar{\tilde{\hat{\sigma}}}}\left( \boldsymbol{k},s\right) =s%
\tilde{\sigma}_{sr}^{\left( c,s\right) }\left( s\right) \left( \mathrm{i}%
\boldsymbol{k}\cdot \boldsymbol{\bar{\tilde{u}}}\left( \boldsymbol{k}%
,s\right) \right) \boldsymbol{\hat{I}}+2s\tilde{\sigma}_{sr}^{\left(
s\right) }\left( s\right) \boldsymbol{\bar{\tilde{\hat{\varepsilon}}}}\left( 
\boldsymbol{k},s\right) ,  \label{SoE-FL-3}
\end{gather}%
since $\mathcal{F}\left[ \nabla \right] \left( \boldsymbol{k}\right) =%
\mathrm{i}\boldsymbol{k}$ implies $\mathcal{F}\left[ \func{div}\boldsymbol{%
\hat{\sigma}}\left( \boldsymbol{r},t\right) \right] \left( \boldsymbol{k}%
\right) =\mathcal{F}\left[ \nabla \boldsymbol{\hat{\sigma}}\left( 
\boldsymbol{r},t\right) \right] \left( \boldsymbol{k}\right) =\mathrm{i}%
\boldsymbol{k\,\bar{\hat{\sigma}}}\left( \boldsymbol{k},t\right) ,$ $%
\mathcal{F}\left[ \func{grad}\boldsymbol{u}\left( \boldsymbol{r},t\right) %
\right] \left( \boldsymbol{k}\right) =$ \newline
$\mathcal{F}\left[ \nabla \otimes \boldsymbol{u}\left( \boldsymbol{r}%
,t\right) \right] \left( \boldsymbol{k}\right) =\mathrm{i}\boldsymbol{k}%
\otimes \boldsymbol{\bar{u}}\left( \boldsymbol{k},t\right) ,$ and $\mathcal{F%
}\left[ \mathrm{\limfunc{tr}}\boldsymbol{\hat{\varepsilon}}\left( 
\boldsymbol{r},t\right) \right] \left( \boldsymbol{k}\right) =\mathcal{F}%
\left[ \func{div}\boldsymbol{u}\left( \boldsymbol{r},t\right) \right] \left( 
\boldsymbol{k}\right) =\mathcal{F}\left[ \nabla \cdot \boldsymbol{u}\left( 
\boldsymbol{r},t\right) \right] \left( \boldsymbol{k}\right) =\mathrm{i}%
\boldsymbol{k}\cdot \boldsymbol{\bar{u}}\left( \boldsymbol{k},t\right) ,$
where $\otimes $ is used to denote the dyadic product, while the system of
equations (\ref{SoE-FL-1}) - (\ref{SoE-FL-3}) reduced to a single equation
containing the displacement field in Fourier and Laplace domain becomes%
\begin{align}
& \boldsymbol{\,}\frac{s\tilde{\sigma}_{sr}^{\left( c,s\right) }\left(
s\right) }{\varrho }\mathrm{i}\boldsymbol{k}\left( \mathrm{i}\boldsymbol{k}%
\cdot \boldsymbol{\bar{\tilde{u}}}\left( \boldsymbol{k},s\right) \right) +%
\frac{s\tilde{\sigma}_{sr}^{\left( s\right) }\left( s\right) }{\varrho }%
\mathrm{i}\boldsymbol{k}\left( \mathrm{i}\boldsymbol{k}\otimes \boldsymbol{%
\bar{\tilde{u}}}\left( \boldsymbol{k},s\right) \right)  \notag \\
& \qquad \qquad \qquad \qquad \qquad +\frac{s\tilde{\sigma}_{sr}^{\left(
s\right) }\left( s\right) }{\varrho }\mathrm{i}\boldsymbol{k}\left( 
\boldsymbol{\bar{\tilde{u}}}\left( \boldsymbol{k},s\right) \otimes \left( 
\mathrm{i}\boldsymbol{k}\right) \right) -s^{2}\boldsymbol{\bar{\tilde{u}}}%
\left( \boldsymbol{k},s\right) =-s\boldsymbol{\bar{u}}_{0}\left( \boldsymbol{%
k}\right) -\boldsymbol{\bar{v}}_{0}\left( \boldsymbol{k}\right) -\boldsymbol{%
\bar{\tilde{f}}}_{b}\left( \boldsymbol{k},s\right) ,  \notag \\
& \left( \left( \frac{s\tilde{\sigma}_{sr}^{\left( s\right) }\left( s\right) 
}{\varrho }k^{2}+s^{2}\right) \boldsymbol{\hat{I}}+\frac{s\left( \sigma
_{sr}^{\left( c\right) }\left( s\right) -\sigma _{sr}^{\left( s\right)
}\left( s\right) \right) }{\varrho }\left( \boldsymbol{k}\otimes \boldsymbol{%
k}\right) \right) \boldsymbol{\bar{\tilde{u}}}\left( \boldsymbol{k},s\right)
=s\boldsymbol{\bar{u}}_{0}\left( \boldsymbol{k}\right) +\boldsymbol{\bar{v}}%
_{0}\left( \boldsymbol{k}\right) +\boldsymbol{\bar{\tilde{f}}}_{b}\left( 
\boldsymbol{k},s\right) ,  \label{Ru=pu}
\end{align}%
since $\mathrm{i}\boldsymbol{k}\left( \mathrm{i}\boldsymbol{k}\otimes 
\boldsymbol{\bar{\tilde{u}}}\left( \boldsymbol{k},s\right) \right) =\left( 
\mathrm{i}\boldsymbol{k}\right) ^{2}\boldsymbol{\bar{\tilde{u}}}\left( 
\boldsymbol{k},s\right) ,$ $\mathrm{i}\boldsymbol{k}\left( \boldsymbol{\bar{%
\tilde{u}}}\left( \boldsymbol{k},s\right) \otimes \left( \mathrm{i}%
\boldsymbol{k}\right) \right) =\left( \mathrm{i}\boldsymbol{k\cdot \bar{%
\tilde{u}}}\left( \boldsymbol{k},s\right) \right) \mathrm{i}\boldsymbol{k}=%
\mathrm{i}\boldsymbol{k}\left( \boldsymbol{\bar{\tilde{u}}}\left( 
\boldsymbol{k},s\right) \cdot \left( \mathrm{i}\boldsymbol{k}\right) \right)
=$\newline
$\left( \left( \mathrm{i}\boldsymbol{k}\right) \otimes \left( \mathrm{i}%
\boldsymbol{k}\right) \right) \boldsymbol{\bar{\tilde{u}}}\left( \boldsymbol{%
k},s\right) ,$ which can be considered as the wave equation of viscoelastic
body in Fourier and Laplace domain.

\subsection{Displacement field via functions $\protect\varphi $ and $%
\boldsymbol{\Omega}$}

The classical three-dimensional wave equations (\ref{we-fi}) and (\ref%
{we-omega}), describing the compressive and shear wave propagation and
expressed in terms of $\varphi \left( \boldsymbol{r},t\right) =\func{div}%
\boldsymbol{u}\left( \boldsymbol{r},t\right) $ and $\boldsymbol{\Omega }%
\left( \boldsymbol{r},t\right) =\func{curl}\boldsymbol{u}\left( \boldsymbol{r%
},t\right) \ $as unknown functions, are generalized for the viscoelastic
body and the solution to the wave equation of viscoelastic body, i.e., to
the system of equations (\ref{SoE-gen-1}) - (\ref{SoE-gen-3}) is calculated
by the use of functions $\varphi $ and $\boldsymbol{\Omega },$ similarly as
in the case of classical wave equation.

The generalization of the wave equation (\ref{we-fi}) is obtained as
follows: by applying the scalar product of wave vector $\mathrm{i}%
\boldsymbol{k}$ with the wave equation of viscoelastic body in Fourier and
Laplace domain (\ref{Ru=pu}), one obtains%
\begin{gather}
\left( \frac{s\tilde{\sigma}_{sr}^{\left( c\right) }\left( s\right) }{%
\varrho }k^{2}+s^{2}\right) \left( \mathrm{i}\boldsymbol{k}\cdot \boldsymbol{%
\bar{\tilde{u}}}\left( \boldsymbol{k},s\right) \right) =\mathrm{i}%
\boldsymbol{k}\cdot \left( s\boldsymbol{\bar{u}}_{0}\left( \boldsymbol{k}%
\right) +\boldsymbol{\bar{v}}_{0}\left( \boldsymbol{k}\right) +\boldsymbol{%
\bar{\tilde{f}}}_{b}\left( \boldsymbol{k},s\right) \right) ,\quad \text{i.e.,%
}  \label{com-we} \\
-\tilde{c}_{c}^{2}\left( s\right) \,k^{2}\left( \mathrm{i}\boldsymbol{k}%
\cdot \boldsymbol{\bar{\tilde{u}}}\left( \boldsymbol{k},s\right) \right) -%
\mathrm{i}\boldsymbol{k}\cdot \left( s^{2}\boldsymbol{\bar{\tilde{u}}}\left( 
\boldsymbol{k},s\right) -s\boldsymbol{\bar{u}}_{0}\left( \boldsymbol{k}%
\right) -\boldsymbol{\bar{v}}_{0}\left( \boldsymbol{k}\right) \right) =-%
\mathrm{i}\boldsymbol{k}\cdot \boldsymbol{\bar{\tilde{f}}}_{b}\left( 
\boldsymbol{k},s\right) ,  \notag
\end{gather}%
since $\mathrm{i}\boldsymbol{k}\cdot \left( \left( \boldsymbol{k}\otimes 
\boldsymbol{k}\right) \boldsymbol{\bar{\tilde{u}}}\left( \boldsymbol{k}%
,s\right) \right) =\mathrm{i}\boldsymbol{k}\cdot \left( \boldsymbol{k}\left( 
\boldsymbol{k\cdot \bar{\tilde{u}}}\left( \boldsymbol{k},s\right) \right)
\right) =k^{2}\left( \mathrm{i}\boldsymbol{k\cdot \bar{\tilde{u}}}\left( 
\boldsymbol{k},s\right) \right) ,$ where%
\begin{equation}
\tilde{c}_{c}\left( s\right) =\sqrt{\frac{s\tilde{\sigma}_{sr}^{\left(
c\right) }\left( s\right) }{\varrho }}  \label{c-tilde-compresive}
\end{equation}%
is the memory function corresponding to compressive waves, so that after
application of the inverse Fourier and Laplace transforms, with $\mathcal{F}%
\left[ \Delta \right] \left( \boldsymbol{k}\right) =\mathcal{F}\left[ \nabla
\cdot \nabla \right] \left( \boldsymbol{k}\right) =\left( \mathrm{i}%
\boldsymbol{k}\right) ^{2}\boldsymbol{=}-k^{2},$ the previous expression
becomes%
\begin{gather}
\tilde{c}_{c}^{2}\left( s\right) \,\Delta \mathrm{\func{div}}\boldsymbol{%
\tilde{u}}\left( \boldsymbol{r},s\right) -\mathrm{\func{div}}\left( s^{2}%
\boldsymbol{\tilde{u}}\left( \boldsymbol{r},s\right) -s\boldsymbol{u}%
_{0}\left( \boldsymbol{r}\right) -\boldsymbol{v}_{0}\left( \boldsymbol{r}%
\right) \right) =-\mathrm{\func{div}}\boldsymbol{\tilde{f}}_{b}\left( 
\boldsymbol{r},s\right) ,\quad \text{i.e.,}  \notag \\
\Delta \left( \mathcal{L}^{-1}\left[ \tilde{c}_{c}^{2}\left( s\right) \,%
\tilde{\varphi}\left( \boldsymbol{r},s\right) \right] \left( \boldsymbol{r}%
,t\right) \right) -\partial _{tt}^{2}\varphi \left( \boldsymbol{r},t\right)
=-\mathrm{\func{div}}\boldsymbol{f}_{b}\left( \boldsymbol{r},t\right) ,\quad 
\text{with}\quad \varphi \left( \boldsymbol{r},t\right) =\func{div}%
\boldsymbol{u}\left( \boldsymbol{r},t\right) ,  \label{eq-grin-fi}
\end{gather}%
reducing to either 
\begin{equation}
\frac{\dot{\sigma}_{sr}^{\left( c\right) }\left( t\right) }{\varrho }\ast
_{t}\Delta \varphi \left( \boldsymbol{r},t\right) +\frac{\sigma
_{sr/g}^{\left( c\right) }}{\varrho }\Delta \varphi \left( \boldsymbol{r}%
,t\right) -\partial _{tt}^{2}\varphi \left( \boldsymbol{r},t\right) =-%
\mathrm{\func{div}}\boldsymbol{f}_{b}\left( \boldsymbol{r},t\right) ,
\label{eq-grin-fi-finite}
\end{equation}%
if glass modulus $\sigma _{sr/g}^{\left( c\right) }=\sigma _{sr}^{\left(
c\right) }\left( 0\right) $ is finite and therefore compressive wave
propagation speed $c_{c}=\sqrt{\frac{\sigma _{sr/g}^{\left( c\right) }}{%
\varrho }}$ as well, due to 
\begin{equation}
\mathcal{L}^{-1}\left[ \tilde{c}_{c}^{2}\left( s\right) \,\tilde{\varphi}%
\left( \boldsymbol{r},s\right) \right] \left( \boldsymbol{r},t\right) =%
\mathcal{L}^{-1}\left[ \frac{s\tilde{\sigma}_{sr}^{\left( c\right) }\left(
s\right) }{\varrho }\,\tilde{\varphi}\left( \boldsymbol{r},s\right) \right]
\left( \boldsymbol{r},t\right) =\left( \frac{1}{\varrho }\dot{\sigma}%
_{sr}^{\left( c\right) }\left( t\right) +\frac{1}{\varrho }\sigma
_{sr/g}^{\left( c\right) }\delta \left( t\right) \right) \ast _{t}\varphi
\left( \boldsymbol{r},t\right) ,  \label{funkcija-c}
\end{equation}%
or reducing to%
\begin{equation}
\frac{\sigma _{sr}^{\left( c\right) }\left( t\right) }{\varrho }\ast
_{t}\Delta \partial _{t}\varphi \left( \boldsymbol{r},t\right) +\frac{\sigma
_{sr}^{\left( c\right) }\left( t\right) }{\varrho }\Delta \varphi _{0}\left( 
\boldsymbol{r}\right) -\partial _{tt}^{2}\varphi \left( \boldsymbol{r}%
,t\right) =-\mathrm{\func{div}}\boldsymbol{f}_{b}\left( \boldsymbol{r}%
,t\right) ,  \label{eq-grin-fi-infinite}
\end{equation}%
with $\varphi _{0}\left( \boldsymbol{r}\right) =\varphi \left( \boldsymbol{r}%
,0\right) ,$ if glass modulus is infinite, so as the compressive wave
propagation speed, due to 
\begin{equation}
\mathcal{L}^{-1}\left[ \tilde{c}_{c}^{2}\left( s\right) \,\tilde{\varphi}%
\left( \boldsymbol{r},s\right) \right] \left( \boldsymbol{r},t\right) =%
\mathcal{L}^{-1}\left[ \frac{\tilde{\sigma}_{sr}^{\left( c\right) }\left(
s\right) }{\varrho }s\tilde{\varphi}\left( \boldsymbol{r},s\right) \right]
\left( \boldsymbol{r},t\right) =\frac{1}{\varrho }\sigma _{sr}^{\left(
c\right) }\left( t\right) \ast _{t}\left( \partial _{t}\varphi \left( 
\boldsymbol{r},t\right) +\varphi _{0}\left( \boldsymbol{r}\right) \delta
\left( t\right) \right) .  \label{f-ja-c}
\end{equation}%
Hence, the general form of wave equation expressed in terms of function $%
\varphi $ is given by (\ref{eq-grin-fi}), while its specific form when
compressive wave propagation speed is finite is given by (\ref%
{eq-grin-fi-finite}), while wave equation (\ref{eq-grin-fi-infinite})
corresponds to the case when compressive wave propagation speed is infinite.
Each of the previously mentioned wave equations requires initial conditions%
\begin{equation}
\varphi _{0}\left( \boldsymbol{r}\right) =\varphi \left( \boldsymbol{r}%
,0\right) =\func{div}\boldsymbol{u}\left( \boldsymbol{r},0\right) =\func{div}%
\boldsymbol{u}_{0}\left( \boldsymbol{r}\right) \quad \text{and}\quad \dot{%
\varphi}_{0}\left( \boldsymbol{r}\right) =\partial _{t}\varphi \left( 
\boldsymbol{r},0\right) =\func{div}\partial _{t}\boldsymbol{u}\left( 
\boldsymbol{r},0\right) =\func{div}\boldsymbol{v}_{0}\left( \boldsymbol{r}%
\right) ,  \label{pu-fi}
\end{equation}%
arising from the initial conditions on displacement field (\ref{poc-usl}).

On the other hand, the generalization of the wave equation (\ref{we-omega})
is obtained as follows: the vector product of wave vector $\mathrm{i}%
\boldsymbol{k}$ with the wave equation of viscoelastic body in Fourier and
Laplace domain (\ref{Ru=pu}) implies%
\begin{gather}
\left( \frac{s\tilde{\sigma}_{sr}^{\left( s\right) }\left( s\right) }{%
\varrho }k^{2}+s^{2}\right) \left( \mathrm{i}\boldsymbol{k}\times 
\boldsymbol{\bar{\tilde{u}}}\left( \boldsymbol{k},s\right) \right) =\mathrm{i%
}\boldsymbol{k}\times \left( s\boldsymbol{\bar{u}}_{0}\left( \boldsymbol{k}%
\right) +\boldsymbol{\bar{v}}_{0}\left( \boldsymbol{k}\right) +\boldsymbol{%
\bar{\tilde{f}}}_{b}\left( \boldsymbol{k},s\right) \right) ,\quad \text{i.e.,%
}  \label{shear-we} \\
-\tilde{c}_{s}^{2}\left( s\right) \,k^{2}\left( \mathrm{i}\boldsymbol{k}%
\times \boldsymbol{\bar{\tilde{u}}}\left( \boldsymbol{k},s\right) \right) -%
\mathrm{i}\boldsymbol{k}\times \left( s^{2}\boldsymbol{\bar{\tilde{u}}}%
\left( \boldsymbol{k},s\right) -s\boldsymbol{\bar{u}}_{0}\left( \boldsymbol{k%
}\right) -\boldsymbol{\bar{v}}_{0}\left( \boldsymbol{k}\right) \right) =-%
\mathrm{i}\boldsymbol{k}\times \boldsymbol{\bar{\tilde{f}}}_{b}\left( 
\boldsymbol{k},s\right) ,  \notag
\end{gather}%
since $\mathrm{i}\boldsymbol{k}\times \left( \left( \boldsymbol{k}\otimes 
\boldsymbol{k}\right) \boldsymbol{\bar{\tilde{u}}}\left( \boldsymbol{k}%
,s\right) \right) =\mathrm{i}\boldsymbol{k}\times \left( \boldsymbol{k}%
\left( \boldsymbol{k\cdot \bar{\tilde{u}}}\left( \boldsymbol{k},s\right)
\right) \right) =\boldsymbol{0},$ where%
\begin{equation}
\tilde{c}_{s}\left( s\right) =\sqrt{\frac{s\tilde{\sigma}_{sr}^{\left(
s\right) }\left( s\right) }{\varrho }}  \label{c-tilde-shear}
\end{equation}%
is the memory function corresponding to shear waves, so that the previous
expression is transformed into%
\begin{gather}
\tilde{c}_{s}^{2}\left( s\right) \,\Delta \left( \func{curl}\boldsymbol{%
\tilde{u}}\left( \boldsymbol{r},s\right) \right) -\mathrm{\func{curl}}\left(
s^{2}\boldsymbol{\tilde{u}}\left( \boldsymbol{r},s\right) -s\boldsymbol{u}%
_{0}\left( \boldsymbol{r}\right) -\boldsymbol{v}_{0}\left( \boldsymbol{r}%
\right) \right) =-\mathrm{\func{curl}}\boldsymbol{\tilde{f}}_{b}\left( 
\boldsymbol{r},s\right) ,\quad \text{i.e.,}  \notag \\
\Delta \left( \mathcal{L}^{-1}\left[ \tilde{c}_{s}^{2}\left( s\right) \,%
\boldsymbol{\tilde{\Omega}}\left( \boldsymbol{r},s\right) \right] \left( 
\boldsymbol{r},t\right) \right) -\partial _{tt}^{2}\boldsymbol{\Omega }%
\left( \boldsymbol{r},t\right) =-\mathrm{\func{curl}}\boldsymbol{f}%
_{b}\left( \boldsymbol{r},t\right) ,\quad \text{with}\quad \boldsymbol{%
\Omega }\left( \boldsymbol{r},t\right) =\func{curl}\boldsymbol{u}\left( 
\boldsymbol{r},t\right) ,  \label{eq-grin-omega}
\end{gather}%
after application of the inverse Fourier and Laplace transforms, reducing to
either 
\begin{equation}
\frac{\dot{\sigma}_{sr}^{\left( s\right) }\left( t\right) }{\varrho }\ast
_{t}\Delta \boldsymbol{\Omega }\left( \boldsymbol{r},t\right) +\frac{\sigma
_{sr/g}^{\left( s\right) }}{\varrho }\Delta \boldsymbol{\Omega }\left( 
\boldsymbol{r},t\right) -\partial _{tt}^{2}\boldsymbol{\Omega }\left( 
\boldsymbol{r},t\right) =-\mathrm{\func{curl}}\boldsymbol{f}_{b}\left( 
\boldsymbol{r},t\right) ,  \label{eq-grin-omega-finite}
\end{equation}%
if glass modulus $\sigma _{sr/g}^{\left( s\right) }=\sigma _{sr}^{\left(
s\right) }\left( 0\right) $ is finite and therefore shear wave propagation
speed $c_{s}=\sqrt{\frac{\sigma _{sr/g}^{\left( s\right) }}{\varrho }}$ as
well, see (\ref{funkcija-c}), or to%
\begin{equation}
\frac{\sigma _{sr}^{\left( s\right) }\left( t\right) }{\varrho }\ast
_{t}\Delta \partial _{t}\boldsymbol{\Omega }\left( \boldsymbol{r},t\right) +%
\frac{\sigma _{sr}^{\left( s\right) }\left( t\right) }{\varrho }\Delta 
\boldsymbol{\Omega }_{0}\left( \boldsymbol{r}\right) -\partial _{tt}^{2}%
\boldsymbol{\Omega }\left( \boldsymbol{r},t\right) =-\mathrm{\func{curl}}%
\boldsymbol{f}_{b}\left( \boldsymbol{r},t\right) ,
\label{eq-grin-omega-infinite}
\end{equation}%
with $\boldsymbol{\Omega }_{0}\left( \boldsymbol{r}\right) =\boldsymbol{%
\Omega }\left( \boldsymbol{r},0\right) ,$ if glass modulus is infinite, so
as the shear wave propagation speed. Hence, the general form of wave
equation expressed in terms of function $\boldsymbol{\Omega }$ is given by (%
\ref{eq-grin-omega}), while its specific form when shear wave propagation
speed is finite is given by (\ref{eq-grin-omega-finite}), while wave
equation (\ref{eq-grin-omega-infinite}) corresponds to the case when shear
wave propagation speed is infinite. Each of the previously mentioned wave
equations requires initial conditions%
\begin{equation}
\boldsymbol{\Omega }_{0}\left( \boldsymbol{r}\right) =\boldsymbol{\Omega }%
\left( \boldsymbol{r},0\right) =\func{curl}\boldsymbol{u}\left( \boldsymbol{r%
},0\right) =\func{curl}\boldsymbol{u}_{0}\left( \boldsymbol{r}\right) \quad 
\text{and}\quad \boldsymbol{\dot{\Omega}}_{0}\left( \boldsymbol{r}\right)
=\partial _{t}\boldsymbol{\Omega }\left( \boldsymbol{r},0\right) =\func{curl}%
\partial _{t}\boldsymbol{u}\left( \boldsymbol{r},0\right) =\func{curl}%
\boldsymbol{v}_{0}\left( \boldsymbol{r}\right) ,  \label{pu-omega}
\end{equation}%
arising from the initial conditions on displacement field (\ref{poc-usl}).

Wave equations for compressive waves (\ref{eq-grin-fi-finite}) and (\ref%
{eq-grin-fi-infinite}), as well as wave equations for shear waves (\ref%
{eq-grin-omega-finite}) and (\ref{eq-grin-omega-infinite}), generalize
classical wave equations (\ref{we-fi}) and (\ref{we-omega}). In particular,
wave propagation speeds of compressive and shear waves in the Laplace
transform of wave equations (\ref{we-fi}) and (\ref{we-omega}) are replaced
by different memory functions (\ref{c-tilde-compresive}) and (\ref%
{c-tilde-shear}) depending on the selected model of viscoelastic body
through the relaxation modulus. Rewriting compressive and shear wave
equations in Fourier and Laplace domain (\ref{com-we}) and (\ref{shear-we})
in the form%
\begin{gather*}
\left( \tilde{c}_{c}^{2}\left( s\right) \,k^{2}+s^{2}\right) \bar{\tilde{%
\varphi}}\left( \boldsymbol{k},s\right) =s\bar{\varphi}_{0}\left( 
\boldsymbol{k}\right) +\bar{\dot{\varphi}}_{0}\left( \boldsymbol{k}\right) +%
\mathrm{i}\boldsymbol{k}\cdot \boldsymbol{\bar{\tilde{f}}}_{b}\left( 
\boldsymbol{k},s\right) , \\
\left( \tilde{c}_{s}^{2}\left( s\right) \,k^{2}+s^{2}\right) \boldsymbol{%
\bar{\tilde{\Omega}}}\left( \boldsymbol{k},s\right) =s\boldsymbol{\bar{\Omega%
}}_{0}\left( \boldsymbol{k}\right) +\boldsymbol{\bar{\dot{\Omega}}}%
_{0}\left( \boldsymbol{k}\right) +\mathrm{i}\boldsymbol{k}\times \boldsymbol{%
\bar{\tilde{f}}}_{b}\left( \boldsymbol{k},s\right) ,
\end{gather*}%
with initial data%
\begin{eqnarray*}
&&\bar{\varphi}_{0}\left( \boldsymbol{k}\right) =\mathrm{i}\boldsymbol{k}%
\cdot \boldsymbol{\bar{u}}_{0}\left( \boldsymbol{k}\right) \quad \text{and}%
\quad \bar{\dot{\varphi}}_{0}\left( \boldsymbol{k}\right) =\mathrm{i}%
\boldsymbol{k}\cdot \boldsymbol{\bar{v}}_{0}\left( \boldsymbol{k}\right)
,\quad \text{as well as} \\
&&\boldsymbol{\bar{\Omega}}_{0}\left( \boldsymbol{k}\right) =\mathrm{i}%
\boldsymbol{k}\times \boldsymbol{\bar{u}}_{0}\left( \boldsymbol{k}\right)
\quad \text{and}\quad \boldsymbol{\bar{\dot{\Omega}}}_{0}\left( \boldsymbol{k%
}\right) =\mathrm{i}\boldsymbol{k}\times \boldsymbol{\bar{v}}_{0}\left( 
\boldsymbol{k}\right) ,
\end{eqnarray*}%
see (\ref{pu-fi}) and (\ref{pu-omega}), for the solution of these wave
equations one has%
\begin{align}
\bar{\tilde{\varphi}}\left( \boldsymbol{k},s\right) & =\bar{\tilde{G}}%
^{\left( c\right) }\left( k,s\right) \left( s\bar{\varphi}_{0}\left( 
\boldsymbol{k}\right) +\bar{\dot{\varphi}}_{0}\left( \boldsymbol{k}\right) +%
\mathrm{i}\boldsymbol{k}\cdot \boldsymbol{\bar{\tilde{f}}}_{b}\left( 
\boldsymbol{k},s\right) \right) ,\quad \text{with}\quad \bar{\tilde{G}}%
^{\left( c\right) }\left( k,s\right) =\frac{1}{\tilde{c}_{c}^{2}\left(
s\right) \,k^{2}+s^{2}},  \label{grin-c} \\
\boldsymbol{\bar{\tilde{\Omega}}}\left( \boldsymbol{k},s\right) & =\bar{%
\tilde{G}}^{\left( s\right) }\left( k,s\right) \left( s\boldsymbol{\bar{%
\Omega}}_{0}\left( \boldsymbol{k}\right) +\boldsymbol{\bar{\dot{\Omega}}}%
_{0}\left( \boldsymbol{k}\right) +\mathrm{i}\boldsymbol{k}\times \boldsymbol{%
\bar{\tilde{f}}}_{b}\left( \boldsymbol{k},s\right) \right) ,\quad \text{with}%
\quad \bar{\tilde{G}}^{\left( s\right) }\left( k,s\right) =\frac{1}{\tilde{c}%
_{s}^{2}\left( s\right) \,k^{2}+s^{2}},  \label{grin-s}
\end{align}%
where $\bar{\tilde{G}}^{\left( c\right) }$ and $\bar{\tilde{G}}^{\left(
s\right) }$ are Green's functions in Fourier and Laplace domain
corresponding to compressive and shear waves respectively, implying the
solution to compressive and shear wave equations (\ref{eq-grin-fi}) and (\ref%
{eq-grin-omega}) in the form%
\begin{align}
\varphi \left( \boldsymbol{r},t\right) & =\partial _{t}G^{\left( c\right)
}\left( r,t\right) \ast _{\boldsymbol{r}}\varphi _{0}\left( \boldsymbol{r}%
\right) +G^{\left( c\right) }\left( r,t\right) \ast _{\boldsymbol{r}}\dot{%
\varphi}_{0}\left( \boldsymbol{r}\right) +G^{\left( c\right) }\left(
r,t\right) \ast _{\boldsymbol{r},t}\func{div}\boldsymbol{f}_{b}\left( 
\boldsymbol{r},t\right) ,  \label{fi-konacno} \\
\boldsymbol{\Omega }\left( \boldsymbol{r},t\right) & =\partial _{t}G^{\left(
s\right) }\left( r,t\right) \ast _{\boldsymbol{r}}\boldsymbol{\Omega }%
_{0}\left( \boldsymbol{r}\right) +G^{\left( s\right) }\left( r,t\right) \ast
_{\boldsymbol{r}}\boldsymbol{\dot{\Omega}}_{0}\left( \boldsymbol{r}\right)
+G^{\left( s\right) }\left( r,t\right) \ast _{\boldsymbol{r},t}\func{curl}%
\boldsymbol{f}_{b}\left( \boldsymbol{r},t\right) ,  \label{omega-konacno}
\end{align}%
if $G^{\left( c\right) }\left( r,0\right) =G^{\left( s\right) }\left(
r,0\right) =0.$

Once the functions $\varphi $ and $\boldsymbol{\Omega }$ are known, see
expressions (\ref{fi-konacno}) and (\ref{omega-konacno}), the displacement
field $\boldsymbol{u}$ can be determined from the wave equation of
viscoelastic body in Fourier and Laplace domain (\ref{Ru=pu}), which
transforms into%
\begin{gather*}
\tilde{c}_{s}^{2}\left( s\right) \,k^{2}\boldsymbol{\bar{\tilde{u}}}\left( 
\boldsymbol{k},s\right) +\left( \tilde{c}_{c}^{2}\left( s\right) -\tilde{c}%
_{s}^{2}\left( s\right) \right) \left( \boldsymbol{k}\otimes \boldsymbol{k}%
\right) \boldsymbol{\bar{\tilde{u}}}\left( \boldsymbol{k},s\right) +s^{2}%
\boldsymbol{\bar{\tilde{u}}}\left( \boldsymbol{k},s\right) =s\boldsymbol{%
\bar{u}}_{0}\left( \boldsymbol{k}\right) +\boldsymbol{\bar{v}}_{0}\left( 
\boldsymbol{k}\right) +\boldsymbol{\bar{\tilde{f}}}_{b}\left( \boldsymbol{k}%
,s\right) , \\
\tilde{c}_{s}^{2}\left( s\right) \,\mathrm{i}\boldsymbol{k}\times 
\boldsymbol{\bar{\tilde{\Omega}}}\left( \boldsymbol{k},s\right) -\tilde{c}%
_{c}^{2}\left( s\right) \,\mathrm{i}\boldsymbol{k}\bar{\tilde{\varphi}}%
\left( \boldsymbol{k},s\right) +s^{2}\boldsymbol{\bar{\tilde{u}}}\left( 
\boldsymbol{k},s\right) =s\boldsymbol{\bar{u}}_{0}\left( \boldsymbol{k}%
\right) +\boldsymbol{\bar{v}}_{0}\left( \boldsymbol{k}\right) +\boldsymbol{%
\bar{\tilde{f}}}_{b}\left( \boldsymbol{k},s\right) ,
\end{gather*}%
due to the memory functions (\ref{c-tilde-compresive}) and (\ref%
{c-tilde-shear}), as well as due to $k^{2}\boldsymbol{\bar{\tilde{u}}}\left( 
\boldsymbol{k},s\right) =\boldsymbol{k}\left( \boldsymbol{k}\cdot 
\boldsymbol{\bar{\tilde{u}}}\left( \boldsymbol{k},s\right) \right) -%
\boldsymbol{k}\times \left( \boldsymbol{k}\times \boldsymbol{\bar{\tilde{u}}}%
\left( \boldsymbol{k},s\right) \right) =-\mathrm{i}\boldsymbol{k}\bar{\tilde{%
\varphi}}\left( \boldsymbol{k},s\right) +\mathrm{i}\boldsymbol{k}\times 
\boldsymbol{\bar{\tilde{\Omega}}}\left( \boldsymbol{k},s\right) $ and $%
\left( \boldsymbol{k}\otimes \boldsymbol{k}\right) \boldsymbol{\bar{\tilde{u}%
}}\left( \boldsymbol{k},s\right) =\boldsymbol{k}\left( \boldsymbol{k}\cdot 
\boldsymbol{\bar{\tilde{u}}}\left( \boldsymbol{k},s\right) \right) =-\mathrm{%
i}\boldsymbol{k}\bar{\tilde{\varphi}}\left( \boldsymbol{k},s\right) ,$
yielding 
\begin{equation*}
\boldsymbol{\bar{\tilde{u}}}\left( \boldsymbol{k},s\right) =\frac{1}{s}%
\boldsymbol{\bar{u}}_{0}\left( \boldsymbol{k}\right) +\frac{1}{s^{2}}%
\boldsymbol{\bar{v}}_{0}\left( \boldsymbol{k}\right) +\frac{1}{s^{2}}\left( 
\boldsymbol{\bar{\tilde{f}}}_{b}\left( \boldsymbol{k},s\right) +\tilde{c}%
_{c}^{2}\left( s\right) \,\mathrm{i}\boldsymbol{k}\bar{\tilde{\varphi}}%
\left( \boldsymbol{k},s\right) -\tilde{c}_{s}^{2}\left( s\right) \,\mathrm{i}%
\boldsymbol{k}\times \boldsymbol{\bar{\tilde{\Omega}}}\left( \boldsymbol{k}%
,s\right) \right) ,
\end{equation*}%
which implies the displacement field in the form%
\begin{equation*}
\boldsymbol{u}\left( \boldsymbol{r},t\right) =\boldsymbol{u}_{0}\left( 
\boldsymbol{r}\right) +\boldsymbol{v}_{0}\left( \boldsymbol{r}\right)
t+t\ast _{t}\left( \boldsymbol{f}_{b}\left( \boldsymbol{r},t\right) +%
\mathcal{L}^{-1}\left[ \tilde{c}_{c}^{2}\left( s\right) \,\func{grad}\tilde{%
\varphi}\left( \boldsymbol{r},s\right) \right] \left( \boldsymbol{r}%
,t\right) -\mathcal{L}^{-1}\left[ \tilde{c}_{s}^{2}\left( s\right) \,\func{%
curl}\boldsymbol{\tilde{\Omega}}\left( \boldsymbol{r},s\right) \right]
\left( \boldsymbol{r},t\right) \right)
\end{equation*}%
after performing the inverse Fourier and Laplace transforms and becoming%
\begin{align}
\boldsymbol{u}\left( \boldsymbol{r},t\right) & =\boldsymbol{u}_{0}\left( 
\boldsymbol{r}\right) +\boldsymbol{v}_{0}\left( \boldsymbol{r}\right) t 
\notag \\
& \qquad +t\ast _{t}\Bigg(\boldsymbol{f}_{b}\left( \boldsymbol{r},t\right)
+\left\{ \!\!\!%
\begin{tabular}{ll}
$\frac{1}{\varrho }\dot{\sigma}_{sr}^{\left( c\right) }\left( t\right) \ast
_{t}\func{grad}\varphi \left( \boldsymbol{r},t\right) +\frac{1}{\varrho }%
\sigma _{sr/g}^{\left( c\right) }\func{grad}\varphi \left( \boldsymbol{r}%
,t\right) ,$ & if $\sigma _{sr/g}^{\left( c\right) }$ is finite, \smallskip
\\ 
$\frac{1}{\varrho }\sigma _{sr}^{\left( c\right) }\left( t\right) \ast
_{t}\left( \partial _{t}\func{grad}\varphi \left( \boldsymbol{r},t\right)
\right) +\frac{1}{\varrho }\sigma _{sr}^{\left( c\right) }\left( t\right) 
\func{grad}\varphi _{0}\left( \boldsymbol{r}\right) ,$ & if $\sigma
_{sr/g}^{\left( c\right) }$ is infinite, \smallskip%
\end{tabular}%
\ \right.  \notag \\
& \qquad \qquad \qquad \qquad -\left\{ \!\!\!%
\begin{tabular}{ll}
$\frac{1}{\varrho }\dot{\sigma}_{sr}^{\left( s\right) }\left( t\right) \ast
_{t}\func{curl}\boldsymbol{\Omega }\left( \boldsymbol{r},t\right) +\frac{1}{%
\varrho }\sigma _{sr/g}^{\left( s\right) }\func{curl}\boldsymbol{\Omega }%
\left( \boldsymbol{r},t\right) ,$ & if $\sigma _{sr/g}^{\left( s\right) }$
is finite, \smallskip \\ 
$\frac{1}{\varrho }\sigma _{sr}^{\left( s\right) }\left( t\right) \ast
_{t}\left( \partial _{t}\func{curl}\boldsymbol{\Omega }\left( \boldsymbol{r}%
,t\right) \right) +\frac{1}{\varrho }\sigma _{sr}^{\left( s\right) }\left(
t\right) \func{curl}\boldsymbol{\Omega }_{0}\left( \boldsymbol{r}\right) ,$
& if $\sigma _{sr/g}^{\left( s\right) }$ is infinite, \smallskip%
\end{tabular}%
\ \right. \Bigg)  \label{u-sol-razgranato}
\end{align}%
when material properties are taken into account, see (\ref{funkcija-c}) and (%
\ref{f-ja-c}), so that expression (\ref{u-sol-razgranato}) contains four
possible forms of the displacement field, due to the possibility of
compressive and shear waves to have finite, determined by%
\begin{equation*}
c_{c}=\sqrt{\frac{\sigma _{sr/g}^{\left( c\right) }}{\varrho }}\quad \text{%
and}\quad c_{s}=\sqrt{\frac{\sigma _{sr/g}^{\left( s\right) }}{\varrho }},
\end{equation*}%
or infinite wave propagation speeds.

Determining the displacement field via the expression (\ref{u-sol-razgranato}%
), which is apparently complicated, requires very complex and demanding
calculations, further complicated by including the expressions for scalar
and vector fields $\varphi $ and $\boldsymbol{\Omega }$,\ given by (\ref%
{fi-konacno}) and (\ref{omega-konacno}), which are also elaborate
expressions. Moreover, the form of displacement field depends on the models
of a viscoelastic body taken into account, having four different forms
depending on the finiteness of the wave propagation speed. Therefore, the
expression (\ref{u-sol-razgranato}) is certainly demanding for both
analytical and numerical calculations.

\subsection{Displacement field via resolvent tensor $\boldsymbol{\hat{R}}$ 
\label{DFRTR}}

Rather than using (\ref{u-sol-razgranato}), which requires laborious and
tedious calculations in order to determine the displacement field as a
solution to the Cauchy initial-value problem for the wave equation of
viscoelastic body, i.e., to the system of equations (\ref{SoE-gen-1}) - (\ref%
{SoE-gen-3}) subject to initial conditions (\ref{poc-usl}), the approach of
resolvent tensor calculation is adopted, so that the resolvent tensor
applied to the initial conditions and body force as prescribed forcing,
yields the displacement field. Namely, the wave equation of viscoelastic
body in Fourier and Laplace domain (\ref{Ru=pu}), rewritten as 
\begin{equation}
\boldsymbol{\bar{\tilde{\hat{R}}}}^{-1}\left( \boldsymbol{k},s\right) \,%
\boldsymbol{\bar{\tilde{u}}}\left( \boldsymbol{k},s\right) =s\boldsymbol{%
\bar{u}}_{0}\left( \boldsymbol{k}\right) +\boldsymbol{\bar{v}}_{0}\left( 
\boldsymbol{k}\right) +\boldsymbol{\bar{\tilde{f}}}_{b}\left( \boldsymbol{k}%
,s\right) ,  \label{Ru=pu-rewritten}
\end{equation}%
where the inverse tensor of the resolvent tensor in Fourier and Laplace
domain takes the form%
\begin{align}
\boldsymbol{\bar{\tilde{\hat{R}}}}^{-1}\left( \boldsymbol{k},s\right) &
=\left( \tilde{c}_{s}^{2}\left( s\right) \,k^{2}+s^{2}\right) \boldsymbol{%
\hat{I}}+\left( \tilde{c}_{c}^{2}\left( s\right) -\tilde{c}_{c}^{2}\left(
s\right) \right) \left( \boldsymbol{k}\otimes \boldsymbol{k}\right)  \notag
\\
& =\frac{1}{\bar{\tilde{G}}^{\left( s\right) }\left( k,s\right) }\left( 
\boldsymbol{\hat{I}}-\frac{\boldsymbol{k}\otimes \boldsymbol{k}}{k^{2}}%
\right) +\frac{1}{\bar{\tilde{G}}^{\left( c\right) }\left( k,s\right) }\frac{%
\boldsymbol{k}\otimes \boldsymbol{k}}{k^{2}},  \label{R-na-minus-prvi}
\end{align}%
according to memory functions (\ref{c-tilde-compresive}) and (\ref%
{c-tilde-shear}), as well as according to Green's functions (\ref{grin-c})$%
_{2}$ and (\ref{grin-s})$_{2}$, implies the displacement field in Fourier
and Laplace domain%
\begin{equation}
\boldsymbol{\bar{\tilde{u}}}\left( \boldsymbol{k},s\right) =\boldsymbol{\bar{%
\tilde{\hat{R}}}}\left( \boldsymbol{k},s\right) \left( s\boldsymbol{\bar{u}}%
_{0}\left( \boldsymbol{k}\right) +\boldsymbol{\bar{v}}_{0}\left( \boldsymbol{%
k}\right) +\boldsymbol{\bar{\tilde{f}}}_{b}\left( \boldsymbol{k},s\right)
\right) ,  \label{u-tilde-hat=R-tilde-hat-pu-tilde-hat}
\end{equation}%
transforming into%
\begin{equation}
\boldsymbol{u}\left( \boldsymbol{r},t\right) =\partial _{t}\boldsymbol{\hat{R%
}}\left( \boldsymbol{r},t\right) \ast _{\boldsymbol{r}}\boldsymbol{u}%
_{0}\left( \boldsymbol{r}\right) +\boldsymbol{\hat{R}}\left( \boldsymbol{r}%
,t\right) \ast _{\boldsymbol{r}}\boldsymbol{v}_{0}\left( \boldsymbol{r}%
\right) +\boldsymbol{\hat{R}}\left( \boldsymbol{r},t\right) \ast _{%
\boldsymbol{r},t}\boldsymbol{f}_{b}\left( \boldsymbol{r},t\right) ,
\label{u-solution-2}
\end{equation}%
assuming $\boldsymbol{\hat{R}}\left( \boldsymbol{r},0\right) =0,$ after the
Fourier and Laplace transforms inversions are performed, so that it is
necessary to determine the resolvent tensor $\boldsymbol{\hat{R}}$ and its
first derivative with respect to time.

In order to do so, the first step is to find the resolvent tensor in Fourier
and Laplace domain $\boldsymbol{\bar{\tilde{\hat{R}}}}\ $in the form 
\begin{equation}
\boldsymbol{\bar{\tilde{\hat{R}}}}\left( \boldsymbol{k},s\right) =\bar{%
\tilde{G}}^{\left( s\right) }\left( k,s\right) \left( \boldsymbol{\hat{I}}-%
\frac{\boldsymbol{k}\otimes \boldsymbol{k}}{k^{2}}\right) +\bar{\tilde{G}}%
^{\left( c\right) }\left( k,s\right) \frac{\boldsymbol{k}\otimes \boldsymbol{%
k}}{k^{2}},  \label{rezolventa-furije-laplas}
\end{equation}%
through its inverse counterpart $\boldsymbol{\bar{\tilde{\hat{R}}}}^{-1},$
given by (\ref{R-na-minus-prvi}), since $\boldsymbol{\bar{\tilde{\hat{R}}}}%
\left( \boldsymbol{k},s\right) \boldsymbol{\bar{\tilde{\hat{R}}}}^{-1}\left( 
\boldsymbol{k},s\right) =\boldsymbol{\hat{I}},$ due to $\frac{\boldsymbol{k}%
\otimes \boldsymbol{k}}{k^{2}}\frac{\boldsymbol{k}\otimes \boldsymbol{k}}{%
k^{2}}=\frac{\boldsymbol{k}\otimes \boldsymbol{k}}{k^{2}},$ so that (\ref%
{rezolventa-furije-laplas}) transforms into%
\begin{align*}
\boldsymbol{\bar{\tilde{\hat{R}}}}\left( \boldsymbol{k},s\right) & =\bar{%
\tilde{G}}^{\left( s\right) }\left( k,s\right) \boldsymbol{\hat{I}}-\frac{1}{%
k^{2}}\bar{\tilde{G}}^{\left( s\right) }\left( k,s\right) \left( \boldsymbol{%
k}\otimes \boldsymbol{k}\right) +\frac{1}{k^{2}}\bar{\tilde{G}}^{\left(
c\right) }\left( k,s\right) \left( \boldsymbol{k}\otimes \boldsymbol{k}%
\right) \\
& =\bar{\tilde{G}}^{\left( s\right) }\left( k,s\right) \boldsymbol{\hat{I}}+%
\frac{\tilde{c}_{s}^{2}\left( s\right) }{s^{2}}\left( \boldsymbol{k}\otimes 
\boldsymbol{k}\right) \bar{\tilde{G}}^{\left( s\right) }\left( k,s\right) -%
\frac{\tilde{c}_{c}^{2}\left( s\right) }{s^{2}}\left( \boldsymbol{k}\otimes 
\boldsymbol{k}\right) \bar{\tilde{G}}^{\left( c\right) }\left( k,s\right)
\end{align*}%
using $\frac{1}{k^{2}}\bar{\tilde{G}}^{\left( x\right) }\left( k,s\right) =%
\frac{1}{k^{2}}\frac{1}{\tilde{c}_{x}^{2}\left( s\right) \,k^{2}+s^{2}}=%
\frac{1}{s^{2}}\frac{1}{k^{2}}-\frac{\tilde{c}_{x}^{2}\left( s\right) }{s^{2}%
}\frac{1}{\tilde{c}_{x}^{2}\left( s\right) \,k^{2}+s^{2}}=\frac{1}{s^{2}}%
\frac{1}{k^{2}}-\frac{\tilde{c}_{x}^{2}\left( s\right) }{s^{2}}\bar{\tilde{G}%
}^{\left( x\right) }\left( k,s\right) .$

In the second step, the inversion of Fourier transform in the previous
expression yields the resolvent tensor in Laplace domain $\boldsymbol{\tilde{%
\hat{R}}}$, taking the form%
\begin{equation}
\boldsymbol{\tilde{\hat{R}}}\left( \boldsymbol{r},s\right) =\tilde{G}%
^{\left( s\right) }\left( r,s\right) \boldsymbol{\hat{I}}-\frac{\tilde{c}%
_{s}^{2}\left( s\right) }{s^{2}}\left( \nabla \otimes \nabla \right) \tilde{G%
}^{\left( s\right) }\left( r,s\right) +\frac{\tilde{c}_{c}^{2}\left(
s\right) }{s^{2}}\left( \nabla \otimes \nabla \right) \tilde{G}^{\left(
c\right) }\left( r,s\right)  \label{Rezolventa-laplas}
\end{equation}%
since $\mathcal{F}\left[ \nabla \otimes \nabla \right] \left( \boldsymbol{k}%
\right) =\mathrm{i}\boldsymbol{k}\otimes \mathrm{i}\boldsymbol{k=}-\left( 
\boldsymbol{k}\otimes \boldsymbol{\boldsymbol{k}}\right) ,$ where Green's
function in Laplace domain is obtained as%
\begin{equation}
\tilde{G}^{\left( x\right) }\left( r,s\right) =\mathcal{F}^{-1}\left[ \bar{%
\tilde{G}}^{\left( x\right) }\left( k,s\right) \right] \left( r,s\right) =%
\frac{1}{4\pi r\tilde{c}_{x}^{2}\left( s\right) }\mathrm{e}^{-\frac{rs}{%
\tilde{c}_{x}\left( s\right) }},\quad x\in \left\{ c,s\right\} ,
\label{Green-f-ld}
\end{equation}%
by calculating the inverse Fourier transform%
\begin{align*}
\tilde{G}^{\left( x\right) }\left( r,s\right) & =\frac{1}{\left( 2\pi
\right) ^{3}}\int_{%
\mathbb{R}
^{3}}\bar{\tilde{G}}^{\left( x\right) }\left( k,s\right) \mathrm{e}^{\mathrm{%
i}\boldsymbol{k}\cdot \boldsymbol{r}}\mathrm{d}_{\boldsymbol{k}}V \\
& =\frac{1}{\left( 2\pi \right) ^{3}}\int_{0}^{2\pi }\int_{0}^{\pi
}\int_{0}^{\infty }\frac{1}{\tilde{c}_{x}^{2}\left( s\right) \,k^{2}+s^{2}}%
\mathrm{e}^{-\mathrm{i}kr\cos \theta }k^{2}\sin \theta \mathrm{d}k\mathrm{d}%
\theta \mathrm{d}\varphi \\
& =\frac{1}{\left( 2\pi \right) ^{2}r}\int_{0}^{\infty }\frac{k}{\tilde{c}%
_{x}^{2}\left( s\right) \,k^{2}+s^{2}}\mathrm{d}k\int_{-kr}^{kr}\mathrm{e}^{-%
\mathrm{i}p}\mathrm{d}p \\
& =\frac{1}{2\pi ^{2}r}\int_{0}^{\infty }\frac{k\sin \left( kr\right) }{%
\tilde{c}_{x}^{2}\left( s\right) \,k^{2}+s^{2}}\mathrm{d}k \\
& =\frac{1}{2\pi ^{2}r\tilde{c}_{x}^{2}\left( s\right) }\int_{0}^{\infty }%
\frac{q\sin q}{q^{2}+\left( \frac{rs}{\tilde{c}_{x}\left( s\right) }\right)
^{2}}\mathrm{d}q \\
& =\frac{1}{4\pi r\tilde{c}_{x}^{2}\left( s\right) }\mathrm{e}^{-\frac{rs}{%
\tilde{c}_{x}\left( s\right) }},
\end{align*}%
using the integral%
\begin{equation*}
\int_{0}^{\infty }\frac{q\sin q}{q^{2}+\lambda }\mathrm{d}q=\frac{\pi }{2}%
\mathrm{e}^{-\sqrt{\lambda }},\quad \text{if}\quad \func{Re}\lambda >0,
\end{equation*}%
obtained by the integration in complex plane, with Green's functions in
Fourier and Laplace domain $\bar{\tilde{G}}^{\left( x\right) },$ given by (%
\ref{grin-c})$_{2}$ and (\ref{grin-s})$_{2}.$ The condition $\func{Re}%
\lambda >0$ implies $\func{Re}\frac{s^{2}}{\tilde{c}_{x}^{2}\left( s\right) }%
>0$ for $\func{Re}s>0,$ whose validity is checked in Appendix \ref%
{Justification}. The terms $\left( \nabla \otimes \nabla \right) \tilde{G}%
^{\left( x\right) }\left( r,s\right) ,$ $x\in \left\{ c,s\right\} ,$
occurring in (\ref{Rezolventa-laplas}), are calculated as 
\begin{align*}
\left( \nabla \otimes \nabla \right) \tilde{G}^{\left( x\right) }\left(
r,s\right) & =\frac{1}{r}\partial _{r}\left( \frac{1}{r}\partial _{r}\tilde{G%
}^{\left( x\right) }\left( r,s\right) \right) \left( \boldsymbol{r}\otimes 
\boldsymbol{r}\right) +\frac{1}{r}\partial _{r}\tilde{G}^{\left( x\right)
}\left( r,s\right) \boldsymbol{\hat{I}} \\
& =\left( \frac{3}{r^{2}}+\frac{3}{r}\frac{s}{\tilde{c}_{x}\left( s\right) }+%
\frac{s^{2}}{\tilde{c}_{x}^{2}\left( s\right) }\right) \tilde{G}^{\left(
x\right) }\left( r,s\right) \frac{\boldsymbol{r}\otimes \boldsymbol{r}}{r^{2}%
}-\frac{1}{r}\left( \frac{1}{r}+\frac{s}{\tilde{c}_{x}\left( s\right) }%
\right) \tilde{G}^{\left( x\right) }\left( r,s\right) \boldsymbol{\hat{I}},
\end{align*}%
using $\left( \nabla \otimes \nabla \right) f\left( r\right) =\nabla \otimes
\left( \nabla f\left( r\right) \right) =\func{grad}\func{grad}f\left(
r\right) =\func{grad}\left( \frac{\mathrm{d}f\left( r\right) }{\mathrm{d}r}%
\func{grad}r\right) =\func{grad}\left( \frac{\mathrm{d}f\left( r\right) }{%
\mathrm{d}r}\frac{\boldsymbol{r}}{r}\right) =\func{grad}\left( \frac{1}{r}%
\frac{\mathrm{d}f\left( r\right) }{\mathrm{d}r}\right) \otimes \boldsymbol{r}%
+\frac{1}{r}\frac{\mathrm{d}f\left( r\right) }{\mathrm{d}r}\boldsymbol{\hat{I%
}}=\frac{1}{r}\frac{\mathrm{d}}{\mathrm{d}r}\left( \frac{1}{r}\frac{\mathrm{d%
}f\left( r\right) }{\mathrm{d}r}\right) \left( \boldsymbol{r}\otimes 
\boldsymbol{r}\right) +\frac{1}{r}\frac{\mathrm{d}f\left( r\right) }{\mathrm{%
d}r}\boldsymbol{\hat{I}}$, as well as using%
\begin{equation*}
\partial _{r}\tilde{G}^{\left( x\right) }\left( r,s\right) =-\left( \frac{1}{%
r}+\frac{s}{\tilde{c}_{x}\left( s\right) }\right) \tilde{G}^{\left( x\right)
}\left( r,s\right) ,
\end{equation*}%
see Green's function in Laplace domain given by (\ref{Green-f-ld}),
transforming (\ref{Rezolventa-laplas}) into 
\begin{align*}
\boldsymbol{\tilde{\hat{R}}}\left( \boldsymbol{r},s\right) & =\tilde{G}%
^{\left( c\right) }\left( r,s\right) \frac{\boldsymbol{r}\otimes \boldsymbol{%
r}}{r^{2}}+\tilde{G}^{\left( s\right) }\left( r,s\right) \left( \boldsymbol{%
\hat{I}}-\frac{\boldsymbol{r}\otimes \boldsymbol{r}}{r^{2}}\right) \\
& \qquad +\frac{1}{r}\left( \frac{\tilde{c}_{s}^{2}\left( s\right) }{s^{2}}%
\left( \frac{1}{r}+\frac{s}{\tilde{c}_{s}\left( s\right) }\right) \tilde{G}%
^{\left( s\right) }\left( r,s\right) -\frac{\tilde{c}_{c}^{2}\left( s\right) 
}{s^{2}}\left( \frac{1}{r}+\frac{s}{\tilde{c}_{c}\left( s\right) }\right) 
\tilde{G}^{\left( c\right) }\left( r,s\right) \right) \left( \boldsymbol{%
\hat{I}}-3\frac{\boldsymbol{r}\otimes \boldsymbol{r}}{r^{2}}\right) ,
\end{align*}%
i.e., 
\begin{equation}
\boldsymbol{\tilde{\hat{R}}}\left( \boldsymbol{r},s\right) =\tilde{G}%
^{\left( c\right) }\left( r,s\right) \frac{\boldsymbol{r}\otimes \boldsymbol{%
r}}{r^{2}}+\tilde{G}^{\left( s\right) }\left( r,s\right) \left( \boldsymbol{%
\hat{I}}-\frac{\boldsymbol{r}\otimes \boldsymbol{r}}{r^{2}}\right) -\frac{1}{%
r}\left( \partial _{r}\tilde{g}^{\left( s\right) }\left( r,s\right)
-\partial _{r}\tilde{g}^{\left( c\right) }\left( r,s\right) \right) \left( 
\boldsymbol{\hat{I}}-3\frac{\boldsymbol{r}\otimes \boldsymbol{r}}{r^{2}}%
\right) ,  \label{R-tilde}
\end{equation}%
where functions $\tilde{g}^{\left( x\right) },$ $x\in \left\{ c,s\right\} ,$
defined by%
\begin{equation}
\tilde{g}^{\left( x\right) }\left( r,s\right) =\frac{\tilde{c}_{x}^{2}\left(
s\right) }{s^{2}}\tilde{G}^{\left( x\right) }\left( r,s\right) =\frac{1}{%
4\pi r}\frac{1}{s^{2}}\mathrm{e}^{-\frac{rs}{\tilde{c}_{x}\left( s\right) }},
\label{g-malo-tilde}
\end{equation}%
see also (\ref{Green-f-ld}), imply%
\begin{equation*}
\partial _{r}\tilde{g}^{\left( x\right) }\left( r,s\right) =-\frac{\tilde{c}%
_{x}^{2}\left( s\right) }{s^{2}}\left( \frac{1}{r}+\frac{s}{\tilde{c}%
_{x}\left( s\right) }\right) \tilde{G}^{\left( x\right) }\left( r,s\right) .
\end{equation*}

In order to find the resolvent tensor $\boldsymbol{\hat{R}},$ in the final
step, it is necessary to invert the Laplace transform of Green's functions
in Laplace domain $\tilde{G}^{\left( x\right) }$, see (\ref{Green-f-ld}), as
well as of functions $\tilde{g}^{\left( x\right) },$ $x\in \left\{
c,s\right\} ,$ see (\ref{g-malo-tilde}), so that the resolvent tensor in
Laplace domain (\ref{R-tilde}) transforms into%
\begin{equation}
\boldsymbol{\hat{R}}\left( \boldsymbol{r},t\right) =G^{\left( c\right)
}\left( r,t\right) \frac{\boldsymbol{r}\otimes \boldsymbol{r}}{r^{2}}%
+G^{\left( s\right) }\left( r,t\right) \left( \boldsymbol{\hat{I}}-\frac{%
\boldsymbol{r}\otimes \boldsymbol{r}}{r^{2}}\right) -\frac{1}{r}\left(
\partial _{r}g^{\left( s\right) }\left( r,t\right) -\partial _{r}g^{\left(
c\right) }\left( r,t\right) \right) \left( \boldsymbol{\hat{I}}-3\frac{%
\boldsymbol{r}\otimes \boldsymbol{r}}{r^{2}}\right) ,  \label{R-tilde-prim}
\end{equation}%
or equivalently into%
\begin{align}
\boldsymbol{\hat{R}}\left( \boldsymbol{r},t\right) & =G^{\left( c\right)
}\left( r,t\right) \left[ 
\begin{array}{c}
\boldsymbol{e}_{r} \\ 
\boldsymbol{0} \\ 
\boldsymbol{0}%
\end{array}%
\right] ^{\mathrm{T}}+G^{\left( s\right) }\left( r,t\right) \left[ 
\begin{array}{c}
\boldsymbol{0} \\ 
\boldsymbol{e}_{\theta } \\ 
\boldsymbol{e}_{\varphi }%
\end{array}%
\right] ^{\mathrm{T}}-\frac{1}{r}\left( \partial _{r}g^{\left( s\right)
}\left( r,t\right) -\partial _{r}g^{\left( c\right) }\left( r,t\right)
\right) \left[ 
\begin{array}{c}
-2\boldsymbol{e}_{r} \\ 
\boldsymbol{e}_{\theta } \\ 
\boldsymbol{e}_{\varphi }%
\end{array}%
\right] ^{\mathrm{T}},\quad \text{i.e.,}  \notag \\
\boldsymbol{\hat{R}}\left( \boldsymbol{r},t\right) & =\left[ 
\begin{array}{c}
G^{\left( c\right) }\left( r,t\right) \boldsymbol{e}_{r} \\ 
G^{\left( s\right) }\left( r,t\right) \boldsymbol{e}_{\theta } \\ 
G^{\left( s\right) }\left( r,t\right) \boldsymbol{e}_{\varphi }%
\end{array}%
\right] ^{\mathrm{T}}-\frac{1}{r}\left( \partial _{r}g^{\left( s\right)
}\left( r,t\right) -\partial _{r}g^{\left( c\right) }\left( r,t\right)
\right) \left[ 
\begin{array}{c}
-2\boldsymbol{e}_{r} \\ 
\boldsymbol{e}_{\theta } \\ 
\boldsymbol{e}_{\varphi }%
\end{array}%
\right] ^{\mathrm{T}},  \label{R-tilde-dvoprim}
\end{align}%
where $\boldsymbol{e}_{r},$ $\boldsymbol{e}_{\theta },$ and $\boldsymbol{e}%
_{\varphi }$ are unit vectors of the spherical coordinate system, due to the
diagonalization of tensors appearing in the resolvent tensor (\ref%
{R-tilde-prim}), namely due to%
\begin{equation*}
\frac{\boldsymbol{r}\otimes \boldsymbol{r}}{r^{2}}=\left[ 
\begin{array}{ccc}
1 & 0 & 0 \\ 
0 & 0 & 0 \\ 
0 & 0 & 0%
\end{array}%
\right] =\left[ 
\begin{array}{c}
\boldsymbol{e}_{r} \\ 
\boldsymbol{0} \\ 
\boldsymbol{0}%
\end{array}%
\right] ^{\mathrm{T}}
\end{equation*}%
implying%
\begin{equation*}
\boldsymbol{\hat{I}}-\frac{\boldsymbol{r}\otimes \boldsymbol{r}}{r^{2}}=%
\left[ 
\begin{array}{ccc}
0 & 0 & 0 \\ 
0 & 1 & 0 \\ 
0 & 0 & 1%
\end{array}%
\right] =\left[ 
\begin{array}{c}
\boldsymbol{0} \\ 
\boldsymbol{e}_{\theta } \\ 
\boldsymbol{e}_{\varphi }%
\end{array}%
\right] ^{\mathrm{T}}\quad \text{and}\quad \boldsymbol{\hat{I}}-3\frac{%
\boldsymbol{r}\otimes \boldsymbol{r}}{r^{2}}=\left[ 
\begin{array}{ccc}
-2 & 0 & 0 \\ 
0 & 1 & 0 \\ 
0 & 0 & 1%
\end{array}%
\right] =\left[ 
\begin{array}{c}
-2\boldsymbol{e}_{r} \\ 
\boldsymbol{e}_{\theta } \\ 
\boldsymbol{e}_{\varphi }%
\end{array}%
\right] ^{\mathrm{T}},
\end{equation*}%
since the eigenvalue problem%
\begin{equation*}
\frac{\boldsymbol{r}\otimes \boldsymbol{r}}{r^{2}}\boldsymbol{u}=\lambda 
\hat{I}\boldsymbol{u}
\end{equation*}%
yields eigenvalues $\lambda _{1}=1$ and $\lambda _{2,3}=0,$ with
eigenvectors $\boldsymbol{u}^{\left( 1\right) }=\frac{\boldsymbol{r}}{r}=%
\boldsymbol{e}_{r}$ and $\boldsymbol{r}\cdot \boldsymbol{u}^{\left(
2,3\right) }=0,$ implying $r\boldsymbol{u}^{\left( 1\right) }\cdot 
\boldsymbol{u}^{\left( 2,3\right) }=0,$ i.e., $r\boldsymbol{e}_{r}\cdot 
\boldsymbol{u}^{\left( 2,3\right) }=0,$ so that eigenvectors $\boldsymbol{u}%
^{\left( 2,3\right) }$ can be identified as the unit vectors $\boldsymbol{e}%
_{\theta }$ and $\boldsymbol{e}_{\varphi }$.

The simple matrix form of resolvent tensor (\ref{R-tilde-dvoprim}) clearly
shows that the compressive wave progresses in the direction of a radius
vector, depending on its intensity, while the shear wave progresses in the
plane perpendicular to the direction of a radius vector, see the first term
in (\ref{R-tilde-dvoprim}), while the second term in (\ref{R-tilde-dvoprim})
accounts for the additional contributions.

\subsection{Green's functions\label{Green's function}}

Green's functions $G^{\left( x\right) },$ $x\in \left\{ c,s\right\} ,$
appearing in the resolvent tensor (\ref{R-tilde-prim}) and (\ref%
{R-tilde-dvoprim}), take similar but different forms depending on the wave
propagation speed character, while functions $g^{\left( x\right) }$ have the
same form regardless of the wave speed. Namely, by examining the short-time
asymptotics of Green's function, one finds that Green's function is actually
a distribution in the case when wave propagation speed is finite, which
implies that one has to regularize Green's function in Laplace domain in
order to be able to apply Laplace inversion formula and obtain Green's
function at arbitrary time instant, while in the case when wave propagation
speed is infinite, no regularization of Green's function is needed, since
its short-time asymptotics implies that Green's function is a function.
Short- and large-time asymptotics of Green's function are obtained from the
asymptotics of Green's function in Laplace domain, governed by the behavior
of memory function $\tilde{c}_{x}\left( s\right) $ as $s\rightarrow \infty $
and $s\rightarrow 0,$ respectively, which is established in Appendix \ref%
{Asimptotika-za-c}.

If wave propagation speed is finite, i.e., if glass modulus $\sigma
_{sr/g}^{\left( x\right) }$ attains a finite value, then Green's function is
actually a distribution, which is apparent from the asymptotics of Green's
function in Laplace domain (\ref{Green-f-ld}), being of the form%
\begin{equation*}
\tilde{G}^{\left( x\right) }\left( r,s\right) \sim \frac{1}{4\pi rc_{x}^{2}}%
\mathrm{e}^{-\frac{rs}{c_{x}}},\quad \text{as}\quad s\rightarrow \infty ,
\end{equation*}%
thus implying the asymptotics of Green's function in time domain 
\begin{equation}
G^{\left( x\right) }\left( r,t\right) \sim \frac{1}{4\pi rc_{x}^{2}}\delta
\left( t-\frac{r}{c_{x}}\right) ,\quad \text{as}\quad t\rightarrow 0,
\label{G-srsha-konacna-brzina}
\end{equation}%
where $\delta $ is a Dirac delta distribution, which is obtained according
to the theorem that if $\tilde{f}\left( s\right) \sim \tilde{g}\left(
s\right) $ as $s\rightarrow \infty ,$ then $f\left( t\right) \sim g\left(
t\right) $ as $t\rightarrow 0,$ since the asymptotics of memory function $%
\tilde{c}_{x},$ according to (\ref{memory-function-asym}), yields%
\begin{equation*}
\tilde{c}_{x}\left( s\right) \sim \kappa ,\quad \text{as}\quad s\rightarrow
\infty ,
\end{equation*}%
with $\kappa $ given in Tables \ref{tbl-1}, \ref{tbl-4}, \ref{tbl-2}, and %
\ref{tbl-3}, on one hand, while the expressions for memory functions (\ref%
{c-tilde-compresive}) and (\ref{c-tilde-shear}) and calculations in Appendix %
\ref{CalcGrin} yield 
\begin{equation*}
\tilde{c}_{x}\left( s\right) =\sqrt{\frac{s\tilde{\sigma}_{sr}^{\left(
x\right) }\left( s\right) }{\varrho }}\sim \sqrt{\frac{\sigma
_{sr/g}^{\left( x\right) }}{\varrho }}=c_{x},\quad \text{as}\quad
s\rightarrow \infty
\end{equation*}%
on the other hand, see also (\ref{ce-iks}). Note, calculation of Green's
function, conducted in Appendix \ref{CalcGrin}, implies that Green's
function has non-zero values if $r<c_{x}t,$ with $c_{x}$ taken as above.
Therefore, materials described by Case I and III of linear fractional-order
models containing derivatives of orders in interval $\left( 0,1\right) $, as
well as fractional Burgers models belonging to the second class, have finite
speed of mechanical disturbance propagation.

If wave propagation speed is finite, then Green's function in time domain is
obtained via regularization in Laplace domain%
\begin{equation}
\tilde{G}_{\varepsilon }^{\left( x\right) }\left( r,s\right) =\tilde{G}%
^{\left( x\right) }\left( r,s\right) \tilde{\delta}_{\varepsilon }\left(
s\right)  \label{G-ld-reg}
\end{equation}%
implying%
\begin{equation}
G^{\left( x\right) }\left( r,t\right) =\lim_{\varepsilon \rightarrow
0}G_{\varepsilon }^{\left( x\right) }\left( r,t\right) =G^{\left( x\right)
}\left( r,t\right) \ast _{t}\lim_{\varepsilon \rightarrow 0}\delta
_{\varepsilon }\left( t\right) =G^{\left( x\right) }\left( r,t\right) \ast
_{t}\delta \left( t\right) ,  \label{racunanje-G-eps}
\end{equation}%
where the regularization of Dirac delta distribution is chosen as 
\begin{equation}
\delta _{\varepsilon }\left( t\right) =\frac{\varepsilon }{2t\sqrt{\pi t}}%
\mathrm{e}^{-\frac{\varepsilon ^{2}}{4t}},\quad \text{with}\quad \tilde{%
\delta}_{\varepsilon }\left( s\right) =\mathrm{e}^{-\varepsilon \sqrt{s}},
\label{delta-reg}
\end{equation}%
since $\lim_{\varepsilon \rightarrow 0}\delta _{\varepsilon }\left( t\right)
=\delta \left( t\right) ,$ due to $\left. \tilde{\delta}_{\varepsilon
}\left( s\right) \right\vert _{\varepsilon =0}=\left. \mathrm{e}%
^{-\varepsilon \sqrt{s}}\right\vert _{\varepsilon =0}=1=\mathcal{L}\left[
\delta \left( t\right) \right] .$ The regularized Green's function is
obtained according to 
\begin{align}
G_{\varepsilon }^{\left( x\right) }\left( r,t\right) & =\mathcal{L}^{-1}%
\left[ \tilde{G}^{\left( x\right) }\left( r,s\right) \tilde{\delta}%
_{\varepsilon }\left( s\right) \right] \left( r,t\right)  \notag \\
& =\mathcal{L}^{-1}\left[ \frac{1}{4\pi r\tilde{c}_{x}^{2}\left( s\right) }%
\mathrm{e}^{-\frac{rs}{\tilde{c}_{x}\left( s\right) }}\mathrm{e}%
^{-\varepsilon \sqrt{s}}\right] \left( r,t\right)  \label{G-f-reg} \\
& =\mathcal{L}^{-1}\left[ \frac{1}{4\pi r}\varrho \frac{\Phi _{\sigma }(s)}{%
\Phi _{\varepsilon }(s)}\mathrm{e}^{-rs\sqrt{\varrho }\sqrt{\frac{\Phi
_{\sigma }(s)}{\Phi _{\varepsilon }(s)}}}\mathrm{e}^{-\varepsilon \sqrt{s}}%
\right] \left( r,t\right) ,  \notag
\end{align}%
see expressions (\ref{Green-f-ld}), (\ref{c-tilde-compresive}), (\ref%
{c-tilde-shear}), and (\ref{sr-cr}), as well as Appendix \ref{CalcGrin} for
calculations, taking the explicit form 
\begin{align}
G_{\varepsilon }^{\left( x\right) }\left( r,t\right) & =\frac{1}{4\pi ^{2}r}%
\int_{0}^{\infty }\frac{1}{\left\vert \tilde{c}_{x}^{2}\left( \rho \mathrm{e}%
^{\mathrm{i}\varphi _{0}}\right) \right\vert }\mathrm{e}^{\rho t\cos \varphi
_{0}-\frac{r\rho }{\left\vert \tilde{c}_{x}\left( \rho \mathrm{e}^{\mathrm{i}%
\varphi _{0}}\right) \right\vert }\cos \left( \varphi _{0}-\arg \tilde{c}%
_{x}\left( \rho \mathrm{e}^{\mathrm{i}\varphi _{0}}\right) \right)
-\varepsilon \sqrt{\rho }\mathrm{\cos }\frac{\varphi _{0}}{2}}  \notag \\
& \qquad \qquad \times \sin \left( \rho t\mathrm{\sin }\varphi _{0}-\frac{%
r\rho }{\left\vert \tilde{c}_{x}\left( \rho \mathrm{e}^{\mathrm{i}\varphi
_{0}}\right) \right\vert }\sin \left( \varphi _{0}-\arg \tilde{c}_{x}\left(
\rho \mathrm{e}^{\mathrm{i}\varphi _{0}}\right) \right) +\varphi _{0}-\arg 
\tilde{c}_{x}^{2}\left( \rho \mathrm{e}^{\mathrm{i}\varphi _{0}}\right)
-\varepsilon \sqrt{\rho }\mathrm{\sin }\frac{\varphi _{0}}{2}\right) \mathrm{%
d}\rho ,  \label{grin-f-finite}
\end{align}%
if $r<c_{x}t,$ with $c_{x}=\sqrt{\frac{\sigma _{sr/g}^{\left( x\right) }}{%
\varrho }},$ while if $r>c_{x}t,$ then%
\begin{equation*}
G_{\varepsilon }^{\left( x\right) }\left( r,t\right) =0,
\end{equation*}%
where $\varphi _{0}=\pi $ if function $\Phi _{\sigma }$ (function $\Phi
_{\varepsilon }$) either has no zeros or has a negative real zero, while $%
\varphi _{0}=\arg s_{0}$ if $s_{0}$ and its complex conjugate $\bar{s}_{0}$
are zeros of function $\Phi _{\sigma }$ (function $\Phi _{\varepsilon }$).
Note, the constant $c_{x}$ represents the speed of compressive/shear wave
propagation, see (\ref{ce-iks}), and it originates from the asymptotics (\ref%
{memory-function-asym}) of the memory function $\tilde{c}_{x}$, being
connected with constant $\kappa $ in cases when $\delta =0,$ see Tables \ref%
{tbl-1}, \ref{tbl-4}, \ref{tbl-2}, and \ref{tbl-3}, by $c_{x}=\kappa $.

The integral representation of the asymptotics of regularized Green's
function (\ref{grin-f-finite}) is calculated in Appendix \ref{Green-asy-calc}
in the form%
\begin{equation}
G_{\varepsilon ,\mathrm{asy}}^{\left( x\right) }\left( r,t\right) =\frac{1}{%
4\pi ^{2}r\kappa ^{2}}\int_{0}^{\infty }\mathrm{e}^{-\rho \left( t-\frac{r}{%
\kappa }\right) }\sin \left( \varepsilon \sqrt{\rho }\right) \mathrm{d}\rho ,
\label{srsha-int-repr}
\end{equation}%
according to (\ref{G-f-reg}), since $\tilde{c}_{x}\left( s\right) \sim
\kappa ,$ as $s\rightarrow \infty .$

On the other hand, if wave propagation speed is infinite, i.e., if glass
modulus $\sigma _{sr/g}^{\left( x\right) }$ tends to infinity, then no
regularization of Green's function is needed, which is apparent from the
asymptotics of Green's function in Laplace domain (\ref{Green-f-ld}), being
of the form%
\begin{equation}
\tilde{G}^{\left( x\right) }\left( r,s\right) \sim \frac{1}{4\pi \kappa ^{2}r%
}\frac{1}{s^{\delta }}\mathrm{e}^{-\frac{r}{\kappa }s^{1-\frac{\delta }{2}%
}},\quad \text{as}\quad s\rightarrow \infty ,  \label{G-f-short-time-infinit}
\end{equation}%
thus, according to the same theorem as above, implying the short-time
asymptotics of Green's function%
\begin{equation}
G^{\left( x\right) }\left( r,t\right) \sim \frac{1}{4\pi ^{2}\kappa ^{2}r}%
\int_{0}^{\infty }\frac{1}{\rho ^{\delta }}\mathrm{e}^{-\rho \left( t-\frac{r%
}{\kappa }\rho ^{-\frac{\delta }{2}}\cos \frac{\delta \pi }{2}\right) }\sin
\left( \frac{r}{\kappa }\rho ^{1-\frac{\delta }{2}}\sin \frac{\delta \pi }{2}%
+\delta \pi \right) \mathrm{d}\rho ,\quad \text{as}\quad t\rightarrow 0,
\label{G-short-time}
\end{equation}%
calculated in Appendix \ref{Green-asy-calc}, since 
\begin{equation*}
\tilde{c}_{x}\left( s\right) =\sqrt{\frac{s\tilde{\sigma}_{sr}^{\left(
x\right) }\left( s\right) }{\varrho }}\sim \kappa s^{\frac{\delta }{2}%
},\quad \text{as}\quad s\rightarrow \infty ,\quad \text{with}\quad \delta
\in \left( 0,1\right) ,
\end{equation*}%
see also (\ref{memory-function-asym}) as well as Tables \ref{tbl-1}, \ref%
{tbl-4}, \ref{tbl-2}, and \ref{tbl-3} for values of $\kappa $ and $\delta .$
Therefore, materials described by Case II and IV of linear fractional-order
models containing derivatives of orders in interval $\left( 0,1\right) $,
fractional Burgers models belonging to the first class, as well as all
fractional anti-Zener and Zener models have infinite speed of mechanical
disturbance propagation.

Green's function, appearing in the resolvent tensor (\ref{R-tilde-prim}) and
(\ref{R-tilde-dvoprim}), is given by (\ref{grin-f-finite}) on the domain $%
r>0 $ and $t>0,$ for $\varepsilon =0,$ i.e., by%
\begin{equation}
G^{\left( x\right) }\left( r,t\right) =\left. G_{\varepsilon }^{\left(
x\right) }\left( r,t\right) \right\vert _{\varepsilon =0}.
\label{G-za-infne}
\end{equation}%
Note, calculation of Green's function, conducted in Appendix \ref{CalcGrin},
implies that Green's function has non-zero values for all points in space $%
r>0$ at any time instant $t>0,$ and moreover the large-time asymptotics, as
proved below, implies that Green's function tends to zero as a power-type
function for any $r>0$.

In order to investigate the large-time asymptotics of Green's function, one
again starts from Green's function in Laplace domain (\ref{Green-f-ld}) and
asymptotics of memory function $\tilde{c}_{x},$ taking the form 
\begin{equation*}
\tilde{c}_{x}\left( s\right) \sim \left\{ \!\!\!%
\begin{tabular}{ll}
$\kappa ,$ & \smallskip \\ 
$\kappa s^{\frac{\delta }{2}},$ & $\delta \in \left( 0,1\right) $, \smallskip%
\end{tabular}%
\ \right. \text{for}\quad s\rightarrow 0,
\end{equation*}%
see also (\ref{memory-function-asym}), as well as Tables \ref{tbl-1}, \ref%
{tbl-4}, \ref{tbl-2}, and \ref{tbl-3} for values of $\kappa $ and $\delta ,$
implying, by (\ref{Green-f-ld}), the asymptotics of Green's function in
Laplace domain%
\begin{equation*}
\tilde{G}^{\left( x\right) }\left( r,s\right) \sim \left\{ \!\!\!%
\begin{tabular}{ll}
$\frac{1}{4\pi r\kappa ^{2}}\mathrm{e}^{-\frac{rs}{\kappa }}\sim \frac{1}{%
4\pi r\kappa ^{2}},$ & \smallskip \\ 
$\frac{1}{4\pi r\kappa ^{2}}\frac{1}{s^{\delta }}\mathrm{e}^{-\frac{r}{%
\kappa }s^{1-\frac{\delta }{2}}}\sim \frac{1}{4\pi r\kappa ^{2}}\frac{1}{%
s^{\delta }},$ & $\delta \in \left( 0,1\right) $, \smallskip%
\end{tabular}%
\ \right. \text{as}\quad s\rightarrow 0.
\end{equation*}%
According to the theorem that if $\tilde{f}\left( s\right) \sim \tilde{g}%
\left( s\right) $ as $s\rightarrow 0,$ then $f\left( t\right) \sim g\left(
t\right) $ as $t\rightarrow \infty ,$ one obtains large-time asymptotics of
Green's function in the form%
\begin{equation}
G^{\left( x\right) }\left( r,t\right) \sim \left\{ \!\!\!%
\begin{tabular}{ll}
$\frac{1}{4\pi r\kappa ^{2}}\delta \left( t\right) =0,$ & \smallskip \\ 
$\frac{1}{4\pi r\kappa ^{2}}\frac{t^{-\left( 1-\delta \right) }}{\Gamma
\left( \delta \right) }\rightarrow 0,$ & $\delta \in \left( 0,1\right) $,
\smallskip%
\end{tabular}%
\ \right. \text{as}\quad t\rightarrow \infty .  \label{G-large-asy}
\end{equation}

Function $g^{\left( x\right) },$ appearing in the resolvent tensor (\ref%
{R-tilde-prim}) and (\ref{R-tilde-dvoprim}), given in the form%
\begin{align}
g^{\left( x\right) }\left( r,t\right) & =\int_{0}^{t}\mathfrak{g}^{\left(
x\right) }\left( r,t^{\prime }\right) \mathrm{d}t^{\prime },\quad \text{where%
}  \notag \\
\mathfrak{g}^{\left( x\right) }\left( r,t\right) & =\frac{1}{4\pi ^{2}r}%
\Bigg(\varphi _{0}+\int_{0}^{\infty }\frac{1}{\rho }\mathrm{e}^{\rho t\cos
\varphi _{0}-\frac{r\rho }{\left\vert \tilde{c}_{x}\left( \rho \mathrm{e}^{%
\mathrm{i}\varphi _{0}}\right) \right\vert }\cos \left( \varphi _{0}-\arg 
\tilde{c}_{x}\left( \rho \mathrm{e}^{\mathrm{i}\varphi _{0}}\right) \right) }
\notag \\
& \qquad \qquad \qquad \qquad \times \sin \left( \rho t\mathrm{\sin }\varphi
_{0}-\frac{r\rho }{\left\vert \tilde{c}_{x}\left( \rho \mathrm{e}^{\mathrm{i}%
\varphi _{0}}\right) \right\vert }\sin \left( \varphi _{0}-\arg \tilde{c}%
_{x}\left( \rho \mathrm{e}^{\mathrm{i}\varphi _{0}}\right) \right) \right) 
\mathrm{d}\rho \Bigg).  \label{g-kaligrafsko-finite}
\end{align}%
with $\varphi _{0}$ taken as above, is calculated in Appendix \ref%
{Calckaligrafskog} according to%
\begin{equation}
g^{\left( x\right) }\left( r,t\right) =\mathcal{L}^{-1}\left[ \frac{1}{s}%
\mathfrak{\tilde{g}}^{\left( x\right) }\left( r,s\right) \right] \left(
r,t\right) ,\quad \text{with}\quad \mathfrak{\tilde{g}}^{\left( x\right)
}\left( r,s\right) =\frac{1}{4\pi r}\frac{1}{s}\mathrm{e}^{-\frac{rs}{\tilde{%
c}_{x}\left( s\right) }},  \label{kaligrafsko-g}
\end{equation}%
see also expression (\ref{g-malo-tilde}) for function $\tilde{g}^{\left(
x\right) }$.

In the case of classical wave equation (\ref{we-rastavljen}), considered on
unbounded domain and subject to initial conditions (\ref{poc-usl}), which
corresponds to the propagation of a mechanical disturbance in an isotropic
and elastic three-dimensional body, the resolvent tensor takes the form%
\begin{align*}
\boldsymbol{\hat{R}}\left( \boldsymbol{r},s\right) & =\frac{1}{4\pi rc_{c}}%
\delta \left( c_{c}t-r\right) \frac{\boldsymbol{r}\otimes \boldsymbol{r}}{%
r^{2}}+\frac{1}{4\pi rc_{s}}\delta \left( c_{s}t-r\right) \left( \boldsymbol{%
\hat{I}}-\frac{\boldsymbol{r}\otimes \boldsymbol{r}}{r^{2}}\right) \\
& \qquad +\frac{t}{4\pi r^{3}}\left( H\left( t-\frac{r}{c_{s}}\right)
-H\left( t-\frac{r}{c_{c}}\right) \right) \left( \boldsymbol{\hat{I}}-3\frac{%
\boldsymbol{r}\otimes \boldsymbol{r}}{r^{2}}\right) ,
\end{align*}%
according to (\ref{R-tilde-prim}), since classical Green's functions $%
G^{\left( x\right) }$, as well as functions $g^{\left( x\right) },$ $x\in
\left\{ c,s\right\} ,$ are obtained as%
\begin{equation}
G^{\left( x\right) }\left( r,t\right) =\frac{1}{4\pi rc_{x}^{2}}\delta
\left( t-\frac{r}{c_{x}}\right) =\frac{1}{4\pi rc_{x}}\delta \left(
c_{x}t-r\right) ,  \label{G-kl}
\end{equation}%
due to $\delta \left( cr\right) =\frac{1}{\left\vert c\right\vert }\delta
\left( r\right) ,$ as well as 
\begin{equation*}
g^{\left( x\right) }\left( r,t\right) =\frac{1}{4\pi r}\left( t-\frac{r}{%
c_{x}}\right) H\left( t-\frac{r}{c_{x}}\right) \quad \text{with}\quad
\partial _{r}g^{\left( x\right) }\left( r,t\right) =-\frac{t}{4\pi r^{2}}%
H\left( t-\frac{r}{c_{x}}\right) ,
\end{equation*}%
due to $r\delta \left( r\right) =0,$ by inverting the Laplace transforms of
classical Green's functions in Laplace domain%
\begin{equation*}
\tilde{G}^{\left( x\right) }\left( r,s\right) =\frac{1}{4\pi rc_{x}^{2}}%
\mathrm{e}^{-\frac{rs}{c_{x}}},
\end{equation*}%
as well as of functions%
\begin{equation*}
\tilde{g}^{\left( x\right) }\left( r,s\right) =\frac{1}{4\pi r}\frac{1}{s^{2}%
}\mathrm{e}^{-\frac{rs}{c_{x}}},
\end{equation*}%
since (\ref{Green-f-ld}) and (\ref{g-malo-tilde}) define classical Green's
function in Laplace domain and function $\tilde{g}^{\left( x\right) }$ if,
instead of the memory functions $\tilde{c}_{x}$, the constant speed of the
wave propagation $c_{x}$ is substituted. This simple substitution is
possible since the generalized wave equations (\ref{eq-grin-fi}) and (\ref%
{eq-grin-omega}) become classical wave equations (\ref{we-fi}) and (\ref%
{we-omega}) using the mentioned substitution.

\section{Numerical examples \label{num-exam}}

Linear fractional-order models, fractional Burgers models, and fractional
anti-Zener and Zener models can be classified into two groups depending on
whether they model a viscoelastic material with infinite or finite wave
propagation speed, as it is stated in Introduction. Case II and IV of linear
fractional-order models, fractional Burgers models of the first class
(Models I - V), and all fractional anti-Zener and Zener models describe
materials with infinite wave propagation speed, while Case I and III of
linear fractional-order models and fractional Burgers models of the second
class (Models VI - VIII) describe materials with finite wave propagation
speed. Fractional anti-Zener and Zener model I$^{{}^{+}}$ID.ID, as well as
fractional Burgers model VII, namely%
\begin{equation}
\left( a_{1}\,_{0}\mathrm{I}_{t}^{\alpha +\beta +\nu }+a_{2}\,_{0}\mathrm{I}%
_{t}^{\nu }+a_{3}\,_{0}\mathrm{D}_{t}^{\alpha +\beta -\nu }\right) \sigma
\left( t\right) =\left( b_{1}\,_{0}\mathrm{I}_{t}^{\alpha }+b_{2}\,_{0}%
\mathrm{D}_{t}^{\beta }\right) \varepsilon \left( t\right) ,
\label{model-ifwps}
\end{equation}%
with narrowed thermodynamical restrictions%
\begin{gather}
0\leqslant \alpha +\beta -\nu \leqslant 1,\quad 1\leqslant \alpha +\beta
+\nu \leqslant 2,\quad \alpha \leqslant \nu \leqslant 1-\beta ,
\label{tdr-ifwps-1} \\
\frac{a_{1}}{a_{2}}\frac{\left\vert \cos \frac{\left( \alpha +2\beta +\nu
\right) \pi }{2}\right\vert }{\cos \frac{\left( \nu -\alpha \right) \pi }{2}}%
\leqslant \frac{a_{1}}{a_{2}}\frac{\left\vert \cos \frac{\left( \alpha
+2\beta +\nu \right) \pi }{2}\right\vert }{\cos \frac{\left( \nu -\alpha
\right) \pi }{2}}\frac{\sin \frac{\left( \alpha +2\beta +\nu \right) \pi }{2}%
}{\sin \frac{\left( \nu -\alpha \right) \pi }{2}}\leqslant \frac{b_{1}}{b_{2}%
},  \notag  \label{tdr-ifwps-2} \\
\frac{b_{1}}{b_{2}}\leqslant \frac{a_{2}}{a_{3}}\frac{\sin \frac{\left(
\beta +\nu \right) \pi }{2}}{\sin \frac{\left( 2\alpha +\beta -\nu \right)
\pi }{2}}\frac{\cos \frac{\left( \beta +\nu \right) \pi }{2}}{\cos \frac{%
\left( 2\alpha +\beta -\nu \right) \pi }{2}}\leqslant \frac{a_{2}}{a_{3}}%
\frac{\sin \frac{\left( \beta +\nu \right) \pi }{2}}{\sin \frac{\left(
2\alpha +\beta -\nu \right) \pi }{2}},  \label{tdr-ifwps-3}
\end{gather}%
and 
\begin{equation}
\left( 1+a_{1}\,{}_{0}\mathrm{D}_{t}^{\alpha }+a_{2}\,{}_{0}\mathrm{D}%
_{t}^{\beta }+a_{3}\,{}_{0}\mathrm{D}_{t}^{2\beta }\right) \sigma \left(
t\right) =\left( b_{1}\,{}_{0}\mathrm{D}_{t}^{\beta }+b_{2}\,{}_{0}\mathrm{D}%
_{t}^{2\beta }\right) \varepsilon \left( t\right) ,  \label{model-fwps}
\end{equation}%
with narrowed thermodynamical restrictions 
\begin{gather*}
0\leq \alpha \leq \beta \leq 1,\;\;\frac{1}{2}\leq \beta \leq \frac{1+\alpha 
}{2},\;\;\frac{a_{3}}{a_{2}}\leq \frac{b_{2}}{b_{1}}\leq a_{2}\frac{1}{%
\left\vert \cos \left( \beta \pi \right) \right\vert }, \\
\frac{a_{3}}{a_{2}}\leq \frac{a_{2}}{2\cos ^{2}\left( \beta \pi \right) }%
\left( 1-\sqrt{1-\frac{4a_{3}\cos ^{2}\left( \beta \pi \right) }{a_{2}^{2}}}%
\right) \leq \frac{b_{2}}{b_{1}}\leq \frac{a_{2}}{\left\vert \cos \left(
\beta \pi \right) \right\vert },
\end{gather*}%
are respectively chosen as representatives of previously mentioned two model
classes in order to plot time evolution of spatial profiles of Green's
function $G^{\left( x\right) },$ $x\in \left\{ c,s\right\} $ and its
asymptotics.

\subsection{Green's function $G^{\left( x\right) }$ - infinite wave
propagation speed}

\begin{figure}[h]
\begin{center}
\begin{minipage}{0.46\columnwidth}
  		\subfloat[Spatial profiles obtained according to analytical expression for different time instanstants.]{
		\includegraphics[width=\columnwidth]{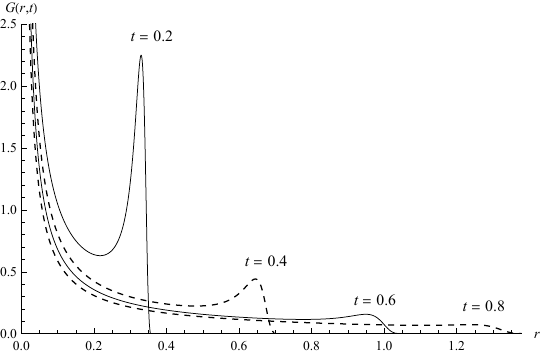}
   		\label{G-primeri}}
	\end{minipage}
\hfil
\begin{minipage}{0.46\columnwidth}
  \subfloat[Comparison of analytically and numerically obtained profiles, respectively depicted by solid and dotted lines.]{
   \includegraphics[width=\columnwidth]{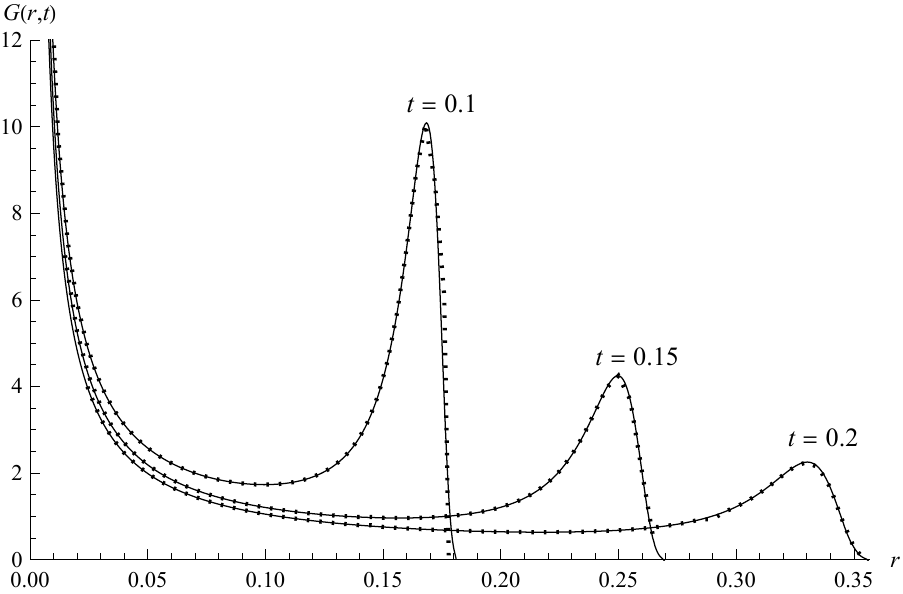}
   \label{G-slaganje-sa-num}}
  \end{minipage}
\end{center}
\caption{Case of infinite wave propagation speed - spatial profiles of
Green's function $G^{\left( x\right) }$ corresponding to model I$^{{}^{+}}$%
ID.ID when Green's function in Laplace domain $\tilde{G}^{\left( x\right) }$
has no other branch points besides $s=0$.}
\label{G-profili-np}
\end{figure}

\begin{figure}[p]
\begin{center}
\begin{minipage}{0.43\columnwidth}
		\subfloat[]{
		\includegraphics[width=\columnwidth]{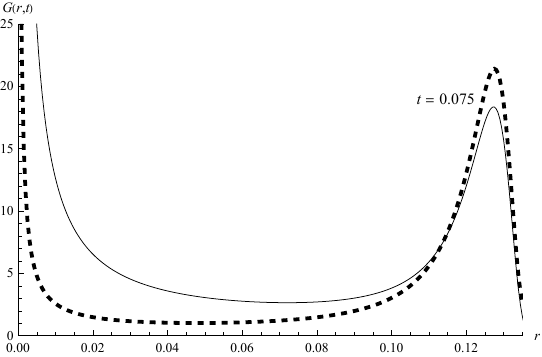}
		\label{Gsrsha0025}}
	\end{minipage}
\hfill 
\begin{minipage}{0.43\columnwidth}
		\subfloat[]{
		\includegraphics[width=\columnwidth]{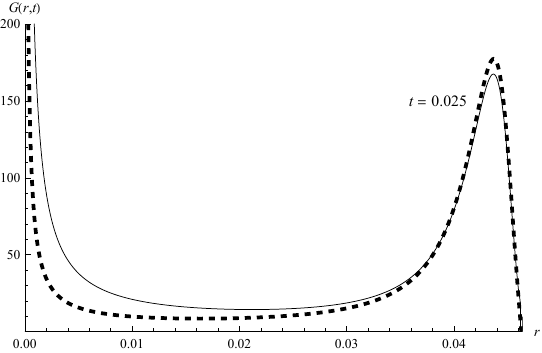}
		\label{Gsrsha0075}}
	\end{minipage}
\vfill
\begin{minipage}{0.43\columnwidth}
  		\subfloat[]{
  		\includegraphics[width=\columnwidth]{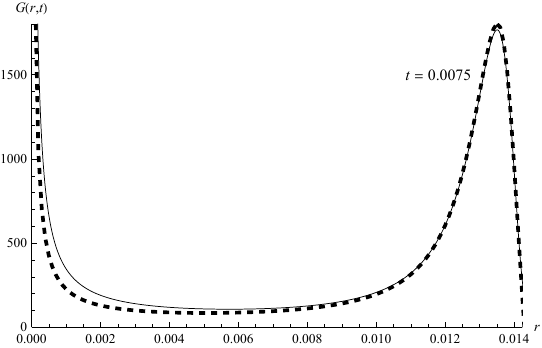}
  		\label{Gsrsha025}}
	\end{minipage}
\hfill 
\begin{minipage}{0.43\columnwidth}
  		\subfloat[]{
  		\includegraphics[width=\columnwidth]{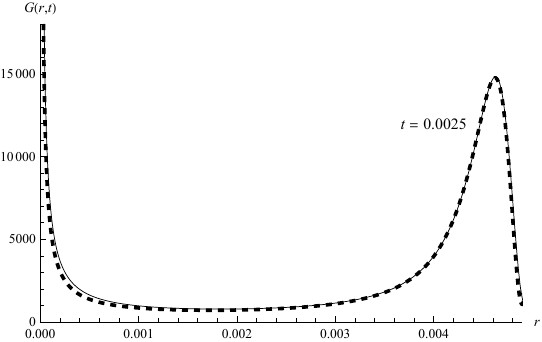}
  		\label{Gsrsha075}}
	\end{minipage}
\end{center}
\caption{Comparison of spatial profiles of Green's function, depicted by
solid lines, with the short time asymptotic profiles, depicted by dashed
lines - case when Green's function in Laplace domain $\tilde{G}^{\left(
x\right) }$ has no other branch points besides $s=0$.}
\label{G-srsha}
\end{figure}

\begin{figure}[p]
\begin{center}
\begin{minipage}{0.43\columnwidth}
		\subfloat[]{
		\includegraphics[width=\columnwidth]{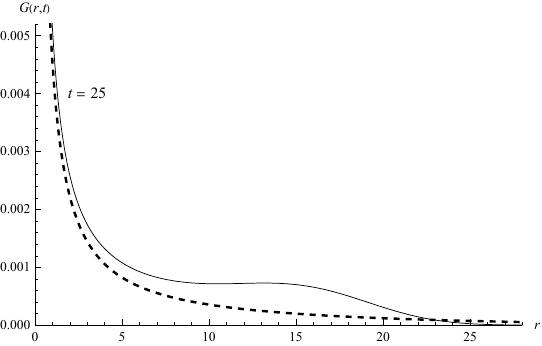}
		\label{G-srla-25}}
	\end{minipage}
\vfill
\begin{minipage}{0.43\columnwidth}
		\subfloat[]{
		\includegraphics[width=\columnwidth]{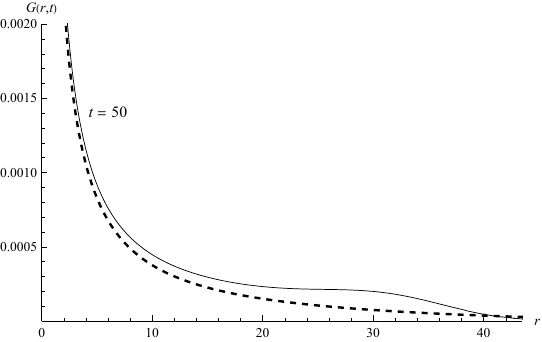}
		\label{G-srla-50}}
	\end{minipage}
\hfill 
\begin{minipage}{0.43\columnwidth}
		\subfloat[]{
		\includegraphics[width=\columnwidth]{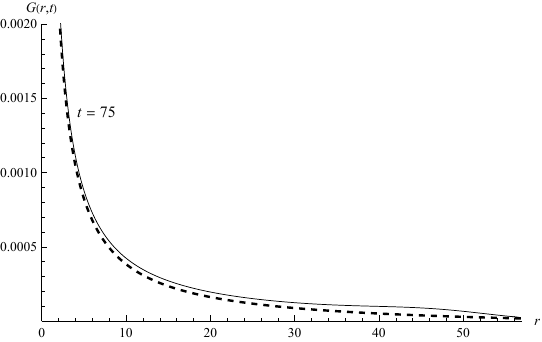}
		\label{G-srla-75}}
	\end{minipage}
\end{center}
\caption{Comparison of spatial profiles of Green's function, depicted by
solid lines, with the large time asymptotic profiles, depicted by dashed
lines - case when Green's function in Laplace domain $\tilde{G}^{\left(
x\right) }$ has no other branch points besides $s=0$.}
\label{G-srla}
\end{figure}

\begin{figure}[p]
\begin{center}
\begin{minipage}{0.43\columnwidth}
			\subfloat[]{
			\includegraphics[width=\columnwidth]{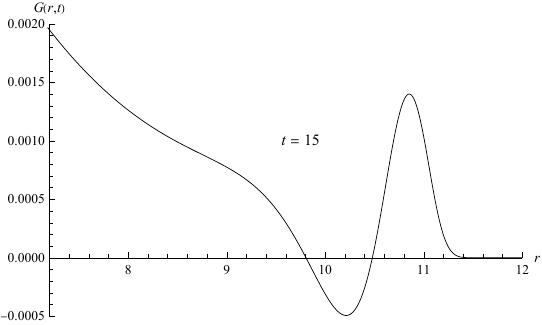}
			\label{G-oscilacije-15-ccp}}
		\end{minipage}
\vfill
\begin{minipage}{0.43\columnwidth}
				\subfloat[]{
				\includegraphics[width=\columnwidth]{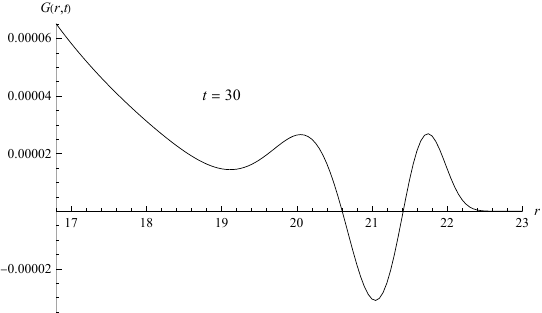}
				\label{G-oscilacije-30-ccp}}
		\end{minipage}
\hfill 
\begin{minipage}{0.43\columnwidth}
				\subfloat[]{
				\includegraphics[width=\columnwidth]{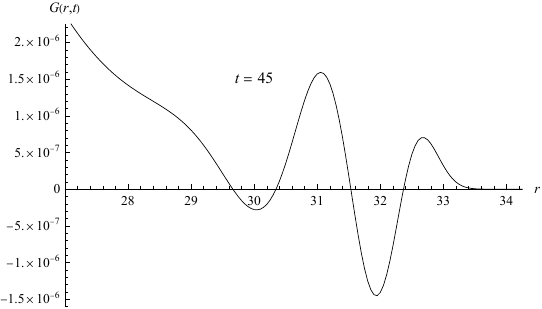}
				\label{G-oscilacije-45-ccp}}
		\end{minipage}
\vfill
\begin{minipage}{0.43\columnwidth}
				\subfloat[]{
				\includegraphics[width=\columnwidth]{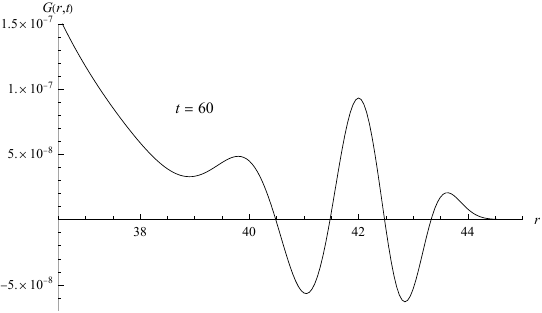}
				\label{G-oscilacije-60-ccp}}
		\end{minipage}
\hfill 
\begin{minipage}{0.43\columnwidth}
				\subfloat[]{
				\includegraphics[width=\columnwidth]{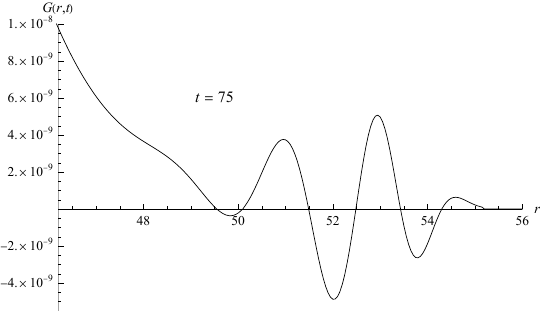}
				\label{G-oscilacije-75-ccp}}
		\end{minipage}
\end{center}
\caption{Case of infinite wave propagation speed - tails of spatial profiles
of Green's function $G^{\left( x\right) }$ corresponding to model I$%
^{{}^{+}} $ID.ID when Green's function in Laplace domain $\tilde{G}^{\left(
x\right) }$ has a pair of complex conjugated branch points $s_0$ and $\bar{s}%
_0$ besides $s=0$.}
\label{oscilacije-ccp}
\end{figure}

\begin{figure}[p]
\begin{center}
\begin{minipage}{0.43\columnwidth}
				\subfloat[]{
				\includegraphics[width=\columnwidth]{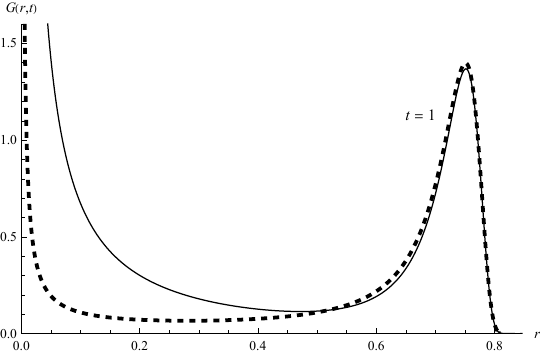}
				\label{Gsrsha01-ccp}}
		\end{minipage}
\hfill 
\begin{minipage}{0.43\columnwidth}
				\subfloat[]{
				\includegraphics[width=\columnwidth]{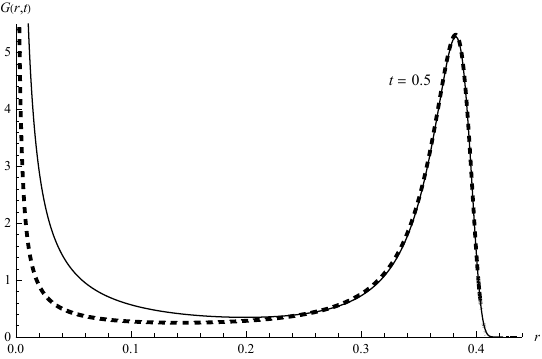}
				\label{Gsrsha1-ccp}}
		\end{minipage}
\vfill
\begin{minipage}{0.43\columnwidth}
  \subfloat[]{
   \includegraphics[width=\columnwidth]{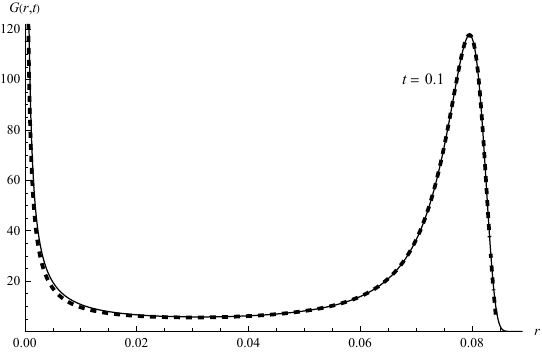}
   \label{Gsrsha05-ccp}}
\end{minipage}
\hfill 
\begin{minipage}{0.43\columnwidth}
  \subfloat[]{
   \includegraphics[width=\columnwidth]{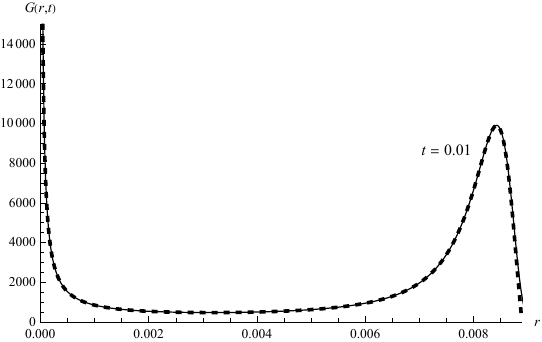}
   \label{Gsrsha10-ccp}}
\end{minipage}
\end{center}
\caption{Comparison of spatial profiles of Green's function, depicted by
solid lines, with the short time asymptotic profiles, depicted by dashed
lines - case when Green's function in Laplace domain $\tilde{G}^{\left(
x\right) }$ has a pair of complex conjugated branch points $s_0$ and $\bar{s}%
_0$ besides $s=0$.}
\label{G-srsha-ccp}
\end{figure}

Model parameters corresponding to fractional anti-Zener and Zener model I$%
^{{}^{+}}$ID.ID, see (\ref{model-ifwps}), are chosen as in Table \ref%
{parametri} in order to plot time evolution of spatial profiles of Green's
function $G^{\left( x\right) }$, either according to analytical expression (%
\ref{grin-f-finite}) with $\varepsilon =0$ and $\varphi _{0}=\pi $, if
function%
\begin{equation}
\phi _{\sigma }\left( s\right) =a_{1}+a_{2}s^{\alpha +\beta
}+a_{3}s^{2\left( \alpha +\beta \right) },  \label{f-fi-iwps}
\end{equation}%
see also Table \ref{skupina}, has no zeros, or according to analytical
expression (\ref{grin-f-finite}) with $\varepsilon =0$ and $\varphi
_{0}=\arg s_{0},$ if $s_{0}$ and its complex conjugate $\bar{s}_{0}$ are
zeros of function $\phi _{\sigma },$ as well as by the numerical Laplace
transform inversion using the expression (\ref{Green-f-ld}) for Green's
function in Laplace domain $\tilde{G}^{\left( x\right) }$ and fixed Talbot
method developed in \cite{AbateValko}, while expressions (\ref{G-short-time}%
) for short time asymptotics, as well as (\ref{G-large-asy})$_{2}$ for large
time asymptotics, are used to produce profiles of the asymptotic behavior of
Green's function. Namely, spatial profiles of Green's function $G^{\left(
x\right) }$ in the case when Green's function in Laplace domain $\tilde{G}%
^{\left( x\right) }$ has no other branch points besides $s=0$ are presented
in Figures \ref{G-profili-np}, \ref{G-srsha}, and \ref{G-srla}, while
spatial profiles of Green's function $G^{\left( x\right) }$ in the case when 
$\tilde{G}^{\left( x\right) }$ has a pair of complex conjugated branch
points $s_{0}$ and $\bar{s}_{0}$ besides $s=0$ are shown in Figures \ref%
{G-profili-ccp-ccp}, \ref{oscilacije-ccp}, \ref{G-srsha-ccp}, and \ref%
{G-srla-ccp}.  \begin{table}[h]
 \begin{center}

\begin{tabular}{|c|c|c|c|c|c|c|c|c|}

 \hline \xrowht{14pt}

Case when $\phi _{\sigma }$ has & $\alpha $ & $\beta $ & $\nu $ & $a_{1}$ & $a_{2}$ & $a_{3}$ & $b_{1}$ & $b_{2}$ \\

 \hhline{|=|=|=|=|=|=|=|=|=|} \xrowht{14pt}

no zeros & \multirow{2}{*}{\multirowcell{2}{$0.35$}} & \multirow{2}{*}{\multirowcell{2}{$0.55$}} & \multirow{2}{*}{\multirowcell{2}{$0.4$}} & $0.05$ & $1.5$ & $0.45$ & $0.7$ & $0.95$ \\ \cline{1-1} \cline{5-9} \xrowht{14pt}
\makecell{a pair of complex\\ conjugated zeros} & &  &  & $11$ & $15$ & $20.27$ & $7$ & $9.5$\\ 

\hline

\end{tabular}

 \end{center}
 \caption{Parameters of model I$^{{}^{+}}$ID.ID used for numerical examples.}
 \label{parametri}
 \end{table}

The spatial profiles of Green's function $G^{\left( x\right) }$,
corresponding to fractional anti-Zener and Zener model I$^{{}^{+}}$ID.ID in
the case when Green's function in Laplace domain $\tilde{G}^{\left( x\right)
}$ has no other branch points besides $s=0$, are presented in Figure \ref%
{G-profili-np} for different time instants, being positive and displaying
decreasing character for smaller distances $r$ from the origin, with a
pronounced peak for medium values of $r$, and finally tending to zero for
large values of $r$. More precisely, the spatial profiles tend to infinity
as $r$ tends to zero and asymptotically settle to zero for $r$ tending to
infinity, presumably since Green's function predominantly behaves as a
hyperbolic function $\frac{1}{r}$ for both $r\rightarrow 0$ and $%
r\rightarrow \infty $, which is supported by the form of the analytical
expression (\ref{grin-f-finite}) for Green's function, as well as by the
short and large time asymptotics (\ref{G-short-time}) and (\ref{G-large-asy}%
). The peak, carrying the information about the initial disturbance and
resembling to the Dirac delta peak in the case of classical wave
propagation, see (\ref{G-kl}), decreases in height and increases in width as
time increases, presumably due to the dissipative character of the model. In
particular, Figure \ref{G-slaganje-sa-num} shows the matching of the
profiles obtained through the analytical expression (\ref{grin-f-finite}) of
Green's function, which are presented by the solid line, with the curves
presented by dotted line, which are obtained by the numerical Laplace
transform inversion of Green's function in the Laplace domain (\ref%
{Green-f-ld}) using the fixed Talbot method. 

Also, it is verified in Figure \ref{G-srsha} that the overlap of spatial
profile of Green's function, depicted by solid line, with the short time
asymptotic profile, depicted by dashed line, becomes better and better as
time decreases, where the solid line profiles are obtained according to the
analytical expression (\ref{grin-f-finite}), with $\varepsilon =0$ and $%
\varphi _{0}=\pi$, while the dotted line profiles are obtained according to
the asymptotic expression (\ref{G-short-time}). Similarly, Figure \ref%
{G-srla} verifies that the overlap of spatial profile of Green's function,
depicted by solid line, with the large time asymptotic profile, depicted by
dashed line, becomes better and better as time increases, where the
expression (\ref{G-large-asy}) is used in order to produce large time
asymptotic profiles of Green's function. 

\begin{figure}[p]
\begin{center}
\begin{minipage}{0.43\columnwidth}
  \subfloat[Comparison of analytically and numerically obtained profiles, respectively depicted by solid and dotted lines.]{
   \includegraphics[width=\columnwidth]{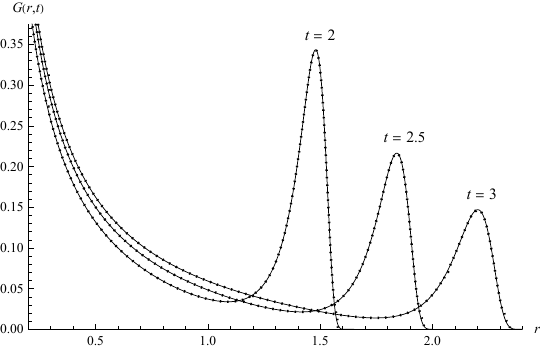}
   \label{G-profili-ccp}}
  \end{minipage}
\hfill 
\begin{minipage}{0.43\columnwidth}
  \subfloat[Spatial profiles obtained according to
analytical expression for different time instanstants.]{
   \includegraphics[width=\columnwidth]{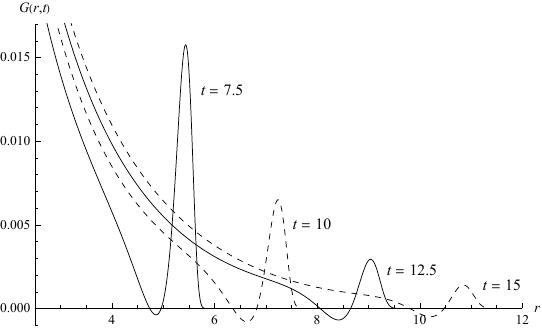}
   \label{G-profili-ccp-dodatak}}
  \end{minipage}
\end{center}
\caption{Case of infinite wave propagation speed - spatial profiles of
Green's function $G^{\left( x\right) }$ corresponding to model I$^{{}^{+}}$%
ID.ID when Green's function in Laplace domain $\tilde{G}^{\left( x\right) }$
has a pair of complex conjugated branch points $s_0$ and $\bar{s}_0$ besides 
$s=0$.}
\label{G-profili-ccp-ccp}
\end{figure}

The spatial profiles that correspond to Green's function $G^{\left( x\right)
}$ in the case when Green's function in Laplace domain $\tilde{G}^{\left(
x\right) }$ has a pair of complex conjugated branch points $s_{0}$ and $\bar{%
s}_{0}$ besides $s=0$, obtained according to the expression (\ref%
{grin-f-finite}), with $\varepsilon =0$ and $\varphi _{0}=\arg s_{0},$ and
shown in Figure \ref{G-profili-ccp}, similarly as in the case when there are
no branch points of $\tilde{G}^{\left( x\right) }$ besides $s=0$, tend to
infinity as $r$ tends to zero and asymptotically settle to zero for $r$
tending to infinity, again as a hyperbolic function $\frac{1}{r}$ for both $%
r \to 0$ and $r\to\infty $, see (\ref{grin-f-finite}) and asymptotic
expressions (\ref{G-short-time}) and (\ref{G-large-asy})$_2$. On one hand,
Figure \ref{G-profili-ccp} presents a single peak for each spatial profile
carrying the information about the initial disturbance, resembling to the
Dirac delta peak in the case of classical wave propagation, and displaying
decrease in height and increase in width as time increases, and on the other
hand, Figure \ref{G-profili-ccp} shows a good agreement between the solid
line curves obtained according to the analytical expression and the dotted
line curves obtained by the numerical Laplace transform inversion of Green's
function in the Laplace domain (\ref{Green-f-ld}), similarly as Figure \ref%
{G-slaganje-sa-num}. 
\begin{figure}[p]
\begin{center}
\begin{minipage}{0.43\columnwidth}
			\subfloat[]{
			\includegraphics[width=\columnwidth]{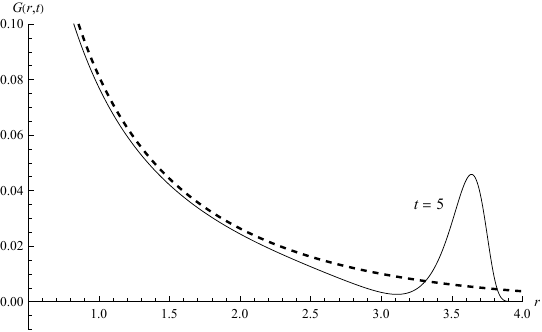}
			\label{G-srla-5-ccp}}
		\end{minipage}
\vfill
\begin{minipage}{0.43\columnwidth}
				\subfloat[]{
				\includegraphics[width=\columnwidth]{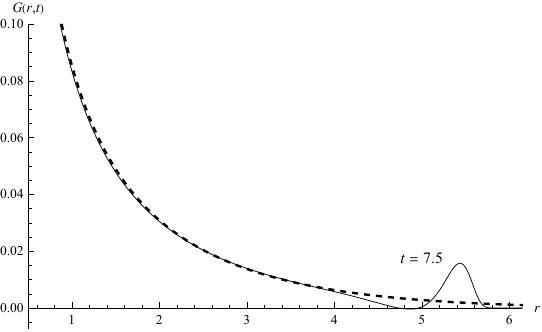}
				\label{G-srla-7-ccp}}
		\end{minipage}
\hfill 
\begin{minipage}{0.43\columnwidth}
				\subfloat[]{
				\includegraphics[width=\columnwidth]{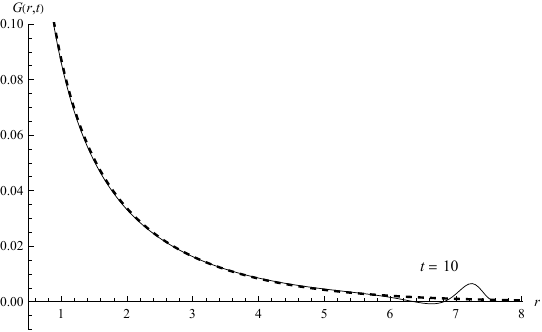}
				\label{G-srla-10-ccp}}
		\end{minipage}
\end{center}
\caption{Comparison of spatial profiles of Green's function, depicted by
solid lines, with the large time asymptotic profiles, depicted by dashed
lines - case when Green's function in Laplace domain $\tilde{G}^{\left(
x\right) }$ has a pair of complex conjugated branch points $s_0$ and $\bar{s}%
_0$ besides $s=0$.}
\label{G-srla-ccp}
\end{figure}

Contrary to the case when Green's function in Laplace domain $\tilde{G}%
^{\left( x\right) }$ has no other branch points besides $s=0$, in which the
spatial profiles do not exhibit oscillatory character, in the case when $%
\tilde{G}^{\left( x\right) }$ has a pair of complex conjugated branch points 
$s_{0}$ and $\bar{s}_{0}$ besides $s=0$, for time large enough, Figure \ref%
{G-profili-ccp-dodatak} clearly shows the damped oscillatory character of
the tails of spatial profiles, that is not prominent for smaller values of
time, see Figure \ref{G-profili-ccp}. The oscillatory character of spatial profiles becomes more pronounced
as time becomes larger, as evident from Figure \ref{oscilacije-ccp}, since
the number of oscillations increases, so instead of at most one oscillation
for time instances $t\leqslant 15$, see Figures \ref{G-profili-ccp-dodatak}
and \ref{G-oscilacije-15-ccp}, there is more than one oscillation, namely a
multiplicative factor of an oscillation belonging to a set $\{3/2, 2,
5/2,3\} $ for respective time instants $t\in \{30,45,60,75\}$, see Figures %
\ref{G-oscilacije-30-ccp} - \ref{G-oscilacije-75-ccp}. Also, as time
increases, the amplitudes of spatial profile tails become smaller,
indicating the damping in both space and time. Therefore, the damped
oscillatory character of Green's function $G^{\left( x\right) }$ in both
space and time is due to the complex value of branch points of Green's
function in Laplace domain $\tilde{G}^{\left( x\right) }$, such that,
presumably, the real part of branch point $s_{0}$ corresponds to the
damping, while its imaginary part corresponds to the oscillatory behavior.

On one hand, the expression for Green's function (\ref{grin-f-finite})
qualitatively corresponds to the short time asymptotic expression (\ref%
{G-short-time}), which can be observed from Figure \ref{G-srsha-ccp}, since
there is a better overlap of the curves corresponding to these expressions
as time takes smaller values, while, on the other hand, there is a better
overlap of the curves from Figure \ref{G-srla-ccp} as time takes larger
values, indicating that the expression for Green's function (\ref%
{grin-f-finite}) qualitatively turns into the large time asymptotic
expression (\ref{G-large-asy}) for a sufficiently large time. Although,
spatial profiles dominantly behave as the asymptotic curves for small and
medium values of $r$, the tail of the spatial profile is oscillatory, while
asymptotic curve is not, however, as can be observed from Figure \ref%
{oscilacije-ccp}, the oscillation amplitudes decrease as time increases.

\subsection{Green's function $G^{\left( x\right) }$ - finite wave
propagation speed}

\begin{figure}[p]
\begin{center}
\begin{minipage}{0.46\columnwidth}
				\subfloat[]{
				\includegraphics[width=\columnwidth]{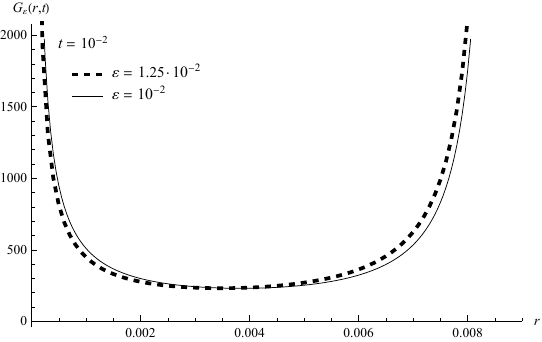}
				\label{G-epsilon-srsha-1}}
		\end{minipage}
\hfill 
\begin{minipage}{0.46\columnwidth}
				\subfloat[]{
				\includegraphics[width=\columnwidth]{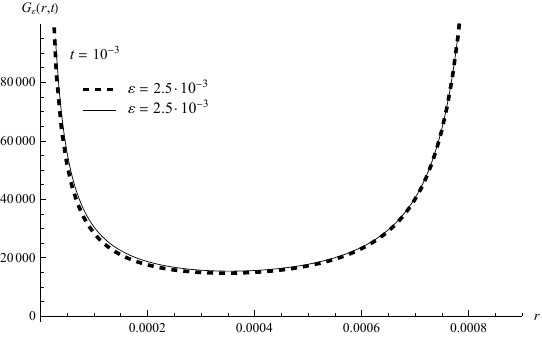}
				\label{G-epsilon-srsha-2}}
		\end{minipage}
\vfill
\begin{minipage}{0.46\columnwidth}
  \subfloat[]{
   \includegraphics[width=\columnwidth]{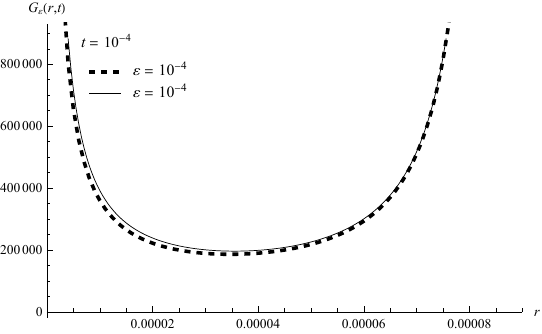}
   \label{G-epsilon-srsha-3}}
\end{minipage}
\hfill 
\begin{minipage}{0.46\columnwidth}
  \subfloat[]{
   \includegraphics[width=\columnwidth]{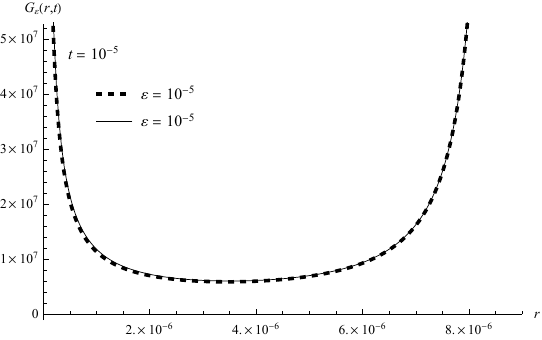}
   \label{G-epsilon-srsha-4}}
\end{minipage}
\end{center}
\caption{Comparison of spatial profiles of regularized Green's function,
depicted by solid lines, with the short time asymptotic profiles, depicted
by dashed lines - case when Green's function in Laplace domain $\tilde{G}%
^{\left( x\right) }$ has no other branch points besides $s=0$.}
\label{G-Burgers-srsha-np}
\end{figure}

Model parameters corresponding to fractional Burgers model VII, see (\ref%
{model-fwps}), are chosen as in Table \ref{tbl} in order to plot time
evolution of spatial profiles of Green's function $G^{\left( x\right) },$
either according to analytical expression (\ref{grin-f-finite}) with $%
\varphi _{0}=\pi $, if function $\Phi _{\sigma }$, see (\ref{fiovi-burgers-7}%
)$_{1}$, has no zeros, or according to analytical expression (\ref%
{grin-f-finite}) with $\varphi _{0}=\arg s_{0},$ if $s_{0}$ and its complex
conjugate $\bar{s}_{0}$ are zeros of function $\Phi _{\sigma }$, as well as
by the numerical Laplace transform inversion using the expression (\ref%
{Green-f-ld}) for Green's function in Laplace domain $\tilde{G}^{\left(
x\right) }$, while the expression (\ref{srsha-int-repr}) is used to produce
profiles of short time asymptotic behavior of Green's function. Namely,
spatial profiles of Green's function $G^{\left( x\right) }$ in the case when
Green's function in Laplace domain $\tilde{G}^{\left( x\right) }$ has no
other branch points besides $s=0$ are presented in Figures \ref%
{G-profili-burgers-np} and \ref{G-Burgers-srsha-np}, while spatial profiles
of Green's function $G^{\left( x\right) }$ in the case when $\tilde{G}%
^{\left( x\right) }$ has a pair of complex conjugated branch points $s_{0}$
and $\bar{s}_{0}$ besides $s=0$ are shown in Figure \ref%
{G-Burgers-ccp-pikovi-med}. 
\begin{table}[h]
\begin{center}
\begin{tabular}{|c|c|c|c|c|c|c|c|} 
\hline \xrowht{14pt}
Case when $\Phi _{\sigma }$ has  & $a_1$ & $a_2$ & $a_3$ & $b_1$ & $b_2$ & $\alpha$ & $\beta$  \\ 
\hhline{=|=|=|=|=|=|=|=} \xrowht{14pt}
no zeros & \multirow{2}{*}{\multirowcell{2}{$0.01$}}& $4.5$ & $4$ & \multirow{2}{*}{\multirowcell{2}{$1$}} & $3$ & \multirow{2}{*}{\multirowcell{2}{$0.7$}} & \multirow{2}{*}{\multirowcell{2}{$0.845$}} \\ \cline{1-1} \cline{3-4} \cline{6-6} \xrowht{14pt} 
\makecell{a pair of complex\\ conjugated zeros}  & & $2.5$ & $5.5$ & & $ 2.21$ &  &    \\ 
\hline
\end{tabular}%
\end{center}
\caption{Parameters of fractional Burgers model VII used for numerical examples.}
\label{tbl}
\end{table}

\begin{figure}[p]
\begin{center}
\begin{minipage}{0.46\columnwidth}
  \subfloat[Comparison of analytically and numerically obtained profiles, respectively depicted by solid/dashed and dotted lines.]{
   \includegraphics[width=\columnwidth]{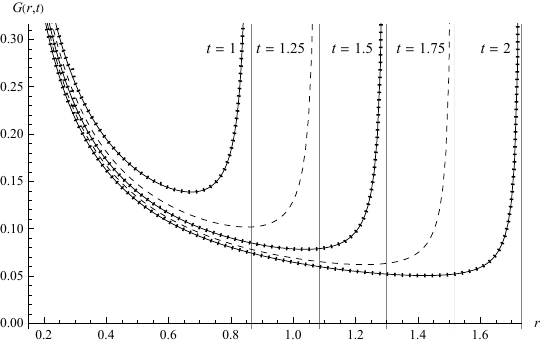}
   \label{G-Burgers-np-sa-num}}
  \end{minipage}
\hfill 
\begin{minipage}{0.46\columnwidth}
  \subfloat[Spatial profiles obtained according to
analytical expression for different values of regularization parameter $\varepsilon$ at $t=0.1$.]{
   \includegraphics[width=\columnwidth]{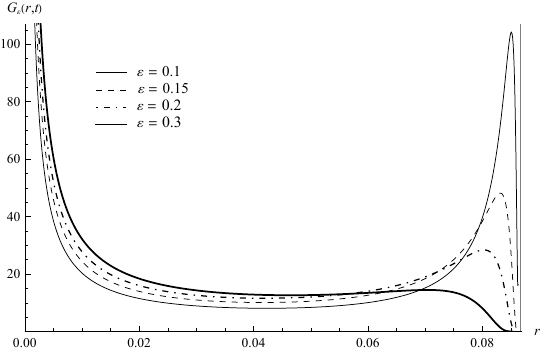}
   \label{G-epsiloniii}}
  \end{minipage}
\end{center}
\caption{Case of finite wave propagation speed - spatial profiles of Green's
function $G^{\left( x\right) }$ corresponding to fractional Burgers model
VII when Green's function in Laplace domain $\tilde{G}^{\left( x\right) }$
has no other branch points besides $s=0$.}
\label{G-profili-burgers-np}
\end{figure}

The spatial profiles of Green's function $G^{\left( x\right)}$,
corresponding to the fractional Burgers model VII in the case when Green's
function in Laplace domain $\tilde{G}^{\left( x\right) }$ has no other
branch points besides $s=0$, are presented for different time instants in
Figure \ref{G-Burgers-np-sa-num} by the solid line curves, obtained
according to analytical expression (\ref{grin-f-finite}) with regularization
parameter $\varepsilon=0$, and moreover compared with spatial profiles
presented by dotted line curves, obtained by the numerical Laplace transform
inversion of Green's function in Laplace domain (\ref{Green-f-ld}) using the
fixed Talbot method. The profiles are positive and decrease from infinity as
the distance $r$ increases from zero up to a medium values, presumably since
Green's function predominantly behaves as a hyperbolic function $\frac{1}{r}$
for $r \to 0$. Unlike the case of spatial profiles, corresponding to Green's
function when the wave propagation speed is infinite, when there is a peak
which translates towards higher values of $r$ as time increases, while
reducing in height and gaining in width, see Figures \ref{G-profili-np} and %
\ref{G-profili-ccp-ccp}, in the case of fractional Burgers model VII, that
assumes a finite value of wave propagation speed $c_x=\frac{1}{\sqrt{\varrho}%
}\sqrt{\frac{b_2}{a_3}}$, the peak is not observed from medium values of $r$
up to $r=c_x t$, referenced by the vertical line, where the profiles show increasing character towards the
infinity. Actually, as indicated by the short time asymptotics of Green's
function (\ref{G-srsha-konacna-brzina}), represented by the Dirac delta
distribution centered at $r=c_x t$, the peak should be centered at point $%
r=c_x t$, which is marked by the vertical line in Figure \ref%
{G-Burgers-np-sa-num}. However, the peak cannot be observed, since the
analytical expression (\ref{grin-f-finite}) holds true only for $r<c_{x}t$,
so the increase of spatial profile towards the infinity is the peak forming
for $r<c_x t$, which cannot be formed for $r\geqslant c_x t$, and therefore
cannot be completely displayed. On the other hand, the expression for
regularized Green's function $G_{\varepsilon}^{\left( x\right)}$ allows for
the formation of peak at the tail of spatial profile, as can be observed
from Figure \ref{G-epsiloniii}, from which is obvious that as the
regularization parameter $\varepsilon$ decreases, the position of peak
shifts towards $r=c_x t$, while peak's width decreases and its height
increases. 

Figure \ref{G-Burgers-srsha-np} presents the comparison of spatial profiles
corresponding to the regularized Green's function $G_{\varepsilon}^{\left(
x\right)}$ with the profiles corresponding to the regularization of Dirac
delta distribution that represents the short time asymptotics of Green's
function, where the expression (\ref{grin-f-finite}) is used for regularized
Green's function calculation, while the expression (\ref{srsha-int-repr}) is
used for the calculation of regularized Dirac delta distribution. One can
notice that the overlap between the curves corresponding to profiles and
asymptotics becomes better as time decreases. Also, note that the
regularization parameter $\varepsilon$ has sufficiently small value, so that
the peak at the tail of profile is not observed.

\begin{figure}[h]
\begin{center}
\includegraphics[width=0.6\columnwidth]{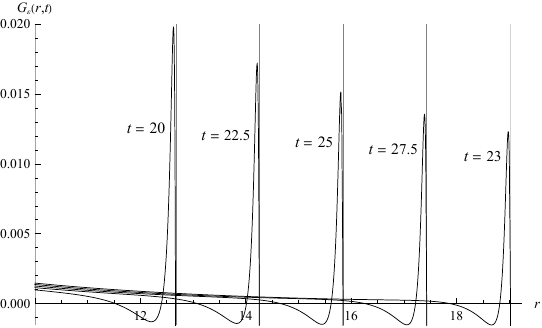}
\end{center}
\caption{Case of finite wave propagation speed - spatial profiles of
regularized Green's function $G_{\protect\varepsilon}^{\left( x\right) }$, $%
\protect\varepsilon=0.5$, corresponding to fractional Burgers model VII when
Green's function in Laplace domain $\tilde{G}^{\left( x\right) }$ has a pair
of complex conjugated branch points $s_0$ and $\bar{s}_0$ besides $s=0$.}
\label{G-Burgers-ccp-pikovi-med}
\end{figure}

Contrary to the case when Green's function in Laplace domain $\tilde{G}%
^{\left( x\right) }$ has no other branch points besides $s=0$, in which, as
the distance $r$ increases, the profiles decrease from the infinity up to
the point in which the peak located at $r=c_x t$ starts to form, see Figure %
\ref{G-profili-burgers-np}, in the case when Green's function in Laplace
domain $\tilde{G}^{\left( x\right) }$ has a pair of complex conjugated
branch points $s_0$ and $\bar{s}_0$ besides $s=0$ an additional negative
peak appears and it precedes the peak forming at $r=c_x t$, as it is obvious
from Figure \ref{G-Burgers-ccp-pikovi-med}. Presumably, the character of
spatial profiles is damped oscillatory, similarly as in the case when wave
propagation speed is infinite, see Figure \ref{G-profili-ccp-dodatak}, which
is not apparent since time is not large enough. More prominent oscillatory
character of tails of spatial profiles would be expected for larger times,
again similarly as in Figure \ref{oscilacije-ccp}. 


\subsection{Function $g^{\left( x\right) }$}

\begin{figure}[h]
\begin{center}
\includegraphics[width=0.6\columnwidth]{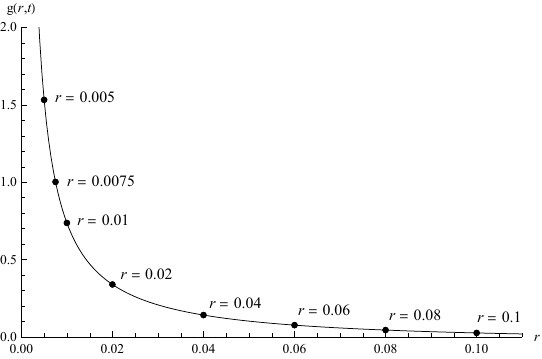}
\end{center}
\caption{Comparison of spatial profiles of function $g$, corresponding to
model I$^{{}^{+}}$ID.ID, obtained according to numerical Laplace transform
inversion (solid line) and by analytical expression (dots) at time instant $%
t=0.1$.}
\label{g}
\end{figure}

Contrary to the case of Green's function, in which it must be taken into
account whether the wave propagation speed is infinite or finite, in the
case of function $g^{\left( x\right) }$, given by the expression (\ref%
{g-kaligrafsko-finite}), wave propagation speed is not important. Spatial
profile of function $g^{\left( x\right) }$ is obtained using numerical
Laplace transform inversion of the expression (\ref{kaligrafsko-g}) and
depicted by the solid line in Figure \ref{g}, displaying quite classical
behavior, since it is positive, decreasing, and convex. Moreover, the solid
line spatial profile is compared with the spatial profile, represented by
dots, that originate form the analytical expression (\ref%
{g-kaligrafsko-finite}). Similarly as Green's function $G^{\left( x\right) }$%
, the function $g^{\left( x\right) }$ is also proportional to $\frac{1}{r}$,
and therefore the spatial profile starts from infinity for $r\rightarrow 0$
and settles at zero for $r\rightarrow \infty $. 

\section{Conclusion}

The starting point of generalization of the classical Hooke's law (\ref{hz}%
), describing three-dimensional isotropic and elastic body, to a
constitutive model (\ref{VEHook}), describing three-dimensional isotropic
and viscoelastic body, is the constitutive equation of one-dimensional
viscoelastic body (\ref{sigma=sr*epsilon}), expressed through the relaxation
moduli $\sigma _{sr}^{\left( c\right) }$ and $\sigma _{sr}^{\left( s\right)
} $, used in order to account for different memory kernels corresponding to
propagation of compressive ($c$) and shear ($s$) waves. Further, in order to
describe the wave propagation in three-dimensional isotropic and
viscoelastic body, the constitutive model (\ref{VEHook}) is coupled with the
equation of motion of deformable solid body (\ref{EoM}) and equation of
infinitesimal strain (\ref{strain}), yielding the system of equations (\ref%
{SoE-gen-1}) - (\ref{SoE-gen-3}), that is subject to initial conditions (\ref%
{poc-usl}) and solved on unbounded domain for displacement field by the use
of Fourier and Laplace transforms. More precisely, the application of
Fourier and Laplace transforms to the system of equations (\ref{SoE-gen-1})
- (\ref{SoE-gen-3}), subject to (\ref{poc-usl}), yielded the wave equation
of viscoelastic body in Fourier and Laplace domain (\ref{Ru=pu}), which is
solved for displacement field using two different approaches.

In the first approach, after scalar and vector multiplication of wave vector
with the wave equation of viscoelastic body in Fourier and Laplace domain (%
\ref{Ru=pu}) and after performing Laplace and Fourier transforms inversions,
wave equations (\ref{eq-grin-fi}) and (\ref{eq-grin-omega}) are obtained,
representing the generalization of classical wave equations (\ref{we-fi})
and (\ref{we-omega}), where speeds of compressive and shear waves, $c_{c}$
and $c_{s}$, are substituted with the memory kernels $\tilde{c}_{c}$ and $%
\tilde{c}_{s},$ given by (\ref{c-tilde-compresive}) and (\ref{c-tilde-shear}%
), which are expressed through the corresponding relaxation moduli $\sigma
_{sr}^{\left( c\right) }$ and $\sigma _{sr}^{\left( s\right) }$, such that
their initial values, i.e., glass moduli, can be either finite or infinite
implying two different cases (\ref{eq-grin-fi-finite}) and (\ref%
{eq-grin-fi-infinite}) for the compressive wave equation (\ref{eq-grin-fi}),
as well as two different cases (\ref{eq-grin-omega-finite}) and (\ref%
{eq-grin-omega-infinite}) for the shear wave equation (\ref{eq-grin-omega}).
Therefore, in this approach, two wave equations (\ref{eq-grin-fi}) and (\ref%
{eq-grin-omega}), expressed in terms of scalar and vector fields $\varphi $
and $\boldsymbol{\Omega }$, defined by (\ref{eq-grin-fi})$_{2}$ and (\ref%
{eq-grin-omega})$_{2}$, are obtained as generalizations of classical wave
equations (\ref{we-fi}) and (\ref{we-omega}), rather then a single wave
equation expressed in terms of displacement field, which would be
generalization of classical wave equation (\ref{we}), or equivalently (\ref%
{we-rastavljen}). After the expressions (\ref{fi-konacno}) and (\ref%
{omega-konacno}) for the scalar and vector fields $\varphi $ and $%
\boldsymbol{\Omega }$ are found dependent on Green's functions $G^{\left(
c\right) }$ and $G^{\left( s\right) }$, the displacement field is expressed
by (\ref{u-sol-razgranato}), accounting for four possibilities due to
Green's functions $G^{\left( c\right) }$ and $G^{\left( s\right) }$ taking
two different forms: (\ref{grin-f-finite}), which holds for $r<c_{x}t$ and
zero otherwise, and (\ref{G-za-infne}), where the first one is valid in the
case of finite glass modulus and therefore wave propagation speed $c_{x},$ $%
x\in \left\{ c,s\right\} ,$ as well, while the second one is valid in the
case of infinite glass modulus and therefore wave propagation speed as well.

Since complex and demanding calculations are required in the first approach,
in the second one, the use of resolvent tensor $\boldsymbol{\hat{R}}$ is
found to be more convenient to determine the displacement field, so that the
wave equation of viscoelastic body in Fourier and Laplace domain (\ref{Ru=pu}%
) is rewritten in matrix representation (\ref{Ru=pu-rewritten}), yielding
the displacement field in Laplace and Fourier domain (\ref%
{u-tilde-hat=R-tilde-hat-pu-tilde-hat}), i.e., yielding the expression (\ref%
{u-solution-2}) for the displacement field as the solution to the system of
equations (\ref{SoE-gen-1}) - (\ref{SoE-gen-3}), subject to (\ref{poc-usl}),
after performing the inverse Laplace and Fourier transforms of the resolvent
tensor (\ref{rezolventa-furije-laplas}), implying that the displacement
field is simply determined as the application of resolvent tensor in the
form (\ref{R-tilde-prim}) to the already known initial conditions (\ref%
{poc-usl}). For the resolvent tensor, besides the expression (\ref%
{R-tilde-prim}) consisting of Green's functions $G^{\left( c\right) }$ and $%
G^{\left( s\right) }$, given by (\ref{grin-f-finite}) and (\ref{G-za-infne}%
), as well as of functions $g^{\left( c\right) }$ and $g^{\left( s\right) },$
obtained as (\ref{g-kaligrafsko-finite}), each of them calculated in
Appendix, one also obtains expression (\ref{R-tilde-dvoprim}) for the
resolvent tensor, clearly showing that the compressive wave progresses in
the direction of a radius vector, while the shear wave progresses in the
plane perpendicular to the direction of a radius vector.

Finally, the qualitative properties of spatial profiles corresponding to
Green's function $G^{\left( x\right) }$, $x\in \left\{ c,s\right\} $, are
investigated through the graphical presentation of numerical calculations.
Fractional anti-Zener and Zener model I$^{{}^{+}}$ID.ID, given by (\ref%
{model-ifwps}), as well as fractional Burgers model VII, given by (\ref%
{model-fwps}), are respectively chosen as representatives of models
describing viscoelastic materials with infinite and finite wave propagation
speed, being $c_{x}=\frac{1}{\sqrt{\varrho }}\sqrt{\frac{b_{2}}{a_{3}}}$, in
order to plot time evolution of spatial profiles of Green's function $%
G^{\left( x\right) }$, as well as to plot its short and large time
asymptotic behavior. Regardless of the fact whether the wave propagation
speed is infinite or finite, the spatial profiles of Green's function are
positive and display decreasing character for smaller distances $r$ from the
origin, see Figures \ref{G-profili-np}, \ref{G-profili-ccp-ccp}, and \ref{G-profili-burgers-np}. However, if
wave propagation speed is infinite, the tails of spatial profiles either
asymptotically tend to zero as distance $r$ tends to infinity, as in Figures \ref{G-profili-np} and \ref{G-profili-ccp}, or, if wave
propagation speed is finite, the spatial profiles tend to infinity as $r$
reaches the point $r=c_{x}t$, forming the peak centered at $r=c_{x}t$, see
Figure \ref{G-Burgers-np-sa-num}, which can actually be observed in Figures %
\ref{G-epsiloniii} and \ref{G-Burgers-ccp-pikovi-med}, where the the plots
corresponding to the regularized Green's function are presented. More
precisely, if wave propagation speed is infinite, there is a single peak for
medium values of $r$, which carries the information about the initial
disturbance, translates towards the higher values of $r$ as time increases,
while reducing in height and gaining in width, and resembles to the Dirac
delta peak in the case of classical wave propagation, as it is shown in
Figures \ref{G-profili-np} and \ref{G-profili-ccp-ccp}, while, if wave
propagation speed is finite, according to the short time asymptotics of
Green's function (\ref{G-srsha-konacna-brzina}), there is the Dirac delta
distribution centered at point $r=c_{x}t$, and therefore for any time
instant one expects the peak forming at point $r=c_{x}t$, which cannot be
completely displayed in Figure \ref{G-Burgers-np-sa-num}, since the
analytical expression (\ref{grin-f-finite}) holds true only for $r<c_{x}t$.
In the case when there are no zeros of functions $\phi _{\sigma }$ and $\Phi
_{\sigma }$, given by (\ref{f-fi-iwps}) and (\ref{fiovi-burgers-7})$_{1}$,
which respectively refer to the model I$^{{}^{+}}$ID.ID and fractional
Burgers model VII, describing a viscoelastic material with infinite and
finite wave propagation speed, i.e., in the case when Green's function in
Laplace domain $\tilde{G}^{\left( x\right) }$ has no other branch points
besides $s=0$, the spatial profiles do not exhibit an oscillatory character,
while if functions $\phi _{\sigma }$ and $\Phi _{\sigma }$ have a pair of
complex conjugated zeros, i.e., if Green's function in Laplace domain $%
\tilde{G}^{\left( x\right) }$ has a pair of complex conjugated branch points 
$s_{0}$ and $\bar{s}_{0}$ besides $s=0$, on one hand, the tails of spatial
profiles in the case of model I$^{{}^{+}}$ID.ID have the damped oscillatory
character, as shown in Figure \ref{G-profili-ccp-dodatak}, and on the other
hand, an additional negative peak, preceding the peak forming at $r=c_{x}t$,
appears in spatial profiles in the case of fractional Burgers model VII, as
obvious from Figure \ref{G-Burgers-ccp-pikovi-med}. Moreover, asymptotic
profiles are compared with spatial profiles of Green's function in Figures %
\ref{G-srsha}, \ref{G-srsha-ccp}, and \ref{G-Burgers-srsha-np} for short
times, as well as in Figures \ref{G-srla} and \ref{G-srla-ccp} for large
times, displaying good agreement, which is also found if the profiles
obtained by the numerical Laplace transform inversion of Green's function in
the Laplace domain are compared with the profiles corresponding to
analytical expression, as it is obvious from Figures \ref{G-slaganje-sa-num}%
, \ref{G-profili-ccp}, and \ref{G-Burgers-np-sa-num}.

\appendix

\section{Asymptotics of memory function $\tilde{c}_{x}$ \label%
{Asimptotika-za-c}}

In order to examine both short- and large-time asymptotic behavior of
Green's functions, as done in Section \ref{Green's function}, it is
necessary to find the asymptotics of memory function $\tilde{c}_{x}$, $x\in
\left\{ c,s\right\} ,$ corresponding to the compressive and shear waves.
Starting from defining relations (\ref{c-tilde-compresive}) and (\ref%
{c-tilde-shear}) for memory function $\tilde{c}_{x}$ and by taking into
account the expression (\ref{sr-cr}) for the relaxation modulus in Laplace
domain, for the memory function one has 
\begin{equation}
\tilde{c}_{x}\left( s\right) =\sqrt{\frac{s\tilde{\sigma}_{sr}^{\left(
x\right) }\left( s\right) }{\varrho }}=\frac{1}{\sqrt{\varrho }}\sqrt{\frac{%
\Phi _{\varepsilon }(s)}{\Phi _{\sigma }(s)}},  \label{memory-function}
\end{equation}%
with functions $\Phi _{\varepsilon }$ and $\Phi _{\sigma }$ given by (\ref%
{Cm-Ld-2}). Memory function $\tilde{c}_{x},$ taken in the form (\ref%
{memory-function}), corresponds to the fractional linear models having
orders in interval $\left( 0,1\right) ,$ given by (\ref{kejsovi}), if
functions $\Phi _{\varepsilon }$ and $\Phi _{\sigma }$ are given by (\ref%
{fiovi, case all}) below, as well as to the fractional Burgers models (\ref%
{burgersi}), if functions $\Phi _{\varepsilon }$ and $\Phi _{\sigma }$ are
given by (\ref{Burgers1-fiovi}) and (\ref{Burgers2-fiovi}) below, while in
the case of fractional anti-Zener and Zener models, listed in Table \ref%
{tbl-5}, the memory function takes the form 
\begin{equation}
\tilde{c}_{x}\left( s\right) =\frac{1}{\sqrt{\varrho }}\sqrt{s^{\xi }\frac{%
\phi _{\varepsilon }\left( s\right) }{\phi _{\sigma }\left( s\right) }}%
,\quad x\in \left\{ c,s\right\} ,  \label{memory-function-aZ/Z}
\end{equation}%
with functions $\phi _{\varepsilon }$ and $\phi _{\sigma }$, as well as with
exponent $\xi $ given in Table \ref{skupina}, due to the form of complex
modulus%
\begin{equation*}
\tilde{E}\left( s\right) =\frac{\Phi _{\varepsilon }(s)}{\Phi _{\sigma }(s)}%
=s^{\xi }\frac{\phi _{\varepsilon }\left( s\right) }{\phi _{\sigma }\left(
s\right) },
\end{equation*}%
see also (\ref{CM-Ld})$_{2}$, specific for the fractional anti-Zener and
Zener models.

Further analysis of the asymptotics of memory function $\tilde{c}_{x},$ see\
(\ref{memory-function}) and (\ref{memory-function-aZ/Z}), is reduced to
finding the asymptotic expressions of functions $\Phi _{\sigma }$ and $\Phi
_{\varepsilon }$, respectively $\phi _{\sigma }$ and $\phi _{\varepsilon },$
whose explicit forms are given below and obtained as consequences of the
operators acting on stress and strain in the constitutive equation in time
domain (\ref{suma}) using the Laplace transform of the constitutive
equation, generally yielding (\ref{CM-Ld})$_{1}$, according to the Laplace
transforms of fractional integral and Riemann-Liouville fractional
derivative (\ref{LT-frac-int}) and (\ref{LT-frac-der}). It is demonstrated
in the sequel that the asymptotic behavior of memory function $\tilde{c}_{x}$%
, given by (\ref{memory-function}) or (\ref{memory-function-aZ/Z}), is
governed by the expression%
\begin{equation}
\tilde{c}_{x}\left( s\right) \sim \left\{ \!\!\!%
\begin{tabular}{ll}
$\kappa ,$ & \smallskip \\ 
$\kappa s^{\frac{\delta }{2}},$ & $\delta \in \left( 0,1\right) $, \smallskip%
\end{tabular}%
\ \right. \text{for both}\quad s\rightarrow \infty \quad \text{and}\quad
s\rightarrow 0,  \label{memory-function-asym}
\end{equation}%
for all considered linear fractional-order constitutive models of
viscoelastic body, where constants $\kappa $ and $\delta $ are given in the
Tables \ref{tbl-1} and \ref{tbl-4} for linear fractional-order models
containing only Riemann-Liouville fractional derivatives of orders in
interval $\left( 0,1\right) ,$ see (\ref{kejsovi}), as well as for
thermodynamically consistent fractional Burgers models, see (\ref{burgersi}%
), while Tables \ref{tbl-2} and \ref{tbl-3} present constants $\kappa $ and $%
\delta $ for symmetric and asymmetric fractional anti-Zener and Zener
models, see Table \ref{tbl-5}.

\begin{landscape}
 \begin{table}[p]
 \begin{center}
 \begin{tabular}{|c|c|c|c|c|c|c|c|}

 \hline \xrowht{14pt}

 & Function  $\phi _{\sigma }$ & Function $\phi _{\varepsilon }$ & Model &  & Order $\xi $ & Order $\lambda $ & Order $\kappa $ \\ 

 \hhline{|=|=|=|=|=|=|=|=|} \xrowht{14pt}

\multirow{8}{*}{\multirowcell{8}{\rotatebox{90}{Symmetric models}}} & \multirow{2}{*}{\multirowcell{2}{$ a_{1}+a_{2}s^{\alpha +\beta}$ }}  & \multirow{2}{*}{\multirowcell{2}{$ b_{1}+b_{2}s^{\alpha +\beta}$} } 
& ID.ID & $*$ & $\alpha -\mu $ & $-$ & $-$ \\

\cline{4-8} \xrowht{14pt}

 & & & ID.DD$^{+}$ &  & $\alpha +\mu <1$ & $-$ & $-$ \\

\Xcline{2-8}{2\arrayrulewidth} \xrowht{14pt}

& \multirow{3}{*}{\multirowcell{3}{$a_{1}+a_{2}s^{\lambda }+a_{3}s^{\alpha +\gamma }$}} &  \multirow{2}{*}{\multirowcell{2}{$b_{1}+b_{2}s^{\lambda }+b_{3}s^{\alpha +\gamma }$}} 
 & IID.IID & $*$ & $\eta-\gamma $ & $\alpha -\beta $ & $-$ \\

\cline{4-8} \xrowht{14pt}

 & & & IDD.IDD & $*$ & $\alpha -\mu $ & $\alpha +\beta <1$ & $-$ \\

\cline{3-8} \xrowht{14pt}

 & & $b_{1}+b_{2}s^{\kappa }+b_{3}s^{\alpha +\gamma }$ & IID.IDD &  & $\alpha -\mu $ & $\mu +\nu <1$ & $\alpha -\beta $ \\ 

\cline{2-8} \xrowht{14pt}

 & \multirow{3}{*}{\multirowcell{3}{$a_{1}+a_{2}s^{\frac{1+\alpha +\gamma }{2}}+a_{3}s^{1+\alpha +\gamma }$}} 
 &\multirow{3}{*}{\multirowcell{3}{$b_{1}+b_{2}s^{\frac{1+\alpha +\gamma }{2}}+b_{3}s^{1+\alpha +\gamma }$}} & I$^{+}$ID.I$^{+}$ID &  & $\alpha -\mu $ & $-$ & $-$ \\

\cline{4-8} \xrowht{14pt} 

 &  &  & IDD$^{+}$.IDD$^{+}$ &  & $\eta -\gamma $ & $-$ & $-$ \\ 

\cline{4-8} \xrowht{14pt}

& & & I$^{+}$ID.IDD$^{+}$ &  & $1-\left( \gamma -\eta \right) $ & $-$ & $-$ \\ 

\Xhline{4\arrayrulewidth} \xrowht{14pt}

 \multirow{7}{*}{\multirowcell{7}{\rotatebox{90}{Asymmetric models}}}  
 & \multirow{2}{*}{\multirowcell{2}{$a_{1}+a_{2}s^{\lambda }+a_{3}s^{\kappa }$}} 
 & \multirow{4}{*}{\multirowcell{4}{$b_{1}+b_{2}s^{\alpha +\beta }$ }}
 & IID.ID &  & $\beta -\gamma $ & $\left( \alpha +\beta \right)-\left( \nu +\gamma \right) $ & $\alpha +\beta <1$ \\ 

\cline{4-8} \xrowht{14pt}

 & & & IDD.DD$^{+}$ &  & $\alpha +\mu <1$ & $\alpha +\beta <1$ & $\alpha +\mu <1$ \\ 

\cline{2-2} \cline{4-8} \xrowht{14pt}

 & \multirow{2}{*}{\multirowcell{2}{$a_{1}+a_{2}s^{\alpha +\beta }+a_{3}s^{2\left( \alpha +\beta \right) }$}}
 & & I$^{+}$ID.ID &  & $\beta +\nu <1$ & $-$ & $-$ \\

\cline{4-8} \xrowht{14pt}

 & & & IDD$^{+}$.DD$^{+}$ &  & $\alpha +\mu <1$ & $-$ & $-$ \\

\Xcline{2-8}{2\arrayrulewidth} \xrowht{14pt}

 & \multirow{3}{*}{\multirowcell{3}{$a_{1}+a_{2}s^{\alpha +\beta }$}} & \multirow{2}{*}{\multirowcell{2}{$b_{1}+b_{2}s^{\lambda }+b_{3}s^{\kappa }$}} & ID.IDD &  & $\alpha-\mu $ & $\mu +\nu <1$ & $\alpha +\beta <1$ \\ 

\cline{4-8} \xrowht{14pt}

 & & & ID.DDD$^{+}$ &  & $\alpha +\mu <1$ & $\nu -\mu $ & $\alpha +\nu-\left( \mu -\beta \right) $ \\

\cline{3-8} \xrowht{14pt} 

  & & $b_{1}+b_{2}s^{\alpha +\beta}+b_{3}s^{2\left( \alpha +\beta \right) }$ & ID.IDD$^{+}$ &  & $\nu -\beta $ & $-$ & $-$ \\

\hline

 \end{tabular}
 \end{center}
 \caption{Summary of constitutive functions $\phi _{\sigma }$ and $\phi _{\varepsilon }$, along with the order $\xi$, corresponding to the thermodynamically consistent fractional anti-Zener and Zener models. The notation $*$ means that the orders $\alpha +\beta$ and $\alpha +\gamma$ belong either to the interval $(0,1)$ or interval $(1,2)$.}
 \label{skupina}
 \end{table}
 
\end{landscape}

The functions $\Phi _{\sigma }$ and $\Phi _{\varepsilon },$ in the case of
linear fractional-order models containing only Riemann-Liouville fractional
derivatives of orders in interval $\left( 0,1\right) ,$ are obtained by
applying the Laplace transform to the unified constitutive equation (\ref%
{kejsovi}) in the form%
\begin{gather}
\Phi _{\sigma }(s)=\sum_{i=1}^{n}a_{i}\,s^{\alpha _{i}}\quad \text{and}\quad
\Phi _{\varepsilon }(s)=\sum_{j=1}^{m}b_{j}\,s^{\beta _{j}},\quad \text{with}
\label{fiovi, case all} \\
0\leqslant \alpha _{1}<\ldots <\alpha _{n}<\beta _{1}<\ldots <\beta _{m}<1,
\label{uredjenje}
\end{gather}%
originating from thermodynamical considerations, respectively reducing to%
\begin{gather}
\Phi _{\sigma }(s)=\sum_{i=1}^{n}a_{i}\,s^{\alpha _{i}}\quad \text{and}\quad
\Phi _{\varepsilon }(s)=\sum_{i=1}^{n}b_{i}\,s^{\alpha _{i}},
\label{fiovi-case1} \\
\Phi _{\sigma }(s)=\sum_{i=1}^{n}a_{i}\,s^{\alpha _{i}}\quad \text{and}\quad
\Phi _{\varepsilon }(s)=\sum_{i=1}^{n}b_{i}\,s^{\alpha
_{i}}+\sum_{i=n+1}^{m}b_{i}\,s^{\beta _{i}},  \label{fiovi-case2} \\
\Phi _{\sigma }(s)=\sum_{i=1}^{n-m}a_{i}\,s^{\alpha
_{i}}+\sum_{i=n-m+1}^{n}a_{i}\,s^{\alpha _{i}}\quad \text{and}\quad \Phi
_{\varepsilon }(s)=\sum_{j=1}^{m}b_{j}{}s^{\alpha _{n-m+j}},
\label{fiovi-case3} \\
\Phi _{\sigma }(s)=\sum_{i=1}^{n}a_{i}\,s^{\alpha _{i}}\quad \text{and}\quad
\Phi _{\varepsilon }(s)=\sum_{j=1}^{m}b_{j}\,s^{\beta _{j}}
\label{fiovi-case4}
\end{gather}%
for models belonging to Case I, II, III, and IV, see (\ref{Case 1}) - (\ref%
{Case 4}) and also Supplementary material. The asymptotics of memory
function $\tilde{c}_{x}$ is governed by (\ref{memory-function-asym}), with
constants $\kappa $ and $\delta $ presented in Table \ref{tbl-1} and
obtained according to (\ref{memory-function}) using functions $\Phi _{\sigma
}$ and $\Phi _{\varepsilon },$ see (\ref{fiovi-case1}) - (\ref{fiovi-case4}%
), and property (\ref{uredjenje}) of fractional differentiation orders. %
 \begin{table}[h]
 \begin{center}
 \begin{tabular}{|c|c||c|c|c|c|} 
 \hline \xrowht{18pt} 

\multirow{2}{*}{\multirowcell{2}{\rotatebox{90}{Linear fractional models }}}&\multirow{2}{*}{\multirowcell{2}{Models}}&\multicolumn{2}{c|}{\makecell{asymptotics of $\tilde{c}_{x}$ as\\ $s \to \infty$}} & \multicolumn{2}{c|}{\makecell{asymptotics of $\tilde{c}_{x}$ as\\ $s \to 0$}}  \\ \cline{3-6}

& & $\kappa$ & $\delta$ &   $\kappa$ & $\delta$  \\ \cline{2-2} \cline{3-6} \xrowht{18pt} 

& Case I  & $\frac{1}{\sqrt{\varrho }}\sqrt{\frac{b_{n}}{a_{n}}}$ & $0$ & \multirow{4}{*}{\multirowcell{4}{$\frac{1}{\sqrt{\varrho }}\sqrt{\frac{b_{1}}{a_{1}}}$}} & $0$ \\ \cline{2-4} \cline{6-6} \xrowht{18pt}

& Case II &\multirow{3}{*}{\multirowcell{3}{$\frac{1}{\sqrt{\varrho }}\sqrt{\frac{b_{m}}{a_{n}}}$}}  & $\beta _{m}-\alpha _{n}$ &  & $0$ \\ \cline{2-2} \cline{4-4} \cline{6-6} \xrowht{18pt} 

 &Case III & & $0$ &  & $\alpha _{n-m}$ \\\cline{2-2} \cline{4-4} \cline{6-6} \xrowht{18pt}

& Case IV & & $\beta _{m}-\alpha _{n}$ &  &  $\beta _{1}-\alpha _{1}$ \\ \hline

 \end{tabular}
 \end{center}
 \caption{Asymptotics of memory function $\tilde{c}_{x}\left( s\right)=\frac{1}{\sqrt{\varrho }}\sqrt{\frac{\Phi _{\varepsilon }(s)}{\Phi _{\sigma }(s)}} \sim \kappa s^{\frac{\delta }{2}}$ for both $s \to \infty$ and $s \to 0$ in the case of linear fractional models having highest differentiation order in interval $(0,1)$.}
 \label{tbl-1}
 \end{table}

In the case of fractional Burgers models, listed in Supplementary material,
grouping into two thermodynamically consistent classes and represented by
the unified constitutive equations (\ref{burgersi}), functions $\Phi
_{\sigma }$ and $\Phi _{\varepsilon },$ for models of the first class take
the form%
\begin{equation}
\Phi _{\sigma }(s)=1+a_{1}s^{\alpha }+a_{2}\,s^{\beta }+a_{3}\,s^{\gamma
}\quad \text{and}\quad \Phi _{\varepsilon }(s)=b_{1}\,s^{\mu }+b_{2}\,s^{\mu
+\eta },  \label{Burgers1-fiovi}
\end{equation}%
with model parameters as in Introduction, reducing to%
\begin{gather}
\Phi _{\sigma }(s)=1+a_{1}s^{\alpha }+a_{2}\,s^{\beta }+a_{3}\,s^{\gamma
}\quad \text{and}\quad \Phi _{\varepsilon }(s)=b_{1}\,s^{\mu }+b_{2}\,s^{\mu
+\eta },  \label{fiovi-burgers-1} \\
\Phi _{\sigma }(s)=1+a_{1}s^{\alpha }+a_{2}\,s^{\beta }+a_{3}\,s^{2\alpha
}\quad \text{and}\quad \Phi _{\varepsilon }(s)=b_{1}\,s^{\mu }+b_{2}\,s^{\mu
+\alpha },  \label{fiovi-burgers-2} \\
\Phi _{\sigma }(s)=1+a_{1}s^{\alpha }+a_{2}\,s^{\beta }+a_{3}\,s^{\alpha
+\beta }\quad \text{and}\quad \Phi _{\varepsilon }(s)=b_{1}\,s^{\mu
}+b_{2}\,s^{\mu +\alpha },  \label{fiovi-burgers-3} \\
\Phi _{\sigma }(s)=1+a_{1}s^{\alpha }+a_{2}\,s^{\beta }+a_{3}\,s^{\alpha
+\beta }\quad \text{and}\quad \Phi _{\varepsilon }(s)=b_{1}\,s^{\mu
}+b_{2}\,s^{\mu +\beta },  \label{fiovi-burgers-4} \\
\Phi _{\sigma }(s)=1+a_{1}s^{\alpha }+a_{2}\,s^{\beta }+a_{3}\,s^{2\beta
}\quad \text{and}\quad \Phi _{\varepsilon }(s)=b_{1}\,s^{\mu }+b_{2}\,s^{\mu
+\beta },  \label{fiovi-burgers-5}
\end{gather}%
for Models I, II, III, IV, and V, respectively, while for models of the
second class, functions $\Phi _{\sigma }$ and $\Phi _{\varepsilon }$ are
given by%
\begin{equation}
\Phi _{\sigma }(s)=1+a_{1}s^{\alpha }+a_{2}\,s^{\beta }+a_{3}\,s^{\beta
+\eta }\quad \text{and}\quad \Phi _{\varepsilon }(s)=b_{1}\,s^{\beta
}+b_{2}\,s^{\beta +\eta },  \label{Burgers2-fiovi}
\end{equation}%
with model parameters as in Introduction, reducing to%
\begin{gather}
\Phi _{\sigma }(s)=1+a_{1}s^{\alpha }+a_{2}\,s^{\beta }+a_{3}\,s^{\alpha
+\beta }\quad \text{and}\quad \Phi _{\varepsilon }(s)=b_{1}\,s^{\beta
}+b_{2}\,s^{\alpha +\beta },  \label{fiovi-burgers-6} \\
\Phi _{\sigma }(s)=1+a_{1}s^{\alpha }+a_{2}\,s^{\beta }+a_{3}\,s^{2\beta
}\quad \text{and}\quad \Phi _{\varepsilon }(s)=b_{1}\,s^{\beta
}+b_{2}\,s^{2\beta },  \label{fiovi-burgers-7} \\
\Phi _{\sigma }(s)=1+\bar{a}_{1}s^{\alpha }+\bar{a}_{2}\,s^{2\alpha }\quad 
\text{and}\quad \Phi _{\varepsilon }(s)=b_{1}\,s^{\alpha }+b_{2}\,s^{2\alpha
},  \label{fiovi-burgers-8}
\end{gather}%
for Models VI, VII, and VIII, respectively. The asymptotics of memory
function $\tilde{c}_{x}$ is governed by (\ref{memory-function-asym}), with
constants $\kappa $ and $\delta $ presented in Table \ref{tbl-4} and
obtained according to (\ref{memory-function}) using functions $\Phi _{\sigma
}$ and $\Phi _{\varepsilon },$ see (\ref{fiovi-burgers-1}) - (\ref%
{fiovi-burgers-5}), in the case of models belonging to the first class, as
well as using functions $\Phi _{\sigma }$ and $\Phi _{\varepsilon },$ see (%
\ref{fiovi-burgers-6}) - (\ref{fiovi-burgers-8}), in the case of models
belonging to the second class.  \begin{table}[h]
 \begin{center}
 \begin{tabular}{|c|c|c||c|c|c|c|} 
 \hline \xrowht{18pt} 

\multirow{9}{*}{\multirowcell{9}{\rotatebox{90}{Fractional Burgers models}}}   &\multicolumn{2}{c||}{\multirow{2}{*}{\makecell{Models}}} &\multicolumn{2}{c|}{\makecell{asymptotics of $\tilde{c}_{x}$ as\\ $s \to \infty$}} & \multicolumn{2}{c|}{\makecell{asymptotics of $\tilde{c}_{x}$ as\\ $s \to 0$}}  \\ \cline{4-7}

&\multicolumn{2}{c||}{}  & $\kappa$ & $\delta$ & $\kappa$ & $\delta$  \\ \cline{2-7} \xrowht{18pt} 
 
& \multirow{5}{*}{\multirowcell{5}{\rotatebox{90}{First class}}}  & Model I$^{*}$ & \multirow{7}{*}{\multirowcell{7}{$\frac{1}{\sqrt{\varrho }}\sqrt{\frac{b_{2}}{a_{3}}}$}}  & $\mu+\eta-\gamma$ & \multirow{8}{*}{\multirowcell{8}{$\frac{1}{\sqrt{\varrho }}\sqrt{b_{1}}$}}  & \multirow{5}{*}{\multirowcell{5}{$\mu$}}     \\ \cline{3-3} \cline{5-5}  \xrowht{18pt}

& &Model II & & $\mu-\alpha$ & &    \\ \cline{3-3} \cline{5-5}   \xrowht{18pt} 

& & Model III   &   & $\mu-\beta$ & &   \\  \cline{3-3} \cline{5-5}  \xrowht{18pt}

& &  Model IV & & $\mu-\alpha$ & &    \\  \cline{3-3} \cline{5-5} \xrowht{18pt} 

& &  Model V  & & $\mu-\beta$ & &   \\ \Xcline{2-3}{2\arrayrulewidth} \Xcline{5-5}{2\arrayrulewidth} \Xcline{7-7}{2\arrayrulewidth}  \xrowht{18pt}

&\multirow{2}{*}{\multirowcell{2}{\rotatebox{90}{Second class}}}&  Model VI  & & \multirow{3}{*}{\multirowcell{3}{$0$}}  & & \multirow{2}{*}{\multirowcell{2}{$\beta$}}  \\ \cline{3-3}  \xrowht{18pt} 

& &  Model VII& & & &   \\ \cline{3-4}   \cline{7-7} \xrowht{18pt} 

& &   Model VIII &  \multirow{1}{*}{\multirowcell{1}{$\frac{1}{\sqrt{\varrho }}\sqrt{\frac{b_{2}}{\bar{a}_{2}}}$}}  &  & & $\alpha$ \\ \hline

 \end{tabular}
 \end{center}
 \caption{Asymptotics of memory function $\tilde{c}_{x}\left( s\right)=\frac{1}{\sqrt{\varrho }}\sqrt{\frac{\Phi _{\varepsilon }(s)}{\Phi _{\sigma }(s)}} \sim \kappa s^{\frac{\delta }{2}}$ for both $s \to \infty$ and $s \to 0$ in the case of fractional Burgers models, where $^{*}$ is used to emphasize that $\eta \in \{\alpha, \beta, \gamma\}$.}
 \label{tbl-4}
 \end{table}

The functions $\phi _{\sigma }$ and $\phi _{\varepsilon },$ as well as
exponent $\xi ,$ appearing in the expression (\ref{memory-function-aZ/Z})
for memory function $\tilde{c}_{x},$ in the case of symmetric and asymmetric
fractional anti-Zener and Zener models, listed in Table \ref{tbl-5} and in
Supplementary material, are given in Table \ref{skupina}, and therefore the
asymptotics of memory function $\tilde{c}_{x}$ is governed by (\ref%
{memory-function-asym}), with constants $\kappa $ and $\delta $ presented in
Table \ref{tbl-2} for symmetric and in Table \ref{tbl-3} for asymmetric
fractional anti-Zener and Zener models.  \begin{table}[p]
 \begin{center}
 \begin{tabular}{|c|c||c|c|c|c|} 
 \hline \xrowht{18pt} 

\multirow{5}{*}{\multirowcell{5}{\rotatebox{90}{Symmetric anti-Zener/Zener models }}}&\multirow{2}{*}{\multirowcell{2}{Models}}&\multicolumn{2}{c|}{\makecell{asymptotics of $\tilde{c}_{x}$ as\\ $s \to \infty$}} & \multicolumn{2}{c|}{\makecell{asymptotics of $\tilde{c}_{x}$ as\\ $s \to 0$}}  \\ \cline{3-6}

&  & $\kappa$ & $\delta=\xi$ & $\kappa$ & $\delta=\xi$  \\ \cline{2-6} \xrowht{18pt} 
 
& ID.ID & \multirow{2}{*}{\multirowcell{2}{$\frac{1}{\sqrt{\varrho }}\sqrt{\frac{b_{2}}{a_{2}}}$}}  & $\alpha-\mu$ & \multirow{8}{*}{\multirowcell{8}{$\frac{1}{\sqrt{\varrho }}\sqrt{\frac{b_{1}}{a_{1}}}$}}  & $\alpha-\mu$    \\ \cline{2-2} \cline{4-4} \cline{6-6} \xrowht{18pt}

& ID.DD$^{+}$  & & $\alpha+\mu<1$ & & $\alpha+\mu<1$   \\ \cline{2-3} \cline{4-4} \cline{6-6}  \xrowht{18pt} 

& IID.IID   &  \multirow{6}{*}{\multirowcell{6}{$\frac{1}{\sqrt{\varrho }}\sqrt{\frac{b_{3}}{a_{3}}}$}}   & $\eta-\gamma$ & &  $\eta-\gamma$  \\ \cline{2-2} \cline{4-4} \cline{6-6}  \xrowht{18pt}

&   IDD.IDD & & \multirow{3}{*}{\multirowcell{3}{$\alpha-\mu$}} & &  \multirow{3}{*}{\multirowcell{3}{$\alpha-\mu$}}   \\  \cline{2-2} \xrowht{18pt} 

&   IID.IDD  & & & &    \\ \cline{2-2} \xrowht{18pt} 

&   I$^{+}$ID.I$^{+}$ID  & & & & \\ \cline{2-2} \cline{4-4} \cline{6-6}  \xrowht{18pt} 

&   IDD$^{+}$.IDD$^{+}$ & & $\eta-\gamma$ & & $\eta-\gamma$  \\ \cline{2-2} \cline{4-4} \cline{6-6} \xrowht{18pt} 

&    I$^{+}$ID.IDD$^{+}$ & & $1-(\gamma-\eta)$ & & $1-(\gamma-\eta)$ \\ \hline

 \end{tabular}
 \end{center}
 \caption{Asymptotics of memory function $\tilde{c}_{x}\left( s\right)=\frac{1}{\sqrt{\varrho }}\sqrt{s^{\xi}\frac{\phi _{\varepsilon }(s)}{\phi _{\sigma }(s)}} \sim \kappa s^{\frac{\delta }{2}}$ for both $s \to \infty$ and $s \to 0$ in the case of symmetric fractional anti-Zener/Zener models.}
 \label{tbl-2}
 \end{table}

  \begin{table}[p]
 \begin{center}
 \begin{tabular}{|c|c||c|c|c|c|} 
 \cline{1-6} \xrowht{18pt} 

\multirow{4}{*}{\multirowcell{4}{\rotatebox{90}{Asymmetric anti-Zener/Zener models }}}&\multirow{2}{*}{\multirowcell{2}{Models}}&\multicolumn{2}{c|}{\makecell{asymptotics of $\tilde{c}_{x}$ as\\ $s \to \infty$}} & \multicolumn{2}{c|}{\makecell{asymptotics of $\tilde{c}_{x}$ as\\ $s \to 0$}}  \\ \cline{3-6}

&  & $\kappa$ & $\delta$ & $\kappa$ & $\delta=\xi$  \\ \cline{2-6} \xrowht{18pt} 
 
 & IID.ID & \multirow{4}{*}{\multirowcell{4}{$\frac{1}{\sqrt{\varrho }}\sqrt{\frac{b_{2}}{a_{3}}}$}}  & $\beta-\gamma=\xi$ & \multirow{7}{*}{\multirowcell{7}{$\frac{1}{\sqrt{\varrho }}\sqrt{\frac{b_{1}}{a_{1}}}$}}  & $\beta-\gamma$    \\ \cline{2-2} \cline{4-4} \cline{6-6} \xrowht{18pt}

& IDD.DD$^{+}$ & & $\alpha+\beta<1$ & & $\alpha+\mu<1$   \\ \cline{2-2} \cline{4-4} \cline{6-6}  \xrowht{18pt} 

&I$^{+}$ID.ID  & & $\nu-\alpha$ & &  $\beta+\nu<1$  \\ \cline{2-2} \cline{4-4} \cline{6-6}  \xrowht{18pt}

&  IDD$^{+}$.DD & & $\mu-\beta$ & & $\alpha+\mu<1$    \\  \cline{2-3} \cline{4-4} \cline{6-6} \xrowht{18pt} 

&   ID.IDD & \multirow{3}{*}{\multirowcell{3}{$\frac{1}{\sqrt{\varrho }}\sqrt{\frac{b_{3}}{a_{2}}}$}}  & $\alpha-\mu=\xi$ & & $\alpha-\mu$   \\ \cline{2-2} \cline{4-4}  \cline{6-6}  \xrowht{18pt} 

&  ID.DDD$^{+}$ & &  \multirow{2}{*}{\multirowcell{2}{$\alpha+\nu<1$}} & & $\alpha+\mu<1$   \\ \cline{2-2} \cline{6-6}  \xrowht{18pt} 

&   ID.IDD$^{+}$ & &  & & $\nu-\beta$  \\ \hline

 \end{tabular}
 \end{center}
 \caption{Asymptotics of memory function $\tilde{c}_{x}\left( s\right)=\frac{1}{\sqrt{\varrho }}\sqrt{s^{\xi}\frac{\phi _{\varepsilon }(s)}{\phi _{\sigma }(s)}} \sim \kappa s^{\frac{\delta }{2}}$ for both $s \to \infty$ and $s \to 0$ in the case of asymmetric fractional anti-Zener/Zener models.}
 \label{tbl-3}
 \end{table}

\section{Calculation of Green's function, its asymptotics, and function $%
\mathfrak{g}^{\left( x\right) }$}

Green's functions $G^{\left( x\right) }$, along with their asymptotics, as
well as functions $g^{\left( x\right) }$, with $x\in \left\{ c,s\right\}$,
appearing in the resolvent tensor (\ref{R-tilde-prim}) and (\ref%
{R-tilde-dvoprim}) and discussed about in Section \ref{Green's function},
are calculated by the application of inverse Laplace transform definition,
using the contour integration and Cauchy integral theorem.

\subsection{Calculation of Green's function\label{CalcGrin}}

Green's function in Laplace domain, given by (\ref{Green-f-ld}), can be
rewritten as%
\begin{equation}
\tilde{G}^{\left( x\right) }\left( r,s\right) =\frac{1}{4\pi r\tilde{c}%
_{x}^{2}\left( s\right) }\mathrm{e}^{-\frac{rs}{\tilde{c}_{x}\left( s\right) 
}}=\frac{\varrho }{4\pi r}\frac{\Phi _{\sigma }(s)}{\Phi _{\varepsilon }(s)}%
\mathrm{e}^{-\sqrt{\varrho }rs\sqrt{\frac{\Phi _{\sigma }(s)}{\Phi
_{\varepsilon }(s)}}},  \label{Green-f-fis}
\end{equation}
using the memory function $\tilde{c}_{x}$ expressed in the form (\ref%
{memory-function}), so that Green's function in Laplace domain (\ref%
{Green-f-fis}) and hence (\ref{Green-f-ld}) as well, has a branch point $%
s=0, $ due to the terms in functions $\Phi _{\sigma }$ and $\Phi
_{\varepsilon }$ containing $s^{\xi },$ $\xi \in \left( 0,2\right) ,$ as
well as a pair of complex conjugated branch points $s_{0}=\rho _{0}\mathrm{e}%
^{\mathrm{i}\varphi _{0}}$ and $\bar{s}_{0},$ with $\varphi _{0}\in \left( 
\frac{\pi }{2},\pi \right) ,$ in the case when $s_{0}$ and $\bar{s}_{0}$ are
zeros of function $\Phi _{\sigma },$ while if $s_{0}$ and $\bar{s}_{0}$ are
zeros of function $\Phi _{\varepsilon },$ then these points are also the
poles of Green's function in Laplace domain. Note, instead of complex
conjugated zeros $s_{0}$ and $\bar{s}_{0}$, either of the functions $\Phi
_{\sigma }$ and $\Phi _{\varepsilon }$ may have a negative real zero as
well, and certainly no zeros at all.

In the case of linear fractional-order models containing only
Riemann-Liouville fractional derivatives of orders in interval $\left(
0,1\right)$, given by the constitutive equation (\ref{kejsovi}), there are
no zeros of functions $\Phi _{\sigma }$ and $\Phi _{\varepsilon }$,
according to Lemma 4.2. in \cite{KOZ19}, since all differentiation orders
are in interval $\left( 0,1\right)$, see (\ref{fiovi, case all}). On the
other hand, in the case of thermodynamically consistent fractional Burgers
models, given by (\ref{burgersi}), the function $\Phi _{\sigma }$ may have a
pair of complex conjugated zeros, or a negative real zero, or no zeros at
all, as proved in Appendix A of \cite{OZ-2}, since function $\Phi _{\sigma }$
is a polynomial function having the highest differentiation order in
interval $\left( 1,2\right)$, see (\ref{Burgers1-fiovi}) and (\ref%
{Burgers2-fiovi}) for models of the first and second class respectively.
Regarding the fractional anti-Zener and Zener models, listed in Table \ref%
{tbl-5}, it is shown in Section 6. of \cite{SD-2} that both functions $\phi
_{\sigma }$ and $\phi _{\varepsilon }$ may have a pair of complex conjugated
zeros, or a negative real zero, or no zeros at all, since both functions are
polynomial functions having the highest order in interval $\left( 1,2\right)$%
, see Table \ref{skupina}.

In the Cauchy integral theorem%
\begin{equation*}
\doint\nolimits_{\Gamma }\tilde{G}_{\varepsilon }^{\left( x\right) }\left(
r,s\right) \mathrm{e}^{st}\mathrm{d}s=0,
\end{equation*}%
the integration is performed along the contour $\Gamma ,$ depicted in Figure %
\ref{komplTG} and taken so that the singular points $s=0$, $s_{0}$, and $%
\bar{s}_{0}$ lie outside the contour $\Gamma ,$ in order to obtain the
regularized Green's function in time domain by the inversion of Laplace
transform of Green's function in Laplace domain, regularized by the smooth
approximation of Dirac delta distribution, yielding%
\begin{equation*}
G_{\varepsilon }^{\left( x\right) }\left( r,t\right) =\frac{1}{2\pi \mathrm{i%
}}\int_{Br}\tilde{G}_{\varepsilon }^{\left( x\right) }\left( r,s\right) 
\mathrm{e}^{st}\mathrm{d}s=\mathcal{L}^{-1}\left[ \tilde{G}^{\left( x\right)
}\left( r,s\right) \tilde{\delta}_{\varepsilon }\left( s\right) \right]
\left( r,t\right) =\mathcal{L}^{-1}\left[ \frac{1}{4\pi r\tilde{c}%
_{x}^{2}\left( s\right) }\mathrm{e}^{-\frac{rs}{\tilde{c}_{x}\left( s\right) 
}}\mathrm{e}^{-\varepsilon \sqrt{s}}\right] \left( r,t\right) ,
\label{calc-grin-reg}
\end{equation*}%
see also (\ref{racunanje-G-eps}), where integration is performed along the
Bromwich contour $Br$, i.e., along the contour $\Gamma _{0}$ in the limit
when $R\rightarrow \infty ,$ having $\Gamma _{0}$ as a part of the closed
contour $\Gamma $ from\ Figure \ref{komplTG}, so that the Cauchy integral
theorem reduces to%
\begin{equation}
2\pi \mathrm{i\,}G_{\varepsilon }^{\left( x\right) }\left( r,t\right) =-\lim 
_{\substack{ R\rightarrow \infty  \\ \check{r}\rightarrow 0}}\left(
I_{\Gamma _{3a}\cup \Gamma _{3b}}+I_{\Gamma _{5a}\cup \Gamma _{5b}}\right) ,
\label{racunanje-G-epsilon}
\end{equation}%
where $I_{\Gamma _{3a}\cup \Gamma _{3b}}\ $and $I_{\Gamma _{5a}\cup \Gamma
_{5b}}$ are integrals along contours $\Gamma _{3a}\cup \Gamma _{3b}$ and $%
\Gamma _{5a}\cup \Gamma _{5b}\ $having non-zero contribution, since for the
left-hand-side of Cauchy integral theorem one has%
\begin{equation}
\doint\nolimits_{\Gamma }\tilde{G}_{\varepsilon }^{\left( x\right) }\left(
r,s\right) \mathrm{e}^{st}\mathrm{d}s=\sum_{i=0}^{9}I_{\Gamma _{i}},\quad 
\text{with}\quad I_{\Gamma _{i}}=\int_{\Gamma _{i}}\frac{1}{4\pi r\tilde{c}%
_{x}^{2}\left( s\right) }\mathrm{e}^{-\frac{rs}{\tilde{c}_{x}\left( s\right) 
}}\mathrm{e}^{-\varepsilon \sqrt{s}}\mathrm{e}^{st}\mathrm{d}s,
\label{int-gama-kompl}
\end{equation}%
where the integrals $I_{\Gamma _{1}},$ $I_{\Gamma _{2}},$ $I_{\Gamma _{4}},$ 
$I_{\Gamma _{6}},$ $I_{\Gamma _{7}},$ $I_{\Gamma _{8}},$ and $I_{\Gamma
_{9}} $ have zero contribution in the limit when $R\rightarrow \infty $ and $%
\check{r}\rightarrow 0.$

\noindent 
\begin{minipage}{\columnwidth}
\begin{minipage}[c]{0.4\columnwidth}
\centering
\includegraphics[width=0.7\columnwidth]{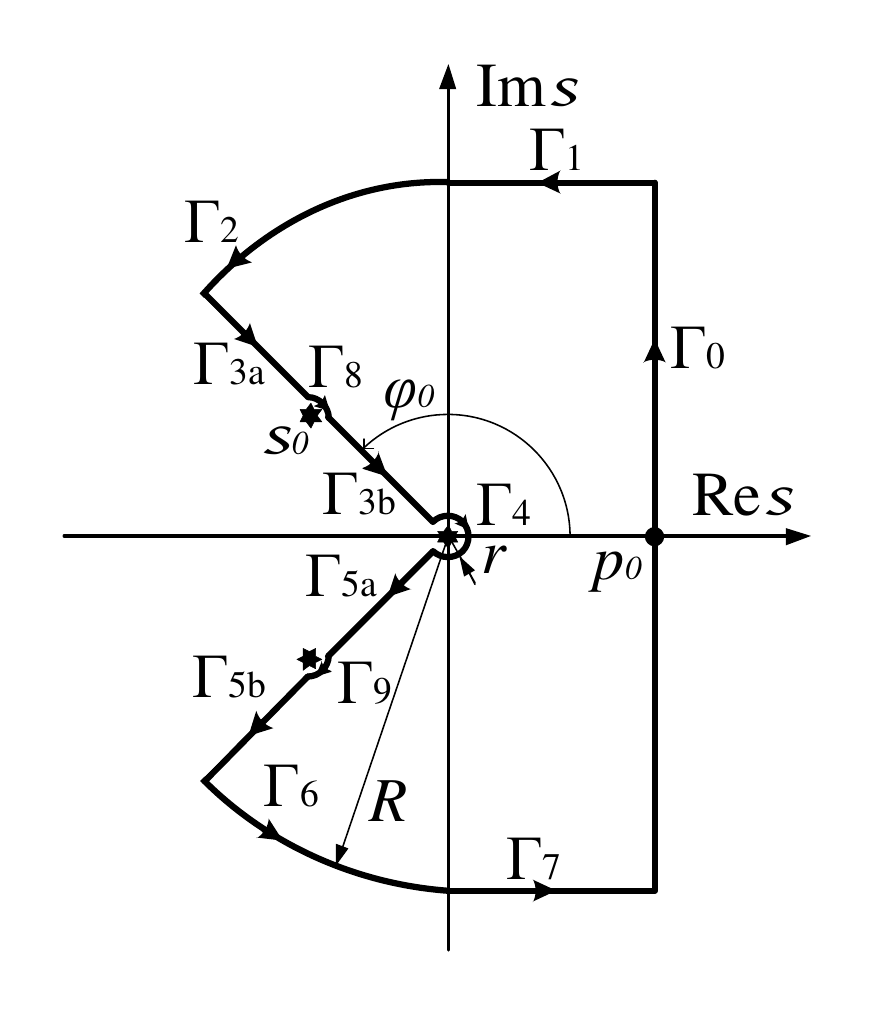}
\captionof{figure}{Integration contour $\Gamma$.}
\label{komplTG}
\end{minipage}
\hfil
\begin{minipage}[c]{0.55\columnwidth}
\centering
\begin{tabular}{rll}
$\Gamma _{0}:$ & Bromwich path, &  \\ 
$\Gamma _{1}:$ & $s=p+\mathrm{i}R,$ & $p\in \left[ 0,p_{0}\right],\, p_0\geq 0$ arbitrary, \\ 
$\Gamma _{2}:$ & $s=R\mathrm{e}^{\mathrm{i}\varphi },$ & $\varphi \in \left[ \frac{\pi }{2},\varphi_0 \right] ,$ \\ 
$\Gamma _{3a}\cup\Gamma _{3b}:$ & $s=\rho \mathrm{e}^{\mathrm{i}\varphi_0},$ & $\rho \in \left[ \check{r},R\right] ,$ \\ 
$\Gamma _{4}:$ & $s=\check{r}\mathrm{e}^{\mathrm{i}\varphi },$ & $\varphi \in \left[ -\varphi_0,\varphi_0\right] ,$ \\ 
$\Gamma _{5a}\cup\Gamma _{5b}:$ & $s=\rho \mathrm{e}^{-\mathrm{i}\varphi_0},$ & $\rho \in \left[ \check{r},R\right] ,$ \\
$\Gamma _{6}:$  & $s=R\mathrm{e}^{\mathrm{i}\varphi },$ & $\varphi \in \left[ -\varphi_0 , -\frac{\pi }{2} \right] ,$ \\
$\Gamma _{7}:$ & $s=p-\mathrm{i}R,$ & $p\in \left[ 0,p_{0}\right],\, p_0\geq 0$ arbitrary,\\
$\Gamma _{8}:$  & $s=s_0+\check{r}\mathrm{e}^{\mathrm{i}\varphi },$ & $\varphi \in \left[ -\pi+\varphi_0, \varphi_0 \right] ,$ \\
$\Gamma _{9}:$  & $s=\bar{s}_0+\check{r}\mathrm{e}^{\mathrm{i}\varphi },$ & $\varphi \in \left[ -\varphi_0,\pi-\varphi_0 \right]$.   
\end{tabular}
\captionof{table}{Parametrization of integration contour $\Gamma$.}
\label{komplTG-param}
\end{minipage}
\end{minipage}\smallskip

By the use of parametrization given in Table \ref{komplTG-param}, the
integrals in (\ref{racunanje-G-epsilon}) are calculated as 
\begin{align*}
2\pi \mathrm{i\,}G_{\varepsilon }^{\left( x\right) }\left( r,t\right) & =-%
\frac{1}{4\pi r}\Bigg(\int_{\infty }^{0}\frac{1}{\left\vert \tilde{c}%
_{x}^{2}\left( \rho \mathrm{e}^{\mathrm{i}\varphi _{0}}\right) \right\vert 
\mathrm{e}^{\mathrm{i}\arg \tilde{c}_{x}^{2}\left( \rho \mathrm{e}^{\mathrm{i%
}\varphi _{0}}\right) }}\mathrm{e}^{-\frac{r\rho \mathrm{e}^{\mathrm{i}%
\varphi _{0}}}{\left\vert \tilde{c}_{x}\left( \rho \mathrm{e}^{\mathrm{i}%
\varphi _{0}}\right) \right\vert \mathrm{e}^{\mathrm{i}\arg \tilde{c}%
_{x}\left( \rho \mathrm{e}^{\mathrm{i}\varphi _{0}}\right) }}}\mathrm{e}%
^{-\varepsilon \sqrt{\rho \mathrm{e}^{\mathrm{i}\varphi _{0}}}}\mathrm{e}%
^{\rho t\mathrm{e}^{\mathrm{i}\varphi _{0}}}\mathrm{e}^{\mathrm{i}\varphi
_{0}}\mathrm{d}\rho \\
& \quad +\int_{0}^{\infty }\frac{1}{\left\vert \tilde{c}_{x}^{2}\left( \rho 
\mathrm{e}^{\mathrm{i}\varphi _{0}}\right) \right\vert \mathrm{e}^{-\mathrm{i%
}\arg \tilde{c}_{x}^{2}\left( \rho \mathrm{e}^{\mathrm{i}\varphi
_{0}}\right) }}\mathrm{e}^{-\frac{r\rho \mathrm{e}^{-\mathrm{i}\varphi _{0}}%
}{\left\vert \tilde{c}_{x}\left( \rho \mathrm{e}^{\mathrm{i}\varphi
_{0}}\right) \right\vert \mathrm{e}^{-\mathrm{i}\arg \tilde{c}_{x}\left(
\rho \mathrm{e}^{\mathrm{i}\varphi _{0}}\right) }}}\mathrm{e}^{-\varepsilon 
\sqrt{\rho \mathrm{e}^{-\mathrm{i}\varphi _{0}}}}\mathrm{e}^{\rho t\mathrm{e}%
^{-\mathrm{i}\varphi _{0}}}\mathrm{e}^{-\mathrm{i}\varphi _{0}}\mathrm{d}%
\rho \Bigg) \\
& =\frac{1}{4\pi r}\int_{0}^{\infty }\frac{1}{\left\vert \tilde{c}%
_{x}^{2}\left( \rho \mathrm{e}^{\mathrm{i}\varphi _{0}}\right) \right\vert }%
\Bigg(\mathrm{e}^{-\frac{r\rho }{\left\vert \tilde{c}_{x}\left( \rho \mathrm{%
e}^{\mathrm{i}\varphi _{0}}\right) \right\vert }\mathrm{e}^{\mathrm{i}\left(
\varphi _{0}-\arg \tilde{c}_{x}\left( \rho \mathrm{e}^{\mathrm{i}\varphi
_{0}}\right) \right) }}\mathrm{e}^{-\varepsilon \sqrt{\rho }\mathrm{e}^{%
\mathrm{i}\frac{\varphi _{0}}{2}}}\mathrm{e}^{\rho t\mathrm{e}^{\mathrm{i}%
\varphi _{0}}}\mathrm{e}^{\mathrm{i}\left( \varphi _{0}-\arg \tilde{c}%
_{x}^{2}\left( \rho \mathrm{e}^{\mathrm{i}\varphi _{0}}\right) \right) } \\
& \quad -\mathrm{e}^{-\frac{r\rho }{\left\vert \tilde{c}_{x}\left( \rho 
\mathrm{e}^{\mathrm{i}\varphi _{0}}\right) \right\vert }\mathrm{e}^{-\mathrm{%
i}\left( \varphi _{0}-\arg \tilde{c}_{x}\left( \rho \mathrm{e}^{\mathrm{i}%
\varphi _{0}}\right) \right) }}\mathrm{e}^{-\varepsilon \sqrt{\rho \mathrm{e}%
^{-\mathrm{i}\varphi _{0}}}}\mathrm{e}^{\rho t\mathrm{e}^{-\mathrm{i}\varphi
_{0}}}\mathrm{e}^{-\mathrm{i}\left( \varphi _{0}-\arg \tilde{c}%
_{x}^{2}\left( \rho \mathrm{e}^{\mathrm{i}\varphi _{0}}\right) \right) }%
\Bigg)\mathrm{d}\rho \\
& =\frac{1}{4\pi r}\int_{0}^{\infty }\frac{1}{\left\vert \tilde{c}%
_{x}^{2}\left( \rho \mathrm{e}^{\mathrm{i}\varphi _{0}}\right) \right\vert }%
\mathrm{e}^{-\frac{r\rho }{\left\vert \tilde{c}_{x}\left( \rho \mathrm{e}^{%
\mathrm{i}\varphi _{0}}\right) \right\vert }\cos \left( \varphi _{0}-\arg 
\tilde{c}_{x}\left( \rho \mathrm{e}^{\mathrm{i}\varphi _{0}}\right) \right) }%
\mathrm{e}^{-\varepsilon \sqrt{\rho }\mathrm{\cos }\frac{\varphi _{0}}{2}}%
\mathrm{e}^{\rho t\mathrm{\cos }\varphi _{0}} \\
& \quad \times \Bigg(\mathrm{e}^{-\mathrm{i}\frac{r\rho }{\left\vert \tilde{c%
}_{x}\left( \rho \mathrm{e}^{\mathrm{i}\varphi _{0}}\right) \right\vert }%
\sin \left( \varphi _{0}-\arg \tilde{c}_{x}\left( \rho \mathrm{e}^{\mathrm{i}%
\varphi _{0}}\right) \right) }\mathrm{e}^{-\mathrm{i}\varepsilon \sqrt{\rho }%
\mathrm{\sin }\frac{\varphi _{0}}{2}}\mathrm{e}^{\mathrm{i}\rho t\mathrm{%
\sin }\varphi _{0}}\mathrm{e}^{\mathrm{i}\left( \varphi _{0}-\arg \tilde{c}%
_{x}^{2}\left( \rho \mathrm{e}^{\mathrm{i}\varphi _{0}}\right) \right) } \\
& \quad -\mathrm{e}^{\mathrm{i}\frac{r\rho }{\left\vert \tilde{c}_{x}\left(
\rho \mathrm{e}^{\mathrm{i}\varphi _{0}}\right) \right\vert }\sin \left(
\varphi _{0}-\arg \tilde{c}_{x}\left( \rho \mathrm{e}^{\mathrm{i}\varphi
_{0}}\right) \right) }\mathrm{e}^{\mathrm{i}\varepsilon \sqrt{\rho }\mathrm{%
\sin }\frac{\varphi _{0}}{2}}\mathrm{e}^{-\mathrm{i}\rho t\mathrm{\sin }%
\varphi _{0}}\mathrm{e}^{-\mathrm{i}\left( \varphi _{0}-\arg \tilde{c}%
_{x}^{2}\left( \rho \mathrm{e}^{\mathrm{i}\varphi _{0}}\right) \right) }%
\Bigg)\mathrm{d}\rho \\
& =2\mathrm{i}\frac{1}{4\pi r}\int_{0}^{\infty }\frac{1}{\left\vert \tilde{c}%
_{x}^{2}\left( \rho \mathrm{e}^{\mathrm{i}\varphi _{0}}\right) \right\vert }%
\mathrm{e}^{\rho t\mathrm{\cos }\varphi _{0}-\frac{r\rho }{\left\vert \tilde{%
c}_{x}\left( \rho \mathrm{e}^{\mathrm{i}\varphi _{0}}\right) \right\vert }%
\cos \left( \varphi _{0}-\arg \tilde{c}_{x}\left( \rho \mathrm{e}^{\mathrm{i}%
\varphi _{0}}\right) \right) -\varepsilon \sqrt{\rho }\mathrm{\cos }\frac{%
\varphi _{0}}{2}} \\
& \quad \times \sin \left( \rho t\mathrm{\sin }\varphi _{0}-\frac{r\rho }{%
\left\vert \tilde{c}_{x}\left( \rho \mathrm{e}^{\mathrm{i}\varphi
_{0}}\right) \right\vert }\sin \left( \varphi _{0}-\arg \tilde{c}_{x}\left(
\rho \mathrm{e}^{\mathrm{i}\varphi _{0}}\right) \right) +\varphi _{0}-\arg 
\tilde{c}_{x}^{2}\left( \rho \mathrm{e}^{\mathrm{i}\varphi _{0}}\right)
-\varepsilon \sqrt{\rho }\mathrm{\sin }\frac{\varphi _{0}}{2}\right) \mathrm{%
d}\rho ,
\end{align*}%
representing the expression (\ref{grin-f-finite}) for the regularized
Green's function $G_{\varepsilon }^{\left( x\right) }.$ It is left to prove
that the remaining integrals in (\ref{int-gama-kompl})$_{1}$ have zero
contributions.

The integral along the contour $\Gamma _{1},$ with the parametrization given
in Table \ref{komplTG-param}, becomes%
\begin{align*}
I_{\Gamma _{1}}& =\frac{1}{4\pi r}\int_{p_{0}}^{0}\frac{1}{\tilde{c}%
_{x}^{2}\left( p+\mathrm{i}R\right) }\mathrm{e}^{-\frac{r\left( p+\mathrm{i}%
R\right) }{\tilde{c}_{x}\left( p+\mathrm{i}R\right) }}\mathrm{e}%
^{-\varepsilon \sqrt{p+\mathrm{i}R}}\mathrm{e}^{\left( p+\mathrm{i}R\right)
t}\mathrm{d}p,\text{ i.e.,} \\
& =\frac{1}{4\pi r}\int_{p_{0}}^{0}\frac{1}{\tilde{c}_{x}^{2}\left( p+%
\mathrm{i}R\right) }\mathrm{e}^{-\frac{\mathrm{i}rR\left( 1+\frac{p}{\mathrm{%
i}R}\right) }{\tilde{c}_{x}\left( p+\mathrm{i}R\right) }}\mathrm{e}%
^{-\varepsilon \sqrt{\mathrm{i}R\left( 1+\frac{p}{\mathrm{i}R}\right) }}%
\mathrm{e}^{pt}\mathrm{e}^{\mathrm{i}Rt}\mathrm{d}p \\
& \sim \frac{1}{4\pi r}\int_{p_{0}}^{0}\frac{1}{\kappa ^{2}R^{\delta }%
\mathrm{e}^{\mathrm{i}\frac{\delta \pi }{2}}}\mathrm{e}^{-\frac{\mathrm{i}rR%
}{\kappa R^{\frac{\delta }{2}}\mathrm{e}^{\mathrm{i}\frac{\delta \pi }{4}}}}%
\mathrm{e}^{-\varepsilon \sqrt{R}\mathrm{e}^{\mathrm{i}\frac{\pi }{4}}}%
\mathrm{e}^{pt}\mathrm{e}^{\mathrm{i}Rt}\mathrm{d}p,\quad \text{as}\quad
R\rightarrow \infty ,
\end{align*}%
since $\tilde{c}_{x}\left( p+\mathrm{i}R\right) \sim \kappa R^{\frac{\delta 
}{2}}\mathrm{e}^{\mathrm{i}\frac{\delta \pi }{4}}$ as $R\rightarrow \infty $
with $\delta \in \left[ 0,1\right) $, see (\ref{memory-function-asym}) with $%
\kappa $ and $\delta $ given in Tables \ref{tbl-1}, \ref{tbl-4}, \ref{tbl-2}%
, and \ref{tbl-3}, so that%
\begin{align*}
\left\vert I_{\Gamma _{1}}\right\vert & \leqslant \frac{1}{4\pi r}%
\int_{0}^{p_{0}}\frac{1}{\kappa ^{2}R^{\delta }}\mathrm{e}^{-\frac{r}{\kappa 
}R^{1-\frac{\delta }{2}}\cos \left( \frac{\pi }{2}-\frac{\delta \pi }{4}%
\right) }\mathrm{e}^{-\varepsilon \sqrt{R}\cos \frac{\pi }{4}}\mathrm{e}^{pt}%
\mathrm{d}p \\
& \leqslant \frac{1}{4\pi r}\int_{0}^{p_{0}}\frac{1}{\kappa ^{2}R^{\delta }}%
\mathrm{e}^{-R\left( \frac{r}{\kappa R^{\frac{\delta }{2}}}\mathrm{\sin }%
\frac{\delta \pi }{4}+\frac{\sqrt{2}}{2}\frac{\varepsilon }{\sqrt{R}}\right)
}\mathrm{e}^{pt}\mathrm{d}p\rightarrow 0,\quad \text{as}\quad R\rightarrow
\infty .
\end{align*}%
Note, when $\delta =0,$ the previous expression becomes%
\begin{equation*}
\left\vert I_{\Gamma _{1}}\right\vert \leqslant \frac{1}{4\pi r\kappa ^{2}}%
\int_{0}^{p_{0}}\mathrm{e}^{-\frac{\sqrt{2}}{2}\varepsilon \sqrt{R}}\mathrm{e%
}^{pt}\mathrm{d}p\rightarrow 0,\quad \text{as}\quad R\rightarrow \infty ,
\end{equation*}%
due to the regularization of Greens' function in Laplace domain introduced
by (\ref{G-f-reg}). Similar argumentation yields that integral $I_{\Gamma
_{7}}$ also tends to zero when $R\rightarrow \infty .$

The integral along the contour $\Gamma _{2},$ with the parametrization given
in Table \ref{komplTG-param}, becomes%
\begin{align*}
I_{\Gamma _{2}}& =\frac{1}{4\pi r}\int_{\frac{\pi }{2}}^{\varphi _{0}}\frac{1%
}{\tilde{c}_{x}^{2}\left( R\mathrm{e}^{\mathrm{i}\varphi }\right) }\mathrm{e}%
^{-\frac{rR\mathrm{e}^{\mathrm{i}\varphi }}{\tilde{c}_{x}\left( R\mathrm{e}^{%
\mathrm{i}\varphi }\right) }}\mathrm{e}^{-\varepsilon \sqrt{R\mathrm{e}^{%
\mathrm{i}\varphi }}}\mathrm{e}^{Rt\mathrm{e}^{\mathrm{i}\varphi }}R\mathrm{e%
}^{\mathrm{i}\varphi }\mathrm{i\,d}\varphi \\
& \sim \frac{1}{4\pi r}\int_{\frac{\pi }{2}}^{\varphi _{0}}\frac{1}{\kappa
^{2}R^{\delta }\mathrm{e}^{\mathrm{i}\frac{\delta \varphi }{2}}}\mathrm{e}^{-%
\frac{rR\mathrm{e}^{\mathrm{i}\varphi }}{\kappa R^{\frac{\delta }{2}}\mathrm{%
e}^{\mathrm{i}\frac{\delta \varphi }{2}}}}\mathrm{e}^{-\varepsilon \sqrt{R}%
\mathrm{e}^{\mathrm{i}\frac{\varphi }{2}}}\mathrm{e}^{Rt\mathrm{e}^{\mathrm{i%
}\varphi }}R\mathrm{e}^{\mathrm{i}\varphi }\mathrm{i\,d}\varphi ,\quad \text{%
as}\quad R\rightarrow \infty ,
\end{align*}%
since $\tilde{c}_{x}\left( R\mathrm{e}^{\mathrm{i}\varphi }\right) \sim
\kappa R^{\frac{\delta }{2}}\mathrm{e}^{\mathrm{i}\frac{\delta \varphi }{4}}$
as $R\rightarrow \infty $ with $\delta \in \left[ 0,1\right) $, see (\ref%
{memory-function-asym}) with $\kappa $ and $\delta $ given in Tables \ref%
{tbl-1}, \ref{tbl-4}, \ref{tbl-2}, and \ref{tbl-3}, so that%
\begin{align*}
\left\vert I_{\Gamma _{2}}\right\vert & \leqslant \frac{1}{4\pi r}\int_{%
\frac{\pi }{2}}^{\varphi _{0}}\frac{1}{\kappa ^{2}}R^{1-\delta }\mathrm{e}^{-%
\frac{r}{\kappa }R^{1-\frac{\delta }{2}}\mathrm{\cos }\left( \left( 1-\frac{%
\delta }{2}\right) \varphi \right) }\mathrm{e}^{-\varepsilon \sqrt{R}\mathrm{%
\cos }\frac{\varphi }{2}}\mathrm{e}^{-Rt\left\vert \mathrm{\cos }\varphi
\right\vert }\mathrm{d}\varphi \\
& \leqslant \frac{1}{4\pi r}\int_{\frac{\pi }{2}}^{\varphi _{0}}\frac{1}{%
\kappa ^{2}}R^{1-\delta }\mathrm{e}^{-R\left( t\left\vert \mathrm{\cos }%
\varphi \right\vert +\frac{\varepsilon }{\sqrt{R}}\mathrm{\cos }\frac{%
\varphi }{2}+\frac{r}{\kappa R^{\frac{\delta }{2}}}\mathrm{\cos }\left(
\left( 1-\frac{\delta }{2}\right) \varphi \right) \right) }\mathrm{d}\varphi
\rightarrow 0,\quad \text{as}\quad R\rightarrow \infty .
\end{align*}%
Note, when $\delta =0,$ the previous expression becomes%
\begin{equation*}
\left\vert I_{\Gamma _{2}}\right\vert \leqslant \frac{1}{4\pi r\kappa ^{2}}%
\int_{\frac{\pi }{2}}^{\varphi _{0}}R\,\mathrm{e}^{-R\left( t-\frac{r}{%
\kappa }\right) \left\vert \mathrm{\cos }\varphi \right\vert }\mathrm{e}%
^{-\varepsilon \sqrt{R}\mathrm{\cos }\frac{\varphi }{2}}\mathrm{d}\varphi
\rightarrow 0,\quad \text{as}\quad R\rightarrow \infty ,\quad \text{for}%
\quad r<\kappa t.
\end{equation*}%
The condition $r<\kappa t$ implies the finite wave propagation speed $%
c_{x}=\kappa $. Similar argumentation yields that integral $I_{\Gamma _{6}}$
also tends to zero when $R\rightarrow \infty .$

The integral along the contour $\Gamma _{4},$ with the parametrization given
in Table \ref{komplTG-param}, becomes%
\begin{align*}
I_{\Gamma _{4}}& =\frac{1}{4\pi r}\int_{\varphi _{0}}^{-\varphi _{0}}\frac{1%
}{\tilde{c}_{x}^{2}\left( \check{r}\mathrm{e}^{\mathrm{i}\varphi }\right) }%
\mathrm{e}^{-\frac{r\check{r}\mathrm{e}^{\mathrm{i}\varphi }}{\tilde{c}%
_{x}\left( \check{r}\mathrm{e}^{\mathrm{i}\varphi }\right) }}\mathrm{e}%
^{-\varepsilon \sqrt{\check{r}\mathrm{e}^{\mathrm{i}\varphi }}}\mathrm{e}^{%
\check{r}t\mathrm{e}^{\mathrm{i}\varphi }}\check{r}\mathrm{e}^{\mathrm{i}%
\varphi }\mathrm{i\,d}\varphi \\
& =\frac{1}{4\pi r}\int_{\varphi _{0}}^{-\varphi _{0}}\frac{1}{\kappa ^{2}%
\check{r}^{\delta }\mathrm{e}^{\mathrm{i}\delta \varphi }}\mathrm{e}^{-\frac{%
r\check{r}\mathrm{e}^{\mathrm{i}\varphi }}{\kappa \check{r}^{\frac{\delta }{2%
}}\mathrm{e}^{\mathrm{i}\frac{\delta \varphi }{2}}}}\mathrm{e}^{-\varepsilon 
\sqrt{\check{r}}\mathrm{e}^{\mathrm{i}\frac{\varphi }{2}}}\mathrm{e}^{\check{%
r}t\mathrm{e}^{\mathrm{i}\varphi }}\check{r}\mathrm{e}^{\mathrm{i}\varphi }%
\mathrm{i\,d}\varphi \\
& =\frac{1}{4\pi r}\int_{\varphi _{0}}^{-\varphi _{0}}\frac{1}{\kappa ^{2}}%
\check{r}^{1-\delta }\mathrm{e}^{\mathrm{i}\left( 1-\delta \right) \varphi }%
\mathrm{e}^{-\frac{\check{r}}{\kappa }\check{r}^{1-\frac{\delta }{2}}\mathrm{%
e}^{\mathrm{i}\left( 1-\frac{\delta }{2}\right) \varphi }}\mathrm{e}%
^{-\varepsilon \sqrt{\check{r}}\mathrm{e}^{\mathrm{i}\frac{\varphi }{2}}}%
\mathrm{e}^{\check{r}t\mathrm{e}^{\mathrm{i}\varphi }}\mathrm{i\,d}\varphi ,
\\
& =\frac{1}{4\pi r}\int_{\varphi _{0}}^{-\varphi _{0}}\frac{1}{\kappa ^{2}}%
\check{r}^{1-\delta }\mathrm{e}^{\mathrm{i}\left( 1-\delta \right) \varphi }%
\mathrm{i\,d}\varphi \rightarrow 0,\quad \text{as}\quad \check{r}\rightarrow
0,
\end{align*}%
since $\tilde{c}_{x}\left( \check{r}\mathrm{e}^{\mathrm{i}\varphi }\right)
\sim \kappa \check{r}^{\frac{\delta }{2}}\mathrm{e}^{\mathrm{i}\frac{\delta
\varphi }{4}}$ as $\check{r}\rightarrow 0$ with $\delta \in \left(
0,1\right) $, see (\ref{memory-function-asym}) with $\kappa $ and $\delta $
given in Tables \ref{tbl-1}, \ref{tbl-4}, \ref{tbl-2}, and \ref{tbl-3}.

The integral along the contour $\Gamma _{8}$, according to parametrization
given in Table \ref{komplTG-param} and memory function (\ref{memory-function}%
), becomes%
\begin{equation*}
I_{\Gamma _{8}}=\frac{\varrho }{4\pi r}\int_{\varphi _{0}}^{-\pi +\varphi
_{0}}\frac{\Phi _{\sigma }\left( s_{0}+\check{r}\mathrm{e}^{\mathrm{i}%
\varphi }\right) }{\Phi _{\varepsilon }\left( s_{0}+\check{r}\mathrm{e}^{%
\mathrm{i}\varphi }\right) }\mathrm{e}^{-\sqrt{\varrho }r\left( s_{0}+\check{%
r}\mathrm{e}^{\mathrm{i}\varphi }\right) \sqrt{\frac{\Phi _{\sigma }(s_{0}+%
\check{r}\mathrm{e}^{\mathrm{i}\varphi })}{\Phi _{\varepsilon }(s_{0}+\check{%
r}\mathrm{e}^{\mathrm{i}\varphi })}}}\mathrm{e}^{-\varepsilon \sqrt{s_{0}+%
\check{r}\mathrm{e}^{\mathrm{i}\varphi }}}\mathrm{e}^{\left( s_{0}+\check{r}%
\mathrm{e}^{\mathrm{i}\varphi }\right) t}\check{r}\mathrm{e}^{\mathrm{i}%
\varphi }\mathrm{i\,d}\varphi ,
\end{equation*}%
so if $\Phi _{\sigma }\left( s_{0}\right) =0,$ then, in the limit when $%
\check{r}\rightarrow 0$, the integral $I_{\Gamma _{8}}$ tends to zero
according to the previous expression, while if $\Phi _{\varepsilon }\left(
s_{0}\right) =0,$ then, in the limit when $\check{r}\rightarrow 0$, the
integral $I_{\Gamma _{8}}$ behaves as%
\begin{align*}
I_{\Gamma _{8}} &\sim \frac{\varrho }{4\pi r}\int_{\varphi _{0}}^{-\pi
+\varphi _{0}}\frac{\Phi _{\sigma }\left( s_{0}\right) }{\Phi _{\varepsilon
}^{\prime }\left( s_{0}\right) \check{r}\mathrm{e}^{\mathrm{i}\varphi
}+O\left( \check{r}^{2}\right) }\mathrm{e}^{-\sqrt{\varrho }rs_{0}\frac{%
\sqrt{\Phi _{\sigma }(s_{0})}}{\sqrt{\Phi _{\varepsilon }^{\prime }\left(
s_{0}\right) \check{r}\mathrm{e}^{\mathrm{i}\varphi }+O\left( \check{r}%
^{2}\right) }}}\mathrm{e}^{-\varepsilon \sqrt{s_{0}}}\mathrm{e}^{s_{0}t}%
\check{r}\mathrm{e}^{\mathrm{i}\varphi }\mathrm{i\,d}\varphi \\
&\sim \mathrm{i}\frac{\varrho }{4\pi r}\Phi _{\sigma }\left( s_{0}\right) 
\mathrm{e}^{-\varepsilon \sqrt{s_{0}}}\mathrm{e}^{s_{0}t}\int_{\varphi
_{0}}^{-\pi +\varphi _{0}}\frac{1}{\Phi _{\varepsilon }^{\prime }\left(
s_{0}\right) +O\left( \check{r}\right) }\mathrm{e}^{-\sqrt{\varrho }rs_{0}%
\sqrt{\frac{\Phi _{\sigma }(s_{0})}{\Phi _{\varepsilon }^{\prime }\left(
s_{0}\right) }}\frac{1}{\sqrt{\check{r}}\mathrm{e}^{\mathrm{i}\frac{\varphi 
}{2}}}+O\left( \sqrt{\check{r}}\right) }\mathrm{\,d}\varphi \rightarrow
0,\quad \text{as}\quad \check{r}\rightarrow 0,
\end{align*}%
due to%
\begin{equation*}
\Phi _{\varepsilon }\left( s_{0}+\check{r}\mathrm{e}^{\mathrm{i}\varphi
}\right) =\Phi _{\varepsilon }\left( s_{0}\right) +\Phi _{\varepsilon
}^{\prime }\left( s_{0}\right) \check{r}\mathrm{e}^{\mathrm{i}\varphi
}+O\left( \check{r}^{2}\right) =\Phi _{\varepsilon }^{\prime }\left(
s_{0}\right) \check{r}\mathrm{e}^{\mathrm{i}\varphi }+O\left( \check{r}%
^{2}\right) ,\quad \text{as}\quad \check{r}\rightarrow 0,
\end{equation*}%
and 
\begin{align*}
\frac{1}{\sqrt{\Phi _{\varepsilon }\left( s_{0}+\check{r}\mathrm{e}^{\mathrm{%
i}\varphi }\right) }} &=\frac{1}{\sqrt{\Phi _{\varepsilon }^{\prime }\left(
s_{0}\right) \check{r}\mathrm{e}^{\mathrm{i}\varphi }+O\left( \check{r}%
^{2}\right) }}=\frac{1}{\sqrt{\Phi _{\varepsilon }^{\prime }\left(
s_{0}\right) }\sqrt{\check{r}}\mathrm{e}^{\mathrm{i}\frac{\varphi }{2}}\sqrt{%
1+O\left( \check{r}\right) }} \\
&=\frac{1}{\sqrt{\Phi _{\varepsilon }^{\prime }\left( s_{0}\right) }\sqrt{%
\check{r}}\mathrm{e}^{\mathrm{i}\frac{\varphi }{2}}}+O\left( \sqrt{\check{r}}%
\right) ,\quad \text{as}\quad \check{r}\rightarrow 0,
\end{align*}%
if%
\begin{equation*}
\func{Re}\left( s_{0}\sqrt{\frac{\Phi _{\sigma }(s_{0})}{\Phi _{\varepsilon
}^{\prime }\left( s_{0}\right) }}\mathrm{e}^{-\mathrm{i}\frac{\varphi }{2}%
}\right) >0\quad \text{for}\quad \varphi \in \left( -\pi +\varphi
_{0},\varphi _{0}\right) ,
\end{equation*}%
transforming into%
\begin{gather*}
\varphi _{0}+\frac{1}{2}\arg \Phi _{\sigma }(s_{0})-\frac{1}{2}\arg \Phi
_{\varepsilon }^{\prime }\left( s_{0}\right) -\frac{\varphi }{2}\in \left( -%
\frac{\pi }{2},\frac{\pi }{2}\right) ,\quad \text{i.e.,} \\
\varphi _{0}+\arg \Phi _{\sigma }(s_{0})-\arg \Phi _{\varepsilon }^{\prime
}\left( s_{0}\right) \in \left( -2\pi ,\pi \right) .
\end{gather*}

\subsection{Calculation of function $\mathfrak{g}^{\left( x\right) }$\label%
{Calckaligrafskog}}

The function $\mathfrak{\tilde{g}}^{\left( x\right) }$, given by (\ref%
{kaligrafsko-g})$_{2},$ takes the form%
\begin{equation*}
\mathfrak{\tilde{g}}^{\left( x\right) }\left( r,s\right) =\frac{1}{4\pi r}%
\frac{1}{s}\mathrm{e}^{-\frac{rs}{\tilde{c}_{x}\left( s\right) }}=\frac{1}{%
4\pi r}\frac{1}{s}\mathrm{e}^{-\sqrt{\varrho }rs\sqrt{\frac{\Phi _{\sigma
}(s)}{\Phi _{\varepsilon }(s)}}},
\end{equation*}%
due to the memory function $\tilde{c}_{x},$ given by (\ref{memory-function}%
), see also (\ref{c-tilde-compresive}) and (\ref{c-tilde-shear}), so that
function $\mathfrak{\tilde{g}}^{\left( x\right) }$ has $s=0$ as a pole, as
well as a branch point, due to the terms in functions $\Phi _{\sigma }$ and $%
\Phi _{\varepsilon }$ containing $s^{\xi },$ $\xi \in \left( 0,2\right)$.
Moreover, function $\mathfrak{\tilde{g}}^{\left( x\right) }$ has a pair of
complex conjugated branch points $s_{0}=\rho _{0}\mathrm{e}^{\mathrm{i}%
\varphi _{0}}$ and $\bar{s}_{0},$ with $\varphi _{0}\in \left( \frac{\pi }{2}%
,\pi \right) ,$ in the case when $s_{0}$ and $\bar{s}_{0}$ are zeros of
function $\Phi _{\sigma }$ (function $\Phi _{\varepsilon }$), with the
possibility that instead of complex conjugated zeros $s_{0}$ and $\bar{s}%
_{0} $, the function $\Phi _{\sigma }$ (function $\Phi _{\varepsilon }$) may
have a negative real zero or no zeros at all.

Therefore, in the Cauchy integral theorem%
\begin{equation*}
\doint\nolimits_{\Gamma }\mathfrak{\tilde{g}}^{\left( x\right) }\left(
r,s\right) \mathrm{e}^{st}\mathrm{d}s=0,
\end{equation*}%
the integration is performed along the contour $\Gamma ,$ depicted in Figure %
\ref{komplTG} and taken so that the singular points $s=0,$ $s_{0},$ and $%
\bar{s}_{0}$ lie outside the contour $\Gamma ,$ in order to obtain the
function $\mathfrak{g}^{\left( x\right) }$ in time domain by the inversion
of Laplace transform of function $\mathfrak{\tilde{g}}^{\left( x\right) }$
in Laplace domain, yielding%
\begin{equation*}
\mathfrak{g}^{\left( x\right) }\left( r,t\right) =\frac{1}{2\pi \mathrm{i}}%
\int_{Br}\mathfrak{\tilde{g}}^{\left( x\right) }\left( r,s\right) \mathrm{e}%
^{st}\mathrm{d}s=\mathcal{L}^{-1}\left[ \frac{1}{4\pi r}\frac{1}{s}\mathrm{e}%
^{-\frac{rs}{\tilde{c}_{x}\left( s\right) }}\right] \left( r,t\right) ,
\end{equation*}%
where the integration is performed along the Bromwich contour $Br$, i.e.,
along the contour $\Gamma _{0}$ in the limit when $R\rightarrow \infty ,$
having $\Gamma _{0}$ as a part of the closed contour $\Gamma $ from\ Figure %
\ref{komplTG}, so that the Cauchy integral theorem reduces to%
\begin{equation}
2\pi \mathrm{i\,}\mathfrak{g}^{\left( x\right) }\left( r,t\right) =-\lim 
_{\substack{ R\rightarrow \infty  \\ \check{r}\rightarrow 0}}\left(
I_{\Gamma _{3a}\cup \Gamma _{3b}}+I_{\Gamma _{5a}\cup \Gamma _{5b}}\right)
-\lim_{\check{r}\rightarrow 0}I_{\Gamma _{4}},
\label{racunanje-g-kaligrafsko}
\end{equation}%
where $I_{\Gamma _{3a}\cup \Gamma _{3b}},\ I_{\Gamma _{5a}\cup \Gamma
_{5b}}, $ and $I_{\Gamma _{4}}$ are integrals along contours $\Gamma
_{3a}\cup \Gamma _{3b},$ $\Gamma _{5a}\cup \Gamma _{5b},\ $and $\Gamma _{4}$
having non-zero contribution, since for the left-hand-side of the Cauchy
integral theorem one has%
\begin{equation}
\doint\nolimits_{\Gamma }\mathfrak{\tilde{g}}^{\left( x\right) }\left(
r,s\right) \mathrm{e}^{st}\mathrm{d}s=\sum_{i=0}^{9}I_{\Gamma _{i}},\quad 
\text{with}\quad I_{\Gamma _{i}}=\int_{\Gamma _{i}}\frac{1}{4\pi r}\frac{1}{s%
}\mathrm{e}^{-\frac{rs}{\tilde{c}_{x}\left( s\right) }}\mathrm{e}^{st}%
\mathrm{d}s,  \label{int-gama-nemaTG}
\end{equation}%
where the integrals $I_{\Gamma _{1}},$ $I_{\Gamma _{2}},$ $I_{\Gamma _{6}},$ 
$I_{\Gamma _{7}},$ $I_{\Gamma _{8}},$ and $I_{\Gamma _{9}}$ have zero
contribution in the limit when $R\rightarrow \infty $ and $\check{r}%
\rightarrow 0.$

By the use of parametrization given in Table \ref{komplTG-param}, the
integrals in (\ref{racunanje-g-kaligrafsko}) are calculated as%
\begin{align*}
2\pi \mathrm{i\,}\mathfrak{g}^{\left( x\right) }\left( r,t\right) & =-\frac{1%
}{4\pi r}\Bigg(\int_{\infty }^{0}\frac{1}{\rho \mathrm{e}^{\mathrm{i}\varphi
_{0}}}\mathrm{e}^{-\frac{r\rho \mathrm{e}^{\mathrm{i}\varphi _{0}}}{%
\left\vert \tilde{c}_{x}\left( \rho \mathrm{e}^{\mathrm{i}\varphi
_{0}}\right) \right\vert \mathrm{e}^{\mathrm{i}\arg \tilde{c}_{x}\left( \rho 
\mathrm{e}^{\mathrm{i}\varphi _{0}}\right) }}}\mathrm{e}^{\rho t\mathrm{e}^{%
\mathrm{i}\varphi _{0}}}\mathrm{e}^{\mathrm{i}\varphi _{0}}\mathrm{d}\rho \\
& \quad +\int_{0}^{\infty }\frac{1}{\rho \mathrm{e}^{-\mathrm{i}\varphi _{0}}%
}\mathrm{e}^{-\frac{r\rho \mathrm{e}^{-\mathrm{i}\varphi _{0}}}{\left\vert 
\tilde{c}_{x}\left( \rho \mathrm{e}^{\mathrm{i}\varphi _{0}}\right)
\right\vert \mathrm{e}^{-\mathrm{i}\arg \tilde{c}_{x}\left( \rho \mathrm{e}^{%
\mathrm{i}\varphi _{0}}\right) }}}\mathrm{e}^{\rho t\mathrm{e}^{-\mathrm{i}%
\varphi _{0}}}\mathrm{e}^{-\mathrm{i}\varphi _{0}}\mathrm{d}\rho +\lim_{%
\check{r}\rightarrow 0}\int_{\varphi _{0}}^{-\varphi _{0}}\frac{1}{\check{r}%
\mathrm{e}^{\mathrm{i}\varphi }}\mathrm{e}^{-\frac{r\check{r}\mathrm{e}^{%
\mathrm{i}\varphi }}{\tilde{c}_{x}\left( \check{r}\mathrm{e}^{\mathrm{i}%
\varphi }\right) }}\mathrm{e}^{\check{r}t\mathrm{e}^{\mathrm{i}\varphi }}%
\check{r}\mathrm{e}^{\mathrm{i}\varphi }\mathrm{i\,d}\varphi \Bigg) \\
& =\frac{1}{4\pi r}\int_{0}^{\infty }\frac{1}{\rho }\Bigg(\mathrm{e}^{-\frac{%
r\rho }{\left\vert \tilde{c}_{x}\left( \rho \mathrm{e}^{\mathrm{i}\varphi
_{0}}\right) \right\vert }\mathrm{e}^{\mathrm{i}\left( \varphi _{0}-\arg 
\tilde{c}_{x}\left( \rho \mathrm{e}^{\mathrm{i}\varphi _{0}}\right) \right)
}}\mathrm{e}^{\rho t\mathrm{e}^{\mathrm{i}\varphi _{0}}} \\
& \quad -\mathrm{e}^{-\frac{r\rho }{\left\vert \tilde{c}_{x}\left( \rho 
\mathrm{e}^{\mathrm{i}\varphi _{0}}\right) \right\vert }\mathrm{e}^{-\mathrm{%
i}\left( \varphi _{0}-\arg \tilde{c}_{x}\left( \rho \mathrm{e}^{\mathrm{i}%
\varphi _{0}}\right) \right) }}\mathrm{e}^{\rho t\mathrm{e}^{-\mathrm{i}%
\varphi _{0}}}\Bigg)\mathrm{d}\rho -\frac{1}{4\pi r}\lim_{\check{r}%
\rightarrow 0}\int_{\varphi _{0}}^{-\varphi _{0}}\mathrm{e}^{-\frac{r\check{r%
}\mathrm{e}^{\mathrm{i}\varphi }}{\tilde{c}_{x}\left( \check{r}\mathrm{e}^{%
\mathrm{i}\varphi }\right) }}\mathrm{e}^{\check{r}t\mathrm{e}^{\mathrm{i}%
\varphi }}\mathrm{i\,d}\varphi \\
& =\frac{1}{4\pi r}\int_{0}^{\infty }\frac{1}{\rho }\mathrm{e}^{-\frac{r\rho 
}{\left\vert \tilde{c}_{x}\left( \rho \mathrm{e}^{\mathrm{i}\varphi
_{0}}\right) \right\vert }\cos \left( \varphi _{0}-\arg \tilde{c}_{x}\left(
\rho \mathrm{e}^{\mathrm{i}\varphi _{0}}\right) \right) }\mathrm{e}^{-\rho
t\left\vert \cos \varphi _{0}\right\vert } \\
& \quad \times \Bigg(\mathrm{e}^{-\mathrm{i}\frac{r\rho }{\left\vert \tilde{c%
}_{x}\left( \rho \mathrm{e}^{\mathrm{i}\varphi _{0}}\right) \right\vert }%
\sin \left( \varphi _{0}-\arg \tilde{c}_{x}\left( \rho \mathrm{e}^{\mathrm{i}%
\varphi _{0}}\right) \right) }\mathrm{e}^{\mathrm{i}\rho t\mathrm{\sin }%
\varphi _{0}} \\
& \quad -\mathrm{e}^{\mathrm{i}\frac{r\rho }{\left\vert \tilde{c}_{x}\left(
\rho \mathrm{e}^{\mathrm{i}\varphi _{0}}\right) \right\vert }\sin \left(
\varphi _{0}-\arg \tilde{c}_{x}\left( \rho \mathrm{e}^{\mathrm{i}\varphi
_{0}}\right) \right) }\mathrm{e}^{-\mathrm{i}\rho t\mathrm{\sin }\varphi
_{0}}\Bigg)\mathrm{d}\rho -\frac{1}{4\pi r}\int_{\varphi _{0}}^{-\varphi
_{0}}\mathrm{i\,d}\varphi \\
& =2\mathrm{i}\frac{1}{4\pi r}\int_{0}^{\infty }\frac{1}{\rho }\mathrm{e}%
^{-\left( \rho t\left\vert \mathrm{\cos }\varphi _{0}\right\vert +\frac{%
r\rho }{\left\vert \tilde{c}_{x}\left( \rho \mathrm{e}^{\mathrm{i}\varphi
_{0}}\right) \right\vert }\cos \left( \varphi _{0}-\arg \tilde{c}_{x}\left(
\rho \mathrm{e}^{\mathrm{i}\varphi _{0}}\right) \right) \right) } \\
& \quad \times \sin \left( \rho t\mathrm{\sin }\varphi _{0}-\frac{r\rho }{%
\left\vert \tilde{c}_{x}\left( \rho \mathrm{e}^{\mathrm{i}\varphi
_{0}}\right) \right\vert }\sin \left( \varphi _{0}-\arg \tilde{c}_{x}\left(
\rho \mathrm{e}^{\mathrm{i}\varphi _{0}}\right) \right) \right) \mathrm{d}%
\rho +\frac{1}{4\pi r}2\mathrm{i}\varphi _{0},
\end{align*}%
representing the expression (\ref{g-kaligrafsko-finite}), where the behavior 
$\tilde{c}_{x}\left( \check{r}\mathrm{e}^{\mathrm{i}\varphi }\right) \sim
\kappa \check{r}^{\frac{\delta }{2}}\mathrm{e}^{\mathrm{i}\frac{\delta
\varphi }{4}} $ as $\check{r}\rightarrow 0$ with $\delta \in \left(
0,1\right) $, see (\ref{memory-function-asym}) with $\kappa $ and $\delta $
given in Tables \ref{tbl-1}, \ref{tbl-4}, \ref{tbl-2}, and \ref{tbl-3}, is
used. One can prove that the remaining integrals in (\ref{int-gama-nemaTG})$%
_1$ have zero contribution as in the case of calculation of Green's function
in Section \ref{CalcGrin}.

\subsection{Calculation of short-time asymptotics of Green's function\label%
{Green-asy-calc}}

The expression for short-time asymptotics of Green's function takes the form
(\ref{G-srsha-konacna-brzina}) if wave propagation speed is finite, i.e., if
glass modulus $\sigma _{sr/g}^{\left( x\right) }$ attains a finite value,
which can be further justified by taking into account the asymptotics of
regularization of Green's function in Laplace domain%
\begin{equation}
\tilde{G}_{\varepsilon }^{\left( x\right) }\left( r,s\right) \sim \tilde{G}%
_{\varepsilon ,\mathrm{asy}}^{\left( x\right) }\left( r,s\right) =\frac{1}{%
4\pi r\kappa ^{2}}\mathrm{e}^{-\frac{rs}{\kappa }}\mathrm{e}^{-\varepsilon 
\sqrt{s}},\quad \text{for}\quad s\rightarrow \infty ,
\label{G-ld-finit-brzina}
\end{equation}%
that follows from the regularized Green's function in Laplace domain (\ref%
{G-ld-reg}), with (\ref{Green-f-ld}) and (\ref{delta-reg}), due to the
asymptotics (\ref{memory-function-asym}) of memory functions $\tilde{c}_{x}$
and parameters $\kappa $ and $\delta $ given in Tables \ref{tbl-1} and \ref%
{tbl-4}, so that the asymptotics (\ref{G-ld-finit-brzina}) implies%
\begin{align}
G_{\varepsilon }^{\left( x\right) }\left( r,s\right) \sim G_{\varepsilon ,%
\mathrm{asy}}^{\left( x\right) }\left( r,t\right) &=\frac{1}{4\pi r\kappa
^{2}}\mathcal{L}^{-1}\left[ \mathrm{e}^{-\varepsilon \sqrt{s}}\right] \left(
t-\frac{r}{\kappa }\right)  \notag \\
&=\frac{1}{4\pi r\kappa ^{2}}\left. \frac{\varepsilon }{2t^{\prime }\sqrt{%
\pi t^{\prime }}}\mathrm{e}^{-\frac{\varepsilon ^{2}}{4t^{\prime }}%
}\right\vert _{t^{\prime }=t-\frac{r}{\kappa }}  \notag \\
&=\frac{1}{4\pi r\kappa ^{2}}\delta _{\varepsilon }\left( t-\frac{r}{\kappa }%
\right) ,\quad \text{for}\quad t\rightarrow 0,  \label{Gf-reg-asy}
\end{align}%
according to (\ref{delta-reg}), yielding%
\begin{equation*}
G^{\left( x\right) }\left( r,s\right) \sim G_{\mathrm{asy}}^{\left( x\right)
}\left( r,t\right) =\lim_{\varepsilon \rightarrow 0}G_{\varepsilon ,\mathrm{%
asy}}^{\left( x\right) }\left( r,t\right) =\frac{1}{4\pi r\kappa ^{2}}\delta
\left( t-\frac{r}{\kappa }\right) ,\quad \text{for}\quad t\rightarrow 0,
\end{equation*}%
which confirms Green's function short-time asymptotics (\ref%
{G-srsha-konacna-brzina}).

The integral representation of the asymptotics of regularized Green's
function (\ref{Gf-reg-asy}) is obtained as the Laplace inversion of (\ref%
{G-ld-finit-brzina}), being of the form%
\begin{equation*}
G_{\varepsilon ,\mathrm{asy}}^{\left( x\right) }\left( r,t\right) =\frac{1}{%
2\pi \mathrm{i}}\int_{Br}\tilde{G}_{\varepsilon ,\mathrm{asy}}^{\left(
x\right) }\left( r,s\right) \mathrm{e}^{st}\mathrm{d}s=\mathcal{L}^{-1}\left[
\frac{1}{4\pi r\kappa ^{2}}\mathrm{e}^{-\frac{rs}{\kappa }}\mathrm{e}%
^{-\varepsilon \sqrt{s}}\right] \left( r,t\right) ,
\end{equation*}%
where the integration is performed along the Bromwich contour $Br$, i.e.,
along the contour $\Gamma _{0}$ in the limit when $R\rightarrow \infty ,$
having $\Gamma _{0}$ as a part of the closed contour $\Gamma $ from\ Figure %
\ref{nemaTG}, which follows from the Cauchy integral theorem%
\begin{equation*}
\doint\nolimits_{\Gamma }\tilde{G}_{\varepsilon ,\mathrm{asy}}^{\left(
x\right) }\left( r,s\right) \mathrm{e}^{st}\mathrm{d}s=0,
\end{equation*}%
with integration performed along the contour $\Gamma ,$ depicted in Figure %
\ref{nemaTG}, yielding%
\begin{equation}
2\pi \mathrm{i\,}G_{\varepsilon ,\mathrm{asy}}^{\left( x\right) }\left(
r,t\right) =-\lim_{\substack{ R\rightarrow \infty  \\ \check{r}\rightarrow 0 
}}\left( I_{\Gamma _{3}}+I_{\Gamma _{5}}\right) ,
\label{racunanje-G-epsilon-asimpt-1}
\end{equation}%
where $I_{\Gamma _{3}}\ $and $I_{\Gamma _{5}}$ are integrals along contours $%
\Gamma _{3}$ and $\Gamma _{5}\ $having non-zero contribution, since for the
left-hand-side of Cauchy integral theorem one has%
\begin{equation}
\doint\nolimits_{\Gamma }\tilde{G}_{\varepsilon ,\mathrm{asy}}^{\left(
x\right) }\left( r,s\right) \mathrm{e}^{st}\mathrm{d}s=\sum_{i=0}^{7}I_{%
\Gamma _{i}},\quad \text{with}\quad I_{\Gamma _{i}}=\int_{\Gamma _{i}}\frac{1%
}{4\pi r\kappa ^{2}}\mathrm{e}^{-\frac{rs}{\kappa }}\mathrm{e}^{-\varepsilon 
\sqrt{s}}\mathrm{e}^{st}\mathrm{d}s,  \label{int-gama-kompl-asimpt-1}
\end{equation}%
where the integrals $I_{\Gamma _{1}},$ $I_{\Gamma _{2}},$ $I_{\Gamma _{4}},$ 
$I_{\Gamma _{6}},$ and $I_{\Gamma _{7}}$ have zero contribution in the limit
when $R\rightarrow \infty $ and $\check{r}\rightarrow 0.$

\noindent 
\begin{minipage}{\columnwidth}
\begin{minipage}[c]{0.4\columnwidth}
\centering
\includegraphics[width=0.7\columnwidth]{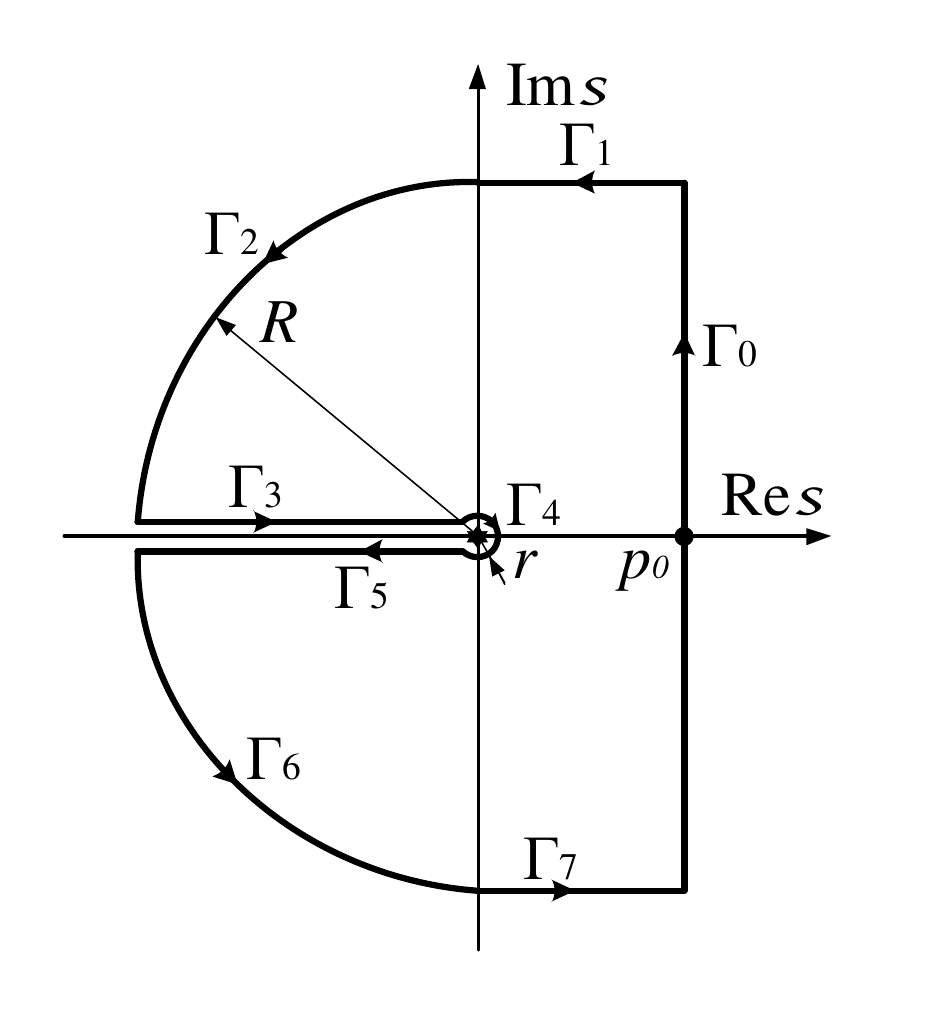}
\captionof{figure}{Integration contour $\Gamma$.}
\label{nemaTG}
\end{minipage}
\hfil
\begin{minipage}[c]{0.55\columnwidth}
\centering
\begin{tabular}{rll}
$\Gamma _{0}:$ & Bromwich path, &  \\ 
$\Gamma _{1}:$ & $s=p+\mathrm{i}R,$ & $p\in \left[ 0,p_{0}\right],\, p_0\geq 0$ arbitrary, \\ 
$\Gamma _{2}:$ & $s=R\mathrm{e}^{\mathrm{i}\varphi },$ & $\varphi \in \left[\frac{\pi }{2},\pi \right] ,$ \\ 
$\Gamma _{3}:$ & $s=\rho \mathrm{e}^{\mathrm{i}\pi },$ & $\rho \in \left[ \check{r},R\right] ,$ \\ 
$\Gamma _{4}:$ & $s=\check{r}\mathrm{e}^{\mathrm{i}\varphi },$ & $\varphi \in \left[ -\pi,\pi \right] ,$ \\ 
$\Gamma _{5}:$ & $s=\rho \mathrm{e}^{-\mathrm{i}\pi },$ & $\rho \in \left[ \check{r},R\right] ,$ \\
$\Gamma _{6}:$  & $s=R\mathrm{e}^{\mathrm{i}\varphi },$ & $\varphi \in \left[ -\pi, -\frac{\pi }{2} \right] ,$ \\
$\Gamma _{7}:$ & $s=p-\mathrm{i}R,$ & $p\in \left[ 0,p_{0}\right],\, p_0\geq 0$ arbitrary.  
\end{tabular}
\captionof{table}{Parametrization of integration contour $\Gamma$.}
\label{nemaTG-param}
\end{minipage}
\end{minipage}\smallskip

By the use of parametrization given in Table \ref{nemaTG-param}, the
integrals in (\ref{racunanje-G-epsilon-asimpt-1}) are calculated as 
\begin{align*}
2\pi \mathrm{i\,}G_{\varepsilon ,\mathrm{asy}}^{\left( x\right) }\left(
r,t\right) & =-\frac{1}{4\pi r\kappa ^{2}}\Bigg(\int_{\infty }^{0}\mathrm{e}%
^{-\frac{r\rho \mathrm{e}^{\mathrm{i}\pi }}{\kappa }}\mathrm{e}%
^{-\varepsilon \sqrt{\rho \mathrm{e}^{\mathrm{i}\pi }}}\mathrm{e}^{\rho t%
\mathrm{e}^{\mathrm{i}\pi }}\mathrm{e}^{\mathrm{i}\pi }\mathrm{d}\rho
+\int_{0}^{\infty }\mathrm{e}^{-\frac{r\rho \mathrm{e}^{-\mathrm{i}\pi }}{%
\kappa }}\mathrm{e}^{-\varepsilon \sqrt{\rho \mathrm{e}^{-\mathrm{i}\pi }}}%
\mathrm{e}^{\rho t\mathrm{e}^{-\mathrm{i}\pi }}\mathrm{e}^{-\mathrm{i}\pi }%
\mathrm{d}\rho \Bigg) \\
& =\frac{1}{4\pi r\kappa ^{2}}\int_{0}^{\infty }\mathrm{e}^{-\rho \left( t-%
\frac{r}{\kappa }\right) }\Bigg(\mathrm{e}^{\mathrm{i}\varepsilon \sqrt{\rho 
}}-\mathrm{e}^{-\mathrm{i}\varepsilon \sqrt{\rho }}\Bigg)\mathrm{d}\rho \\
& =2\mathrm{i}\frac{1}{4\pi r\kappa ^{2}}\int_{0}^{\infty }\mathrm{e}^{-\rho
\left( t-\frac{r}{\kappa }\right) }\sin \left( \varepsilon \sqrt{\rho }%
\right) \mathrm{d}\rho ,
\end{align*}%
yielding the integral representation of the asymptotics of regularized
Green's function (\ref{Gf-reg-asy}), given by (\ref{srsha-int-repr}). One
can prove that the remaining integrals in (\ref{int-gama-kompl-asimpt-1})$%
_{1}$ have zero contributions as in the case of calculation of Green's
function in Section \ref{CalcGrin}.

If wave propagation speed is infinite, i.e., if glass modulus $\sigma
_{sr/g}^{\left( x\right) }$ attains a infinite value, then the asymptotic
expression takes the form (\ref{G-short-time}), which is obtained using the
Cauchy integral theorem%
\begin{equation*}
\doint\nolimits_{\Gamma }\tilde{G}_{\mathrm{asy}}^{\left( x\right) }\left(
r,s\right) \mathrm{e}^{st}\mathrm{d}s=0,
\end{equation*}%
where integration is performed along the contour $\Gamma ,$ depicted in
Figure \ref{nemaTG}, since%
\begin{equation}
\tilde{G}^{\left( x\right) }\left( r,s\right) \sim \tilde{G}_{\mathrm{asy}%
}^{\left( x\right) }\left( r,s\right) =\frac{1}{4\pi r\kappa ^{2}}\frac{1}{%
s^{\delta }}\mathrm{e}^{-\frac{r}{\kappa }s^{1-\frac{\delta }{2}}},\quad 
\text{for}\quad s\rightarrow \infty ,  \label{Green-srsha-ld}
\end{equation}%
being short-time asymptotics of Green's function in Laplace domain, see also
(\ref{G-f-short-time-infinit}), and originating from the expression of
Green's functions in Laplace domain (\ref{Green-f-ld}) with asymptotics (\ref%
{memory-function-asym}) of memory functions $\tilde{c}_{x}$ and parameters $%
\kappa $ and $\delta $ given in Tables \ref{tbl-1}, \ref{tbl-4}, \ref{tbl-2}%
, and \ref{tbl-3}, has a branch point $s=0,$ due to the term $s^{\delta },$ $%
\delta \in \left( 0,1\right) $.

Therefore, in order to obtain the asymptotic behavior of Green's function in
time domain, the Laplace inversion of Green's function asymptotics (\ref%
{Green-srsha-ld}) is carried out, yielding%
\begin{equation*}
G_{\varepsilon }^{\left( x\right) }\left( r,s\right) \sim G_{\mathrm{asy}%
}^{\left( x\right) }\left( r,t\right) =\frac{1}{2\pi \mathrm{i}}\int_{Br}%
\tilde{G}_{\mathrm{asy}}^{\left( x\right) }\left( r,s\right) \mathrm{e}^{st}%
\mathrm{d}s=\mathcal{L}^{-1}\left[ \frac{1}{4\pi r\kappa ^{2}}\frac{1}{%
s^{\delta }}\mathrm{e}^{-\frac{r}{\kappa }s^{1-\frac{\delta }{2}}}\right]
\left( r,t\right) ,
\end{equation*}%
where integration is performed along the Bromwich contour $Br$, i.e., along
the contour $\Gamma _{0}$ in the limit when $R\rightarrow \infty ,$ having $%
\Gamma _{0}$ as a part of the closed contour $\Gamma $ from\ Figure \ref%
{nemaTG}, so that the Cauchy integral theorem reduces to%
\begin{equation}
2\pi \mathrm{i\,}G_{\mathrm{asy}}^{\left( x\right) }\left( r,t\right) =-\lim 
_{\substack{ R\rightarrow \infty  \\ \check{r}\rightarrow 0}}\left(
I_{\Gamma _{3}}+I_{\Gamma _{5}}\right) ,  \label{racunanje-G-epsilon-asimpt}
\end{equation}%
where $I_{\Gamma _{3}}\ $and $I_{\Gamma _{5}}$ are integrals along contours $%
\Gamma _{3}$ and $\Gamma _{5}\ $having non-zero contribution, since for the
left-hand-side of Cauchy integral theorem one has%
\begin{equation}
\doint\nolimits_{\Gamma }\tilde{G}_{\mathrm{asy}}^{\left( x\right) }\left(
r,s\right) \mathrm{e}^{st}\mathrm{d}s=\sum_{i=0}^{7}I_{\Gamma _{i}},\quad 
\text{with}\quad I_{\Gamma _{i}}=\int_{\Gamma _{i}}\frac{1}{4\pi r\kappa ^{2}%
}\frac{1}{s^{\delta }}\mathrm{e}^{-\frac{r}{\kappa }s^{1-\frac{\delta }{2}}}%
\mathrm{e}^{st}\mathrm{d}s,  \label{int-gama-kompl-asimpt}
\end{equation}%
where the integrals $I_{\Gamma _{1}},$ $I_{\Gamma _{2}},$ $I_{\Gamma _{4}},$ 
$I_{\Gamma _{6}},$ and $I_{\Gamma _{7}}$ have zero contribution in the limit
when $R\rightarrow \infty $ and $\check{r}\rightarrow 0.$

By the use of parametrization given in Table \ref{nemaTG-param}, the
integrals in (\ref{racunanje-G-epsilon-asimpt}) are calculated as 
\begin{align*}
2\pi \mathrm{i\,}G_{\mathrm{asy}}^{\left( x\right) }\left( r,t\right) & =-%
\frac{1}{4\pi r\kappa ^{2}}\Bigg(\int_{\infty }^{0}\frac{1}{\rho ^{\delta }%
\mathrm{e}^{\mathrm{i}\delta \pi }}\mathrm{e}^{-\frac{r}{\kappa }\rho ^{1-%
\frac{\delta }{2}}\mathrm{e}^{\mathrm{i}\left( 1-\frac{\delta }{2}\right)
\pi }}\mathrm{e}^{\rho t\mathrm{e}^{\mathrm{i}\pi }}\mathrm{e}^{\mathrm{i}%
\pi }\mathrm{d}\rho +\int_{0}^{\infty }\frac{1}{\rho ^{\delta }\mathrm{e}^{-%
\mathrm{i}\delta \pi }}\mathrm{e}^{-\frac{r}{\kappa }\rho ^{1-\frac{\delta }{%
2}}\mathrm{e}^{-\mathrm{i}\left( 1-\frac{\delta }{2}\right) \pi }}\mathrm{e}%
^{\rho t\mathrm{e}^{-\mathrm{i}\pi }}\mathrm{e}^{-\mathrm{i}\pi }\mathrm{d}%
\rho \Bigg) \\
& =-\frac{1}{4\pi r\kappa ^{2}}\int_{0}^{\infty }\frac{1}{\rho ^{\delta }}%
\Bigg(\mathrm{e}^{\frac{r}{\kappa }\rho ^{1-\frac{\delta }{2}}\mathrm{e}^{-%
\mathrm{i}\frac{\delta \pi }{2}}}\mathrm{e}^{-\mathrm{i}\delta \pi }-\mathrm{%
e}^{\frac{r}{\kappa }\rho ^{1-\frac{\delta }{2}}\mathrm{e}^{\mathrm{i}\frac{%
\delta \pi }{2}}}\mathrm{e}^{\mathrm{i}\delta \pi }\Bigg)\mathrm{e}^{-\rho t}%
\mathrm{d}\rho \\
& =\frac{1}{4\pi r\kappa ^{2}}\int_{0}^{\infty }\frac{1}{\rho ^{\delta }}%
\mathrm{e}^{\frac{r}{\kappa }\rho ^{1-\frac{\delta }{2}}\mathrm{\cos }\frac{%
\delta \pi }{2}}\mathrm{e}^{-\rho t}\Bigg(\mathrm{e}^{\mathrm{i}\left( \frac{%
r}{\kappa }\rho ^{1-\frac{\delta }{2}}\mathrm{\sin }\frac{\delta \pi }{2}%
+\delta \pi \right) }-\mathrm{e}^{-\mathrm{i}\left( \frac{r}{\kappa }\rho
^{1-\frac{\delta }{2}}\mathrm{\sin }\frac{\delta \pi }{2}+\delta \pi \right)
}\Bigg)\mathrm{d}\rho \\
& =2\mathrm{i}\frac{1}{4\pi r\kappa ^{2}}\int_{0}^{\infty }\frac{1}{\rho
^{\delta }}\mathrm{e}^{-\rho \left( t-\frac{r}{\kappa }\rho ^{-\frac{\delta 
}{2}}\mathrm{\cos }\frac{\delta \pi }{2}\right) }\sin \left( \frac{r}{\kappa 
}\rho ^{1-\frac{\delta }{2}}\mathrm{\sin }\frac{\delta \pi }{2}+\delta \pi
\right) \mathrm{d}\rho ,
\end{align*}%
yielding the expression (\ref{G-short-time}) for short-time asymptotic
behavior of Green's functions $G_{\mathrm{asy}}^{\left( x\right) }$ in the
case when wave propagation speed is infinite. One can prove that the
remaining integrals in (\ref{int-gama-kompl-asimpt})$_{1}$ have zero
contributions as in the case of calculation of Green's function in Section %
\ref{CalcGrin}.

\section{Validity of condition needed for Fourier transform inversion \label%
{Justification}}

In order to obtain Green's function in Laplace domain $\tilde{G}^{\left(
x\right) },$ see (\ref{Green-f-ld}), the inverse Fourier transform can be
applied to Green's function in Fourier and Laplace domain $\bar{\tilde{G}}%
^{\left( x\right) },$ see (\ref{grin-c})$_{2}$ and (\ref{grin-s})$_{2},$
provided that the condition%
\begin{equation}
\func{Re}\frac{s^{2}}{\tilde{c}_{x}^{2}\left( s\right) }>0\quad \text{for}%
\quad \func{Re}s>0,  \label{condition-lambda}
\end{equation}%
is satisfied, as noted in Section \ref{DFRTR}. The condition (\ref%
{condition-lambda}) is equivalent to the request that the function%
\begin{equation}
s^{2}+k^{2}\,\tilde{c}_{x}^{2}\left( s\right) =\tilde{c}_{x}^{2}\left(
s\right) \left( k^{2}+\frac{s^{2}}{\tilde{c}_{x}^{2}\left( s\right) }\right)
\neq 0,\quad \text{for}\quad k\in 
\mathbb{R}
,\;\func{Re}s>0,  \label{condition-lambda-2}
\end{equation}%
since $\tilde{c}_{x}^{2}\left( s\right) =\frac{1}{\varrho }\frac{\Phi
_{\varepsilon }(s)}{\Phi _{\sigma }(s)},$ see (\ref{memory-function}), has
neither zeros nor poles for $\func{Re}s>0,$ while the term in brackets is
non-zero if $\func{Re}\frac{s^{2}}{\tilde{c}_{x}^{2}\left( s\right) }>0$ for 
$\func{Re}s>0.$

The validity of condition (\ref{condition-lambda}), through its equivalent
form (\ref{condition-lambda-2}), is checked in Section 4 of \cite{KOZ19} for
linear fractional-order models having differentiation orders up to one,
given by (\ref{kejsovi}), while condition's validity in the case of
thermodynamically consistent fractional Burgers models, given by (\ref%
{burgersi}), is proved in Appendix A of \cite{OZO}.

In the case of fractional anti-Zener and Zener models, listed in Table \ref%
{tbl-5}, the condition (\ref{condition-lambda-2}) reads%
\begin{equation}
s^{2}+\frac{k^{2}}{\varrho }\,s^{\xi }\frac{\phi _{\varepsilon }\left(
s\right) }{\phi _{\sigma }\left( s\right) }\neq 0,\quad \text{for}\quad k\in 
\mathbb{R}
,\;\func{Re}s>0,  \label{condition-lambda-3}
\end{equation}%
according to (\ref{memory-function-aZ/Z}), with functions $\phi
_{\varepsilon }$ and $\phi _{\sigma }$ listed in Table \ref{skupina}.
Considering the model I$^{{}^{+}}$ID.ID, used in numerical examples in
Section \ref{num-exam} and given by (\ref{model-ifwps}), one finds that
condition (\ref{condition-lambda-3}) becomes%
\begin{equation}
s^{2}+\frac{k^{2}}{\varrho }\,s^{\beta +\nu }\frac{b_{1}+b_{2}s^{\alpha
+\beta }}{a_{1}+a_{2}s^{\alpha +\beta }+a_{3}s^{2\left( \alpha +\beta
\right) }}\neq 0,\quad \text{for}\quad k\in 
\mathbb{R}
,\;\func{Re}s>0,  \label{kondisn-5}
\end{equation}%
so that, using the substitution $s=\rho \mathrm{e}^{\mathrm{i}\varphi },$ $%
\rho >0,$ $\varphi \in \left[ -\frac{\pi }{2},\frac{\pi }{2}\right] ,$ and
by separating real and imaginary parts, condition's imaginary part takes the
form%
\begin{equation}
F\left( \rho ,\varphi \right) =\rho ^{2}\sin \left( 2\varphi \right) +\frac{%
k^{2}}{\varrho }\,\frac{\rho ^{\beta +\nu }}{\left\vert a_{1}+a_{2}s^{\alpha
+\beta }+a_{3}s^{2\left( \alpha +\beta \right) }\right\vert ^{2}}f\left(
\rho ,\varphi \right) \neq 0,  \label{kondisn-4}
\end{equation}%
with%
\begin{align}
f\left( \rho ,\varphi \right) & =a_{1}b_{1}\sin \left( \left( \beta +\nu
\right) \varphi \right) +a_{1}b_{2}\rho ^{\alpha +\beta }\sin \left( \left(
\alpha +2\beta +\nu \right) \varphi \right)   \notag \\
& \quad +a_{2}b_{1}\rho ^{\alpha +\beta }\sin \left( \left( \nu -\alpha
\right) \varphi \right) +a_{2}b_{2}\rho ^{2\left( \alpha +\beta \right)
}\sin \left( \left( \beta +\nu \right) \varphi \right)   \notag \\
& \quad -a_{3}b_{1}\rho ^{2\left( \alpha +\beta \right) }\sin \left( \left(
2\alpha +\beta -\nu \right) \varphi \right) +a_{3}b_{2}\rho ^{3\left( \alpha
+\beta \right) }\sin \left( \left( \nu -\alpha \right) \varphi \right) .
\label{f-I+ID.ID}
\end{align}%
Note, if $\varphi \rightarrow -\varphi ,$ then function $F,$ defined in (\ref%
{kondisn-4}), satisfies $F\left( \rho ,\varphi \right) =-F\left( \rho
,\varphi \right) $ and therefore it is sufficient to consider the interval $%
\varphi \in \left[ 0,\frac{\pi }{2}\right] .$ Moreover, if $\varphi =0$,
then $s=\rho $ and hence 
\begin{equation*}
\rho ^{2}+\frac{k^{2}}{\varrho }\,\rho ^{\beta +\nu }\frac{b_{1}+b_{2}\rho
^{\alpha +\beta }}{a_{1}+a_{2}\rho ^{\alpha +\beta }+a_{3}\rho ^{2\left(
\alpha +\beta \right) }}>0,\quad \text{for}\quad \rho >0,
\end{equation*}%
by (\ref{kondisn-5}), so that one considers the interval $\varphi \in \left(
0,\frac{\pi }{2}\right] $. 

If function $f\left( \rho ,\varphi \right) >0$ for $\rho >0$ and $\varphi
\in \left( 0,\frac{\pi }{2}\right] ,$ then one has that function $F\left(
\rho ,\varphi \right) >0$ and hence different from zero, since all other
terms in $F$ are positive. Rewriting the function $f,$ given by (\ref%
{f-I+ID.ID}), as%
\begin{align*}
f\left( \rho ,\varphi \right) & =a_{1}b_{1}\sin \left( \left( \beta +\nu
\right) \varphi \right) +a_{1}b_{2}\rho ^{\alpha +\beta }\sin \left( \left(
\alpha +2\beta +\nu \right) \varphi \right)  \\
& \quad +a_{2}b_{1}\rho ^{\alpha +\beta }\sin \left( \left( \nu -\alpha
\right) \varphi \right) +a_{3}b_{2}\rho ^{3\left( \alpha +\beta \right)
}\sin \left( \left( \nu -\alpha \right) \varphi \right)  \\
& \quad +a_{3}b_{2}\rho ^{2\left( \alpha +\beta \right) }\sin \left( \left(
2\alpha +\beta -\nu \right) \varphi \right) \left( \frac{a_{2}}{a_{3}}\frac{%
\sin \left( \left( \beta +\nu \right) \varphi \right) }{\sin \left( \left(
2\alpha +\beta -\nu \right) \varphi \right) }-\frac{b_{1}}{b_{2}}\right) 
\end{align*}%
one has that $f\left( \rho ,\varphi \right) >0$ for $\rho >0$ and $\varphi
\in \left( 0,\frac{\pi }{2}\right] $ if 
\begin{equation}
\frac{b_{1}}{b_{2}}< \frac{a_{2}}{a_{3}}\frac{\sin \left( \left(
\beta +\nu \right) \varphi \right) }{\sin \left( \left( 2\alpha +\beta -\nu
\right) \varphi \right) },  \label{kondisn-6}
\end{equation}%
since all other terms are non-negative according to the thermodynamical
restrictions (\ref{tdr-ifwps-1}). In order to prove (\ref{kondisn-6}),
becoming%
\begin{equation*}
\frac{b_{1}}{b_{2}}\leqslant \frac{a_{2}}{a_{3}}\frac{\sin \frac{\left(
\beta +\nu \right) \pi }{2}}{\sin \frac{\left( 2\alpha +\beta -\nu \right)
\pi }{2}}< \frac{a_{2}}{a_{3}}\frac{\sin \left( \left( \beta +\nu
\right) \varphi \right) }{\sin \left( \left( 2\alpha +\beta -\nu \right)
\varphi \right) },
\end{equation*}%
due to the thermodynamical restrictions (\ref{tdr-ifwps-3}) and property of function $g$. Namely, one considers
function $g$ and its first derivative $g^{\prime }$:%
\begin{equation*}
g\left( \varphi \right) =\frac{\sin \left( \zeta \varphi \right) }{\sin
\left( \xi \varphi \right) }>0\quad \text{and}\quad g^{\prime }\left(
\varphi \right) =\frac{\xi \varphi \,\zeta \varphi \,\cos \left( \xi \varphi
\right) \cos \left( \zeta \varphi \right) }{\varphi \sin ^{2}\left( \xi
\varphi \right) }\left( \frac{\tan \left( \xi \varphi \right) }{\xi \varphi }%
-\frac{\tan \left( \zeta \varphi \right) }{\zeta \varphi }\right) <0,
\end{equation*}%
with $0<\xi <\zeta <1$ on the interval $\varphi \in \left( 0,\frac{\pi }{2}%
\right]$, for which one has 
\begin{equation*}
g\left( \varphi \right) >g\left( \frac{\pi }{2}\right) ,\quad \text{for}%
\quad \varphi \in \left( 0,\frac{\pi }{2}\right] ,
\end{equation*}%
since function $g\left( \varphi \right) $ is a decreasing function for $%
\varphi \in \left( 0,\frac{\pi }{2}\right] ,$ due to the function $\frac{%
\tan x}{x}$ being monotonically increasing for $x\in \left( 0,\frac{\pi }{2}%
\right] .$ 

Note, the same procedure can be followed for proving the condition (\ref%
{condition-lambda-2}) in the case of all other fractional anti-Zener and
Zener models, listed in Table \ref{tbl-5}.

\bigskip

%
%
%

\noindent \textbf{Acknowledgements} \smallskip

\noindent The work is supported by the Ministry of Science, Technological
Development and Innovation of the Republic of Serbia under grant
451-03-47/2023-01/200125 (SJ and DZ).


\end{document}